\documentclass[11pt,a4paper]{article}
\pdfoutput=1
\usepackage{jheppub}
\usepackage[vcentermath]{youngtab}
\usepackage{ytableau}
\usepackage{amsfonts,amsmath,amssymb,mathalfa,mathtools,tabularx,multirow,array,color, bbold, physics, empheq, xcolor}
\usepackage{caption}
\usepackage{subcaption}

\usepackage{tikz}
\usetikzlibrary{decorations.pathreplacing}
\newcommand{\tikznode}[3][inner sep=0pt]{\tikz[remember
picture,baseline=(#2.base)]{\node(#2)[#1]{$#3$};}}

\newcommand{\sym}[1]{\ytableausetup{centertableaux, boxsize=1.1em}
\underbrace{\begin{ytableau}
~ & ~ & ~ & \none[\ldots] & ~ \end{ytableau}}_{#1}}

\newcommand{\symC}[1]{\ytableausetup{centertableaux, boxsize=1.1em}
\underbrace{\begin{ytableau}
\bullet & \bullet & \bullet & \none[\dots] & \bullet \end{ytableau}}_{#1}}

\newcommand{\mixed}[2]{\ytableausetup{centertableaux, boxsize=1.1em} \begin{ytableau}
\tikznode{a1}{} \bullet & \bullet & \none[\ldots] & \tikznode{a2}{} \bullet & \tikznode{a3}{} ~ & ~ & \none[\ldots] & \tikznode{a4}{} ~ \end{ytableau} 
\tikz[overlay,remember picture]{%
\draw[decorate,decoration={brace},thick] ([yshift=-2mm,xshift=2.7mm]a2.south east) -- 
([yshift=-2mm,xshift=-1mm]a1.south west) node[midway,below]{\small{$#1$}};
\draw[decorate,decoration={brace},thick] ([yshift=-2mm,xshift=2.7mm]a4.south east) -- 
([yshift=-2mm,xshift=-1mm]a3.south west) node[midway,below]{\small{$#2$}};
}}

\title{\boldmath Entanglement on multiple $S^2$ boundaries in Chern-Simons theory}

\author[a]{Siddharth Dwivedi}
\author[b]{, Vivek Kumar Singh}
\author[c]{, P. Ramadevi}
\author[d]{, Yang Zhou}
\author[c]{, Saswati Dhara}

\affiliation[a]{Center for Theoretical Physics, College of Physical Science and Technology, Sichuan University,\\ Chengdu, 610064, China}
\affiliation[b]{Faculty of Physics, University of Warsaw, ul. Pasteura 5, 02-093 Warsaw, Poland}
\affiliation[c]{Department of Physics, Indian Institute of Technology Bombay, Mumbai 400076, India}
\affiliation[d]{Department of Physics and Center for Field Theory and Particle Physics, Fudan University,\\ Shanghai 200433, China}

\emailAdd{sdwivedi@scu.edu.cn}
\emailAdd{ vivek.singh@fuw.edu.pl}
\emailAdd{ramadevi@phy.iitb.ac.in}
\emailAdd{yang\_zhou@fudan.edu.cn}
\emailAdd{saswati123@phy.iitb.ac.in}

\abstract{Topological entanglement structure amongst disjoint torus boundaries of three
manifolds have already been studied within the context of Chern-Simons theory. In this work,
we study the topological entanglement due to interaction between the quasiparticles inside three-manifolds with one or more disjoint $S^2$ boundaries in SU($N$) Chern-Simons theory. We
focus on the world-lines of quasiparticles (Wilson lines), carrying SU($N$) representations,
creating four punctures on every $S^2$. We compute the entanglement entropy by
partial tracing some of the boundaries. In fact, the entanglement entropy depends on
the
 SU($N$) representations on these four-punctured $S^2$ boundaries. Further, we observe
interesting features on the GHZ-like and W-like entanglement structures. Such a distinction
crucially depends on the multiplicity of the irreducible representations in the tensor product of SU($N$) representations.
}

\begin{document} 
\maketitle

\section{Introduction}
\label{sec1}
\emph{Quantum entanglement} \cite{Horodecki:2009zz} is one of the important fundamental aspect of quantum physics and is an active area of research. It is a physical phenomenon which occurs when two particles or systems interact in a way such that the state of one cannot be described independently of the state of the other, even when they are separated by a large distance. Such systems are called as entangled systems. The concept of entanglement has many important applications in various areas such as in quantum information theory, quantum computing \cite{2003RSPSA.459.2011J}, quantum teleportation \cite{1997Natur.390..575B} etc. It can be understood in a formal framework as following: Consider two systems or parties Alice ($A$) and Bob ($B$) with corresponding Hilbert spaces $\mathcal{H}_A$ and $\mathcal{H}_B$ spanned by the basis vectors $|e_i^{A}\rangle$ and $|e_j^{B}\rangle$ respectively. The Hilbert space of the composite system is the direct product space $\mathcal{H}_A \otimes \mathcal{H}_B$. A normalized state $|\Psi \rangle \in \mathcal{H}_A \otimes \mathcal{H}_B$ can be written as: 
\begin{equation}
|\Psi\rangle = \sum_{i, j}c_{ij} |e_i^{A}\rangle \otimes |e_j^{B}\rangle \equiv \sum_{i, j}c_{ij} |e_i^{A}, e_j^{B}\rangle ~.
\label{AB-state}
\end{equation} 
If we can write the coefficient $c_{ij} = c_i^A c_j^B$, then $|\Psi\rangle$ will be a tensor product of the states describing $A$ and $B$. These are called separable or non-entangled state. The quantum states that can not be separated in this fashion are precisely the entangled states. One of the famous measure of the entanglement between the systems is the \emph{von Neumann entropy}, also known as the \emph{entanglement entropy}. In order to compute this, we associate a projection operator called as \emph{density matrix} to the state $|\Psi\rangle$ given as $\rho = |\Psi\rangle \langle{\Psi}|$. Tracing out the Hilbert space of one system (say $B$) gives a reduced density matrix acting on $\mathcal{H}_A$:
\begin{equation}
\rho_{A} = \text{Tr}_B (\rho) = \sum_{i}\langle e_i^B| \rho |e_i^B \rangle ~.
\end{equation}
The entanglement entropy can be calculated from the reduced density matrix as,
\begin{equation}
\text{EE} = -\text{Tr}(\rho_{A} \ln \rho_{A}) = -\sum_{i} \lambda_i \ln \lambda_i ~,
\end{equation} 
where $\lambda_i$ are the eigenvalues of $\rho_A$.

An important question in information theory is to understand the possible patterns of entanglement and to classify the states into different entanglement classes. For example, in two qubit system with Hilbert space $\mathbb{C}^2 \otimes \mathbb{C}^2$, there is a class of states called `Bell states' defined up to SLOCC (stochastic local operations and classical communication)  which are maximally entangled:
\begin{equation}
|\Psi_{\text{Bell}}\rangle = \frac{1}{\sqrt{2}}\left( |0^A, 0^B\rangle + |1^A, 1^B\rangle \right) 
\label{Bell-state} ~.
\end{equation}
The entanglement entropy for Bell state can be calculated as EE$_{\text{Bell}} = \ln 2$ which is the maximum possible entanglement entropy for the reduced density matrix defined on two dimensional Hilbert space. Similarly, in three qubit system $\mathbb{C}^2 \otimes \mathbb{C}^2 \otimes \mathbb{C}^2$, there are two non-trivial classes of tri-partite entanglement \cite{2000PhRvA..62f2314D}: the `GHZ' state (where partial trace over one of the system results in a separable reduced density matrix on the other two systems) and the `W' state (where a partial trace still leaves a non-separable reduced density matrix on the other two systems). A mixed state defined by a density matrix $\rho$ acting on $\mathcal{H}_A \otimes \mathcal{H}_B$ is said to be separable if it can be decomposed as a convex combination of pure states:
\begin{equation}
\rho = \sum_i \alpha_i |\Psi_i^{A}\rangle \langle {\Psi_i^{A}} | \otimes |{\Phi_i^{B}}\rangle  \langle {\Phi_i^{B}}| = \sum_i \alpha_i\, \rho_i^A \otimes \rho_i^B ~,
\end{equation}
where $\alpha_i > 0$ and $\sum_i \alpha_i = 1$. But deciding whether such a decomposition exists is a NP hard problem \cite{gurvits2003classical}. However there are other separability criteria available which provide necessary but not sufficient condition for separability. One widely used concept is the positive partial transpose (PPT) criterion \cite{Peres:1996dw} which says that if $\rho$ is separable then its partial transpose $\rho^{\Gamma_B}$ (with respect to, say B) must be positive semidefinite. The partial transpose can be obtained from the density matrix as following:
\begin{equation}
\langle e_i^A, e_j^B | \rho^{\Gamma_B} |e_k^A, e_l^B \rangle = \langle e_i^A, e_l^B | \rho |e_k^A, e_j^B \rangle ~.
\label{par-transpose}
\end{equation}
If $\rho^{\Gamma_B}$ is not positive semidefinite, it necessarily implies that the density matrix $\rho$ is non-separable (i.e. entangled). To capture this information, we define \emph{entanglement negativity} $\mathcal{N}$ as,
\begin{equation}
\mathcal{N} = \frac{||\rho^{\Gamma}||-1}{2} = \sum_{i} \frac{\abs{\mu_i} - \mu_i}{2} ~,
\label{negativity}
\end{equation} 
where $\mu_i$ are the eigenvalues of $\rho^{\Gamma}$. A non-zero value of $\mathcal{N}$ means that $\rho^{\Gamma_B}$ has negative eigenvalue and is not positive semidefinite. Thus $\mathcal{N} \neq 0$ implies that $\rho$ is entangled.

So far, we reviewed the entanglement entropy for any quantum mechanical system  with finite degrees of freedom. Analyzing the entanglement structures in a quantum field theory is a difficult task due to large number of degrees of freedom. However, there is a class of exactly solvable quantum field theories called `topological quantum field theories' where entanglement structures can be examined. In particular, there has been a lot of interest to study entanglement in three dimensional Chern-Simons theory. 

In \cite{Witten:1988hf}, the SU($N$) Chern-Simons functional integral on a three-dimensional manifold with a $S^2$ boundary is given by a state in the Hilbert space $\mathcal{H}$ spanned by $n$-point correlator conformal blocks of SU$(N)_k$ Wess-Zumino conformal field theory on $S^2$ with $n$ punctures. The Chern-Simons coupling constant $k$ denotes the level of the SU$(N)_k$ Wess-Zumino conformal field theory. The Wilson lines can be viewed as the world line of quasiparticles creating punctures on the $S^2$ boundary. Using the correspondence of Chern-Simons theory with the conformal field theory, we associate a primary field $\phi_{R_i} \equiv R_i$ with every puncture where $\phi_{R_i}$ transform under the representation $R_i$ of SU($N$). The fusion rules of these primary fields determine the dimensionality of the space of correlator conformal blocks.

\begin{figure}[h]
\centerline{\includegraphics[width=4.4in]{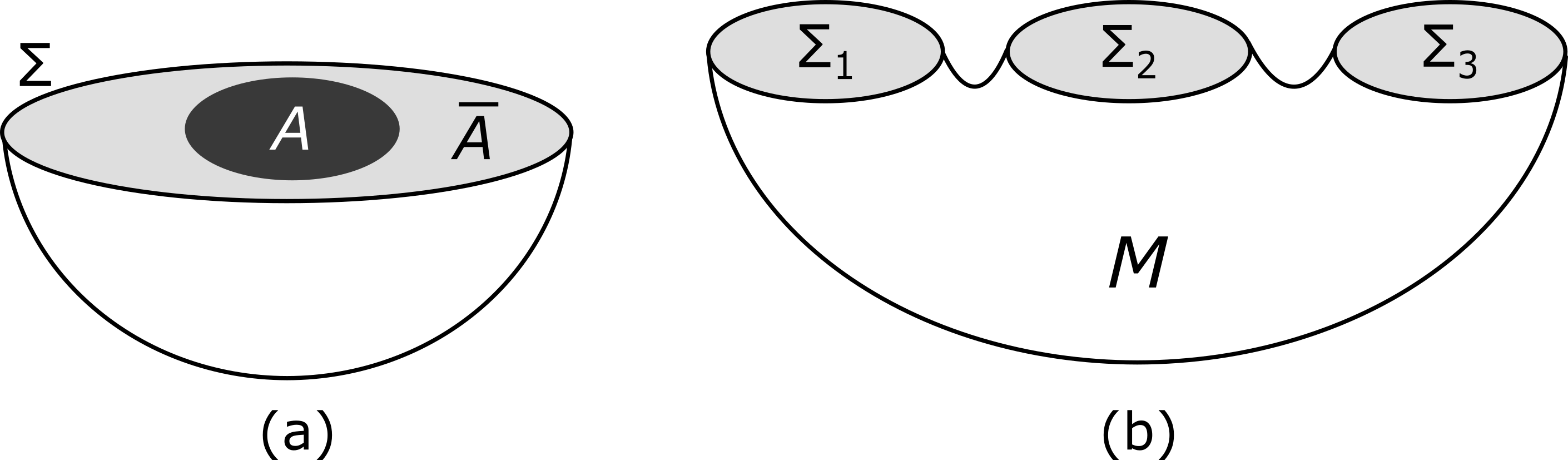}}
\caption[]{Figure (a) shows a manifold having a single boundary $\Sigma$ which is partitioned into a sub-region $A$ and its complement $\bar{A}$. Figure (b) shows the manifold $M$ whose boundary has three disconnected components, i.e. $\Sigma_1 \cup \Sigma_2 \cup \Sigma_3$.}
\label{Multi-boundary}
\end{figure}

In a typical set-up to study entanglement, one considers a boundary $\Sigma$ with a Hilbert space $\mathcal{H}_{\Sigma}$ associated with it. The $\Sigma$ is then partitioned into region $A$ and its complement $\bar{A}$ as shown in figure \ref{Multi-boundary}(a). If the Hilbert space can be factorized as $\mathcal{H}_{\Sigma} = \mathcal{H}_{A} \otimes \mathcal{H}_{\bar{A}}$, then by tracing over one of the factors and computing the reduced density matrix, one can obtain the entanglement structure. The entanglement between connected spatial regions in Chern-Simons theory was studied in \cite{Levin:2006zz, Dong:2008ft, Kitaev:2005dm}.
Another set-up is to consider the quantum field theory on a manifold whose boundary consists of disconnected components, i.e. $\Sigma = \Sigma_1 \cup \Sigma_2 \ldots \cup \Sigma_n$ like the one shown in figure \ref{Multi-boundary}(b). Thus the Hilbert space is the tensor product of Hilbert spaces associated with each component:
\begin{equation}
\mathcal{H}_{\Sigma} = \mathcal{H}_{\Sigma_1} \otimes \mathcal{H}_{\Sigma_2} \otimes \ldots \otimes \mathcal{H}_{\Sigma_n} ~.
\end{equation}
One can then study the entanglement structure by tracing out one or more boundary components which is usually referred as \emph{multi-boundary entanglement}. 

The multi-boundary entanglement in the context of 1+1 dimensional conformal field theory where $\Sigma$ was disjoint union of $n$ circles ($\Sigma = S^1 \cup S^1 \cup \ldots \cup S^1$) was investigated in \cite{Balasubramanian:2014hda, Marolf:2015vma}. Recently, the study of entanglement for the states on multiple copies of $T^2$ (i.e. $\Sigma = T^2 \cup T^2 \cup \ldots \cup T^2$) was initiated in \cite{Balasubramanian:2016sro, Salton:2016qpp}. The quantum states in these works are obtained by performing the path integral of Chern-Simons theory on link complements with $n$ torus boundaries. This set-up shows some interesting features like periodic behavior of the R\'enyi entropy for torus link states \cite{Dwivedi:2017rnj}. Further the entanglement structure of torus/hyperbolic link states is similar as that of a GHZ/W state as reflected from the partial tracing giving separable/entangled reduced density matrix \cite{Balasubramanian:2018por}. We refer the readers for other related works in this set-up \cite{Chun:2017hja,Hung:2018rhg,Schnitzer:2019icr,Camilo:2019bbl}. 

The entanglement between two $S^2$ boundaries related by cobordism \cite{Melnikov:2018zfn} gives a nice pictorial understanding on the entanglement entropy. The qualitative remarks given in \cite{Melnikov:2018zfn} on the entanglement in the presence of the Wilson lines in Chern-Simons theory creating punctured $S^2$ boundaries needs to be quantified.

\begin{figure}[h]
\centerline{\includegraphics[width=5.0in]{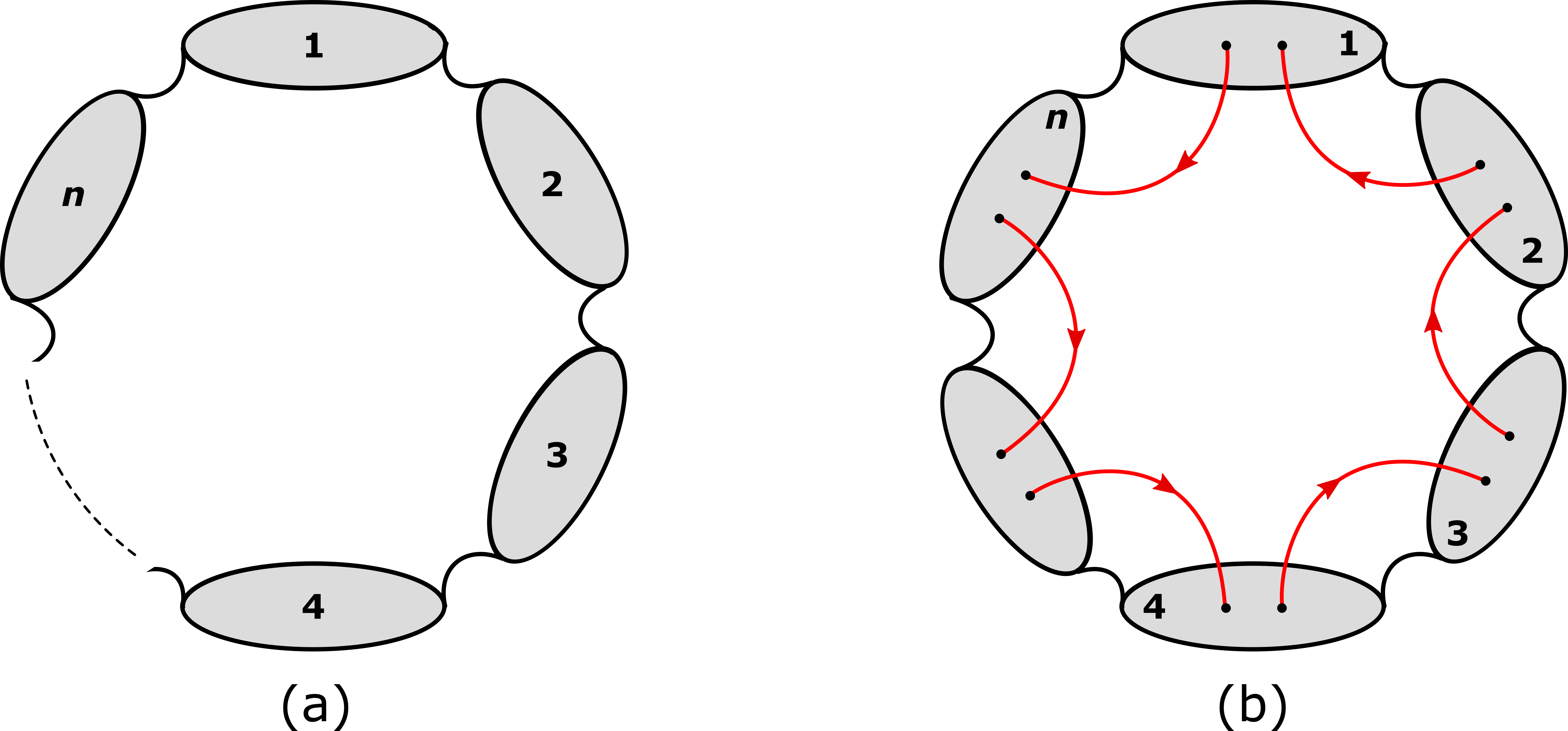}}
\caption[]{Manifolds whose boundary consists of $n$ disjoint $S^2$ components are shown in figures (a) and (b). In figure $(b)$, each $S^2$ has some punctures with Wilson lines stretching from one puncture to other.}
\label{multi-S2}
\end{figure}

In the present work, we will extend the idea of multi-boundary entanglement to study the structure of the states prepared on $n$ copies of $S^2$, i.e. we consider a manifold $M$, obtained by slicing $S^3$, such that the boundary consists of $n$ number of disconnected $S^2$ components (shown in figure \ref{multi-S2}(a)):
\begin{equation}
\Sigma = S^2 \cup S^2 \cup \ldots \cup S^2 = \cup_{i=1}^n S^2  ~.
\end{equation}
The Hilbert space associated with $\Sigma$ can thus be given as the tensor product of Hilbert spaces associated with each $S^2$:
\begin{equation}
\mathcal{H}_{\Sigma} = \mathcal{H}_{S^2} \otimes \mathcal{H}_{S^2} \otimes \ldots \otimes \mathcal{H}_{S^2} ~.
\end{equation}
It is also possible to have Wilson lines in $M$ which intersect the $S^2$ boundaries creating marked points or punctures on $S^2$. The primary fields of SU$(N)_k$ Wess-Zumino conformal field theory transforming in the representation $R_i$ are assigned to each of these marked points. One typical example is shown in figure \ref{multi-S2}(b). In such cases, the physical Hilbert space is the space of conformal blocks with the primary fields inserted at the marked points \cite{Witten:1988hf}. In the case of four punctured $S^2$ with representations $R_1, R_2, R_3, R_4$, the Hilbert space is the SU($N$)-invariant subspace:
\begin{equation}
\mathcal{H}_{S^2} = \text{Inv}(R_1 \otimes R_2 \otimes R_3 \otimes R_4) ~.
\end{equation}
In order to have a non-trivial space, the $R_i$ should correspond to the integrable representations of the affine extension of SU$(N)$ i.e. SU$(N)_k$ where $k$ denotes the level. Thus these Hilbert spaces are finite dimensional spaces and are given by the number of times the trivial representation of SU$(N)$ appears in the tensor decomposition of $R_1 \otimes R_2 \otimes R_3 \otimes R_4$. The quantum states $|{\Psi}\rangle  \in \mathcal{H}_{\Sigma}$ can be constructed from the so called `basic building blocks' by appropriately applying braiding and gluing techniques \cite{Nawata:2013qpa, Gu:2014nva}. We shall analyze some of these states in this work which are given in terms of the intermediate representations $t$ taking values in $(R_1 \otimes R_2) \cap (\bar{R}_3 \otimes \bar{R}_4)$ (more details are given in section \ref{sec2}). Some of the representations $t$ can appear more than once and the number of times they appear denotes their multiplicity. We will show that the multiplicity of such intermediate representations plays a very crucial role in determining the entanglement structure of the states. When each $t$ appears once (multiplicity-free case), the states are GHZ-like and the reduced density matrix is separable. Further if any of the $t$ appears with multiplicity, the reduced density matrix is entangled and the states will have a W-like structure.

As remarked earlier, the authors of \cite{Melnikov:2018zfn} have studied the entanglement structure on two disjoint $S^2 \cup S^2$ boundaries. Basically, they invoke the replica trick to obtain the entropy for the states in $\mathcal{H}_{S^2} \otimes \mathcal{H}_{S^2}$. They show that the states in which the Wilson lines stretch from punctures on one $S^2$ to the  punctures on other $S^2$ without braiding (as shown in figure \ref{Results}(a)) are maximally entangled states and are analogous to the Bell states. Any braiding between the Wilson lines in these states, like the one in figure \ref{Results}(b), will not affect the entropy. If the Wilson lines only stretch locally as in figure \ref{Results}(c), the state is a product (unentangled) state. However if the Wilson lines connecting the punctures on one boundary undergo non-
trivial braiding with the Wilson lines connecting the punctures on other boundary as shown in figure \ref{Results}(d), the state can be an entangled state. It would be interesting to study the effect of such braiding (which are referred to as necklaces in \cite{Melnikov:2018zfn}) on the entropy. In this work, we will systematically perform the calculations of the entanglement entropy by explicitly writing the pure quantum state $|{\Psi}\rangle $ on $n$ copies of $S^2$. The total Hilbert space can be bi-partitioned into two parts as:
\begin{figure}[h]
\centerline{\includegraphics[width=5.0in]{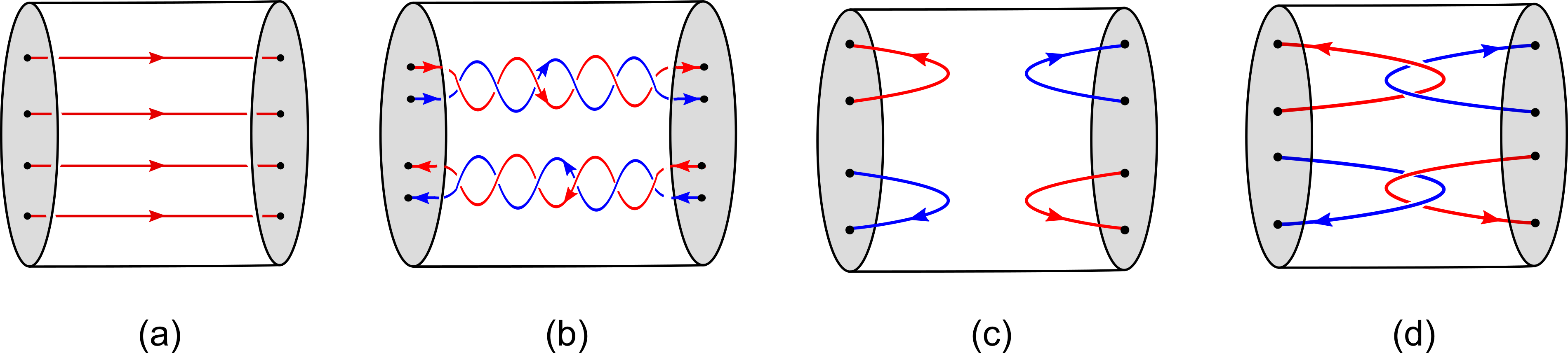}}
\caption[]{Some of the entanglement results presented in \cite{Melnikov:2018zfn} using the replica trick. States (a) and (b) are maximally entangled and hence the Bell states. State (c) is non-entangled or a product state while state (d) is non-trivially entangled state.}
\label{Results}
\end{figure}
\begin{equation}
\mathcal{H}_{\Sigma} = \mathcal{H}_A \otimes \mathcal{H}_B = \left(\bigotimes_{i=1}^{m} \mathcal{H}_i(S^2)\right) \, \, \otimes \, \, \left(\bigotimes_{i=m+1}^{n} \mathcal{H}_i(S^2)\right) ~.
\label{parts-HS}
\end{equation}
We will denote this partition as $(\mathcal{H}_A|\mathcal{H}_B)$ or $(m|n-m)$. By tracing out $\mathcal{H}_B$, we compute the reduced density matrix and the entanglement entropy.

The paper is organized as follows. In section \ref{sec2}, we will explain the set-up including a brief review of the Hilbert space and the basis of conformal blocks. In section \ref{sec3}, we consider the states which are a $n$-boundary generalization of the state shown in figure \ref{Results}(a). We compute the entropy and study their large $k$ and large $N$ behavior and give the separability criterion of the reduced density matrices. In section \ref{sec4}, we will consider the states in which the Wilson lines undergo horizontal braiding, which is a generalization of the state in figure \ref{Results}(d) to $n$ boundaries. We will show that its entanglement structure shows a periodic behavior as we increase the braiding between the Wilson lines. We give explicit calculation of the entropy and negativity using various examples. We conclude and discuss some future prospects in section \ref{sec5}.
\section{Set-up}
\label{sec2}
\begin{figure}[h]
\centerline{\includegraphics[width=5.8in]{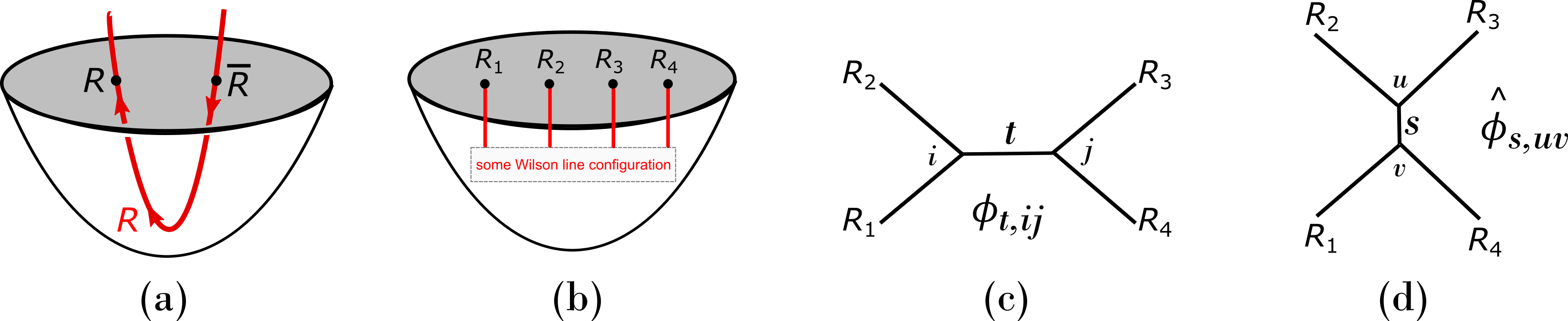}}
\caption[]{A Wilson line carrying representation $R$ intersects $S^2$ creating punctures marked with $R$ or $\bar{R}$ depending on whether Wilson line is going out or going into the manifold. For four punctured boundary shown in figure (b), there are two types of basis of Hilbert space denoted as $|\phi_{t,ij}\rangle$ and $|\hat{\phi}_{s,uv}\rangle$ respectively which depend on the two ways of fusing various representations as shown in (c) and (d).}
\label{S2-Basis}
\end{figure}

The canonical quantization of the Chern-Simons theory on a three-manifold $M$ with boundary $\Sigma$ associates a Hilbert space $\mathcal{H}_{\Sigma}$ to the boundary. If there are Wilson lines embedded in $M$ which intersect $\Sigma$ creating marked points or punctures on it, the Hilbert space is the space of conformal blocks with primary fields (transforming in the representations of the gauge group) inserted at these punctures. The usual convention followed is the following. Consider the Wilson line carrying representation $R$ which intersects $\Sigma$ at point $P$. We associate the representation $R$ or $\bar{R}$ to the puncture $P$, depending upon whether the Wilson line is going out or going into the manifold as shown in figure \ref{S2-Basis}(a). Since in the current set-up we are only working when $\Sigma$ is a Riemann sphere with four punctures, we will briefly explain the Hilbert space associated with it.
\subsection{Hilbert space associated with Riemann sphere and the conformal basis}
Consider $S^2$ with four punctures with four primary fields inserted at these points. Suppose they transform under the representations $R_1, R_2, R_3, R_4$ respectively of the gauge group $G=\text{SU($N$)}$ as shown in the figure \ref{S2-Basis}(b). Thus the Hilbert space is simply given as the invariant subspace of SU($N$):
\begin{equation}
\mathcal{H}_{S^2} = \text{Inv}( R_1 \otimes R_2 \otimes R_3 \otimes R_4 ) ~.
\end{equation} 
This Hilbert space is isomorphic to the space of four-point conformal blocks. The latter has two types of basis as shown in figure \ref{S2-Basis}(c) and \ref{S2-Basis}(d), usually called as $t$-channel and $s$-channel respectively. The basis states are labeled by the intermediate representations which can be obtained as given in the following equation:
\begin{equation}
\text{$t$-channel} =\, \underbrace{R_1 \otimes R_2}_{t} \, \otimes \, \underbrace{R_3 \otimes R_4}_{t'} \quad;\quad \text{$s$-channel} =\, \underbrace{R_2 \otimes R_3}_{s} \, \otimes \, \underbrace{R_1 \otimes R_4}_{s'} ~.
\label{fusion-channel}
\end{equation} 
In $t$-channel, we first fuse $R_1$ and $R_2$ which gives intermediate representations $t \in (R_1 \otimes R_2)$. Further $R_3$ and $R_4$ are fused which gives $t' \in (R_3 \otimes R_4)$. Now since we want $R_1 \otimes R_2 \otimes R_3 \otimes R_4$ to give the trivial representation of SU($N$), we must have $t' = \bar{t}$ otherwise the Hilbert space will be zero dimensional. Also since $t' \in (R_3 \otimes R_4)$ is equivalent to $\bar{t'} \in (\bar{R}_3 \otimes \bar{R}_4)$, the Hilbert space can be given as
\begin{equation}
\mathcal{H}_{S^2} = \{t \mid t \in (R_1 \otimes R_2) \cap (\bar{R}_3 \otimes \bar{R}_4) \}
~.
\end{equation} 
This is true when $t$ appears in $(R_1 \otimes R_2)$ and $(\bar{R}_3 \otimes \bar{R}_4)$ only once. But we can also have $t$ appearing multiple times in these fusion and this will give many more possibilities of getting the trivial representation. For a generic case, the decomposition is given as, 
\begin{equation}
R_{1} \otimes R_{2} = \bigoplus_{t} (N_{R_{1} R_{2}}^{t}) \,t \quad;\quad \bar{R}_3 \otimes \bar{R}_4 = \bigoplus_{t} (N_{\bar{R}_3 \bar{R}_4}^{t}) \,t ~,
\end{equation}
where $N_{ab}^{c}$ is the Verlinde coefficient which tells how many times $c$ appears in the tensor decomposition of $a$ and $b$. Thus following \cite{Gu:2014nva}, we will use extra labels $i$ and $j$ to keep track of these multiple occurrences of $t$ where $i$ and $j$ take values $1,2,\ldots,N_{R_{1} R_{2}}^{t}$ and $1,2,\ldots,N_{\bar{R}_3 \bar{R}_4}^{t}$ respectively. Thus the conformal block basis of the Hilbert space can be given as:
\begin{equation}
\text{basis}(\mathcal{H}_{S^2}) = \{\, |{\phi_{t,ij}}\rangle  \,\mid \, t \in (R_1 \otimes R_2) \cap (\bar{R}_3 \otimes \bar{R}_4) \,,\, 1 \leq i \leq N_{R_{1} R_{2}}^{t}\,,\, 1 \leq j \leq N_{\bar{R}_3 \bar{R}_4}^{t} \, \} ~.
\end{equation}
For example, if the representation $t_0$ appears twice in $R_1 \otimes R_2$ and thrice in $\bar{R}_3 \otimes \bar{R}_4$, there will be 6 basis states corresponding to $t_0$ given as $|{\phi_{t_0,11}}\rangle ,|{\phi_{t_0,12}}\rangle , |{\phi_{t_0,13}}\rangle , |{\phi_{t_0,21}}\rangle , |{\phi_{t_0,22}}\rangle , |{\phi_{t_0,23}}\rangle $. In general, the four punctured $S^2$ boundary like the one shown in figure \ref{S2-Basis}(b) has the Hilbert space with the following dimension:
\begin{equation}
\left(R_{1} \otimes R_{2} = \bigoplus_{\ell=1}^{\textsf{d}} N_{R_{1} R_{2}}^{t_\ell} \,t_\ell \right) \,\, \& \,\, \left(\bar{R}_3 \otimes \bar{R}_4 = \bigoplus_{\ell=1}^{\textsf{d}} N_{\bar{R}_3 \bar{R}_4}^{t_\ell} \,t_\ell \right) \Longrightarrow \text{dim}(\mathcal{H}_{S^2}) = \sum_{\ell=1}^{\textsf{d}} N_{R_{1} R_{2}}^{t_\ell} N_{\bar{R}_3 \bar{R}_4}^{t_\ell}  ~.
\label{HS-Dimension}
\end{equation}
The other conformal block basis is the $s$-channel of figure \ref{S2-Basis}(d) and is given as, 
\begin{equation}
\text{basis}(\mathcal{H}_{S^2}) = \{\, |{\hat{\phi}_{s,uv}}\rangle  \,\mid \, s \in (R_2 \otimes R_3) \cap (\bar{R}_1 \otimes \bar{R}_4) \,,\, 1 \leq u \leq N_{R_{2} R_{3}}^{s}\,,\, 1 \leq v \leq N_{\bar{R}_1 \bar{R}_4}^{s} \, \} ~,
\end{equation} 
where $u$ and $v$ are the labels to keep track of multiplicities of $s$ in $(R_2 \otimes R_3)$ and $(\bar{R}_1 \otimes \bar{R}_4)$ respectively. If the representation occurs once in a tensor product, we will omit the corresponding multiplicity label. Note that we need to make sure that $R_1, R_2, R_3, R_4$ must be the integrable representations of SU$(N)_k$ for a non-trivial Hilbert space.\footnote{See Appendix of \cite{Dwivedi:2017rnj} for a brief review of integrable representations. A representation of SU$(N)$ is integrable if it satisfies $l_1 \leq k$, where $l_1$ is the number of boxes in the first row of its Young tableau.} Moreover the representations $t$ and $s$ in the above discussion should also be integrable to get the full Hilbert space.

The two types of conformal basis given above are orthonormal, i.e.
\begin{equation}
\langle \phi_{t_1,\,i_1 j_1} | \phi_{t_2,\,i_2 j_2} \rangle = \delta_{t_1 t_2} \delta_{i_1 i_2} \delta_{j_1 j_2} \quad;\quad \langle \hat{\phi}_{s_1,\,u_1 v_1} | \hat{\phi}_{s_2,\,u_2 v_2} \rangle = \delta_{s_1 s_2} \delta_{u_1 u_2} \delta_{v_1 v_2}
\end{equation}
and are related to each other by unitary transformation which can be achieved by using the fusion matrices (also known as \emph{Racah matrices}). The transformation rule is given as following: 
\begin{alignat}{2}
&|{\phi_{t,ij}}\rangle  &&= \sum_s \sum_{u,v}\, a_{s,uv}^{t,ij}\left[
\begin{array}{cc}
R_1 & R_2 \\
R_3 & R_4
\end{array}
\right] |{\hat{\phi}_{s,uv}}\rangle  \nonumber \\
&|{\hat{\phi}_{s,uv}}\rangle  &&= \sum_t \sum_{i,j}\, a_{s,uv}^{t,ij}\left[
\begin{array}{cc}
 R_1 & R_2 \\
R_3 & R_4
\end{array}
\right]^{*} |{\phi_{t,ij}}\rangle  ~,
\label{basis-transform}
\end{alignat}   
where $s,uv$ and $t,ij$ label the row and the column of these matrices. The unitary property is given as,
\begin{equation} 
\sum_{s,\, u,\, v} a_{s,\, uv}^{t_1,\, i_1 j_1}\left[
\begin{array}{cc}
 R_1 & R_2 \\
 R_3 & R_4 \\
\end{array}
\right] a_{s,\, uv}^{t_2,\, i_2 j_2}\left[
\begin{array}{cc}
 R_1 & R_2 \\
 R_3 & R_4 \\
\end{array}
\right]^{*} = \delta_{t_1 t_2} \delta_{i_1 i_2} \delta_{j_1 j_2} ~.
\label{Racah-unitary}
\end{equation}
\subsection{Braiding operators}
The braiding operators or the \emph{half-monodromy operators}, usually denoted as $b_i^{(\pm)}$ act on the neighboring $i^{\text{th}}$ and $(i+1)^{\text{th}}$ strands of the Wilson lines by winding the two strands around each other producing a right-handed crossing.\footnote{A crossing is called right-handed, if the index finger and thumb of the right hand point in the directions of strands going above and below respectively at the crossing. Similarly, if the index finger and thumb of the left hand point in the directions of strands going above and below respectively, then the crossing is left-handed.} The inverse operator $(b_i^{(\pm)})^{-1}$ winds the $i^{\text{th}}$ and $(i+1)^{\text{th}}$ strands resulting in a left-handed crossing. The superscripts $(+)$ and $(-)$ are used depending on whether the strands are parallel or anti-parallel as given in the following:
\begin{equation}
{\begin{array}{c}
\includegraphics[width=0.8\linewidth]{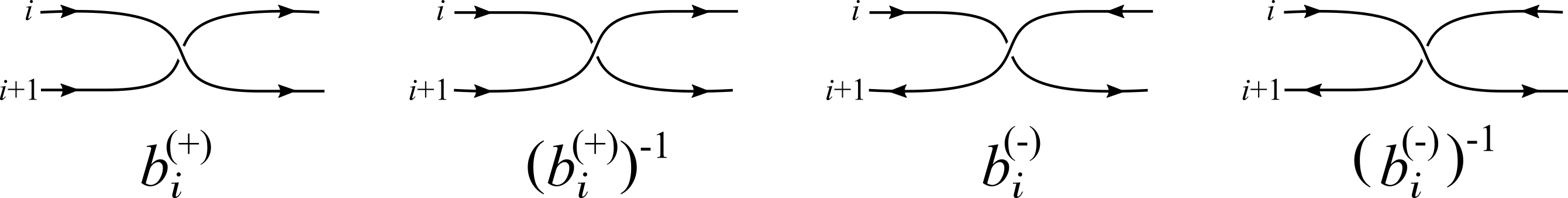}
\end{array}}
\label{braiding-operators}
\end{equation}
The basis state $|{\phi_{t,\,ij}}\rangle $ is an eigenstate of operators $b_1^{(\pm)}$ and $b_3^{(\pm)}$. Similarly, $|{\hat{\phi}_{s,\,uv}}\rangle $ is an eigenstate of $b_2^{(\pm)}$. The action of the braiding operators on these basis states are given as following:
\begin{alignat}{2}
&b_1^{(\pm)}|{\phi_{t,ij}}\rangle   &&= \{R_1, R_2, t, i\} \exp\left(\pm\frac{i\pi (C_{R_1}+C_{R_2}-C_{t})}{k+N}\right) |{\phi_{t,ij}}\rangle  \nonumber \\
&b_2^{(\pm)} |{\hat{\phi}_{s,uv}}\rangle   &&= \{R_2, R_3, s, u\} \exp\left(\pm\frac{i\pi (C_{R_2}+C_{R_3}-C_{s})}{k+N}\right) |{\hat{\phi}_{s,uv}}\rangle  \nonumber \\
&b_3^{(\pm)} |{\phi_{t,ij}}\rangle  &&= \{\bar{R}_3, \bar{R}_4, t, j\} \exp\left(\pm\frac{i\pi (C_{R_3}+C_{R_4}-C_{t})}{k+N}\right) |{\phi_{t,ij}}\rangle ~.
\label{B-eigenvalue}
\end{alignat}
Here 3j-phases $\{R_a, R_b, t, r\}$ only take values $\pm 1$ and are the symmetry phases of the Clebsch-Gordon coefficients when the two coupling representations $R_a$ and $R_b$ are exchanged \cite{butler2012point}. The term $C_R$ denotes the quadratic Casimir of the representation $R$ given as
\begin{equation}
C_R = \frac{1}{2} \left( N l + l + \sum_{i=1}^{N-1}(l_i^2 - 2i l_i) - \frac{l^2}{N} \right)~,
\end{equation}
where $l_i$ are the number of boxes in the $i$-th row of Young tableau for $R$ and $l$ is the total number of boxes.
\begin{figure}[h]
\centerline{\includegraphics[width=4.5in]{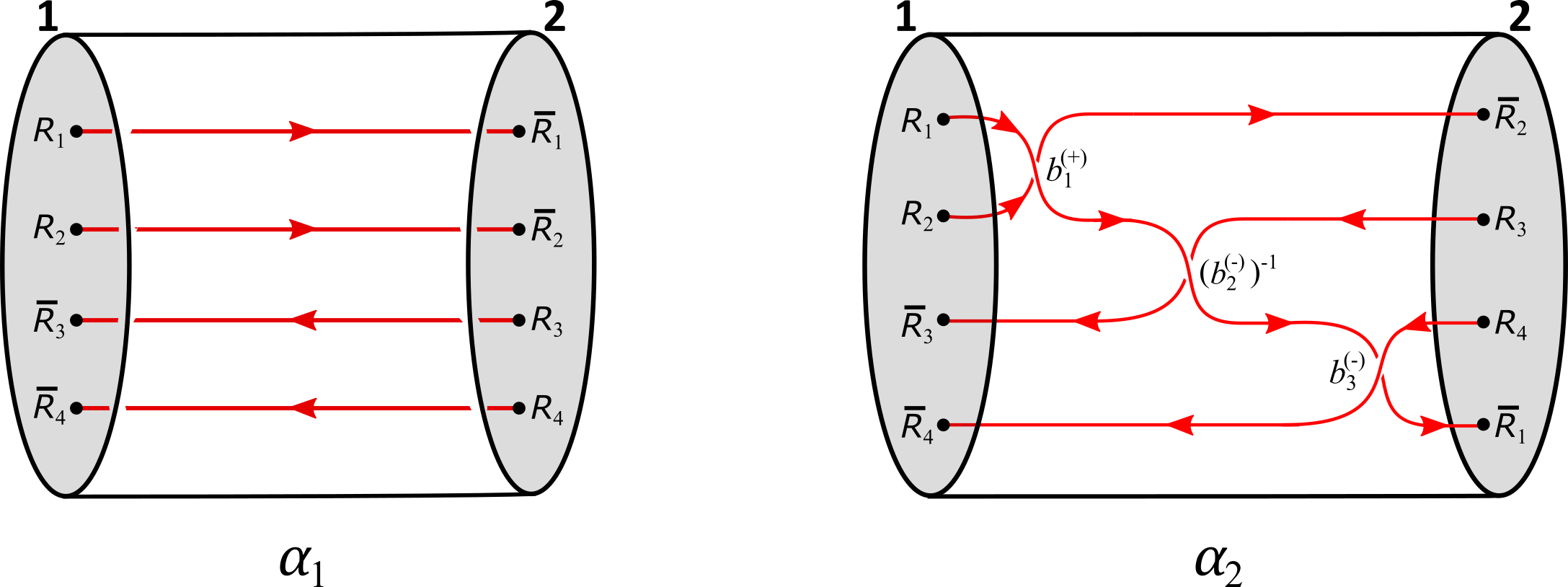}}
\caption[]{The state $|{\alpha_2}\rangle$ can be obtained from state $|{\alpha_1}\rangle$ by applying a series of braiding operators as shown.}
\label{Alpha-States}
\end{figure}
\subsection{Constructing the quantum states using braiding operators and gluing method}
\begin{figure}[h]
\centerline{\includegraphics[width=5.9in]{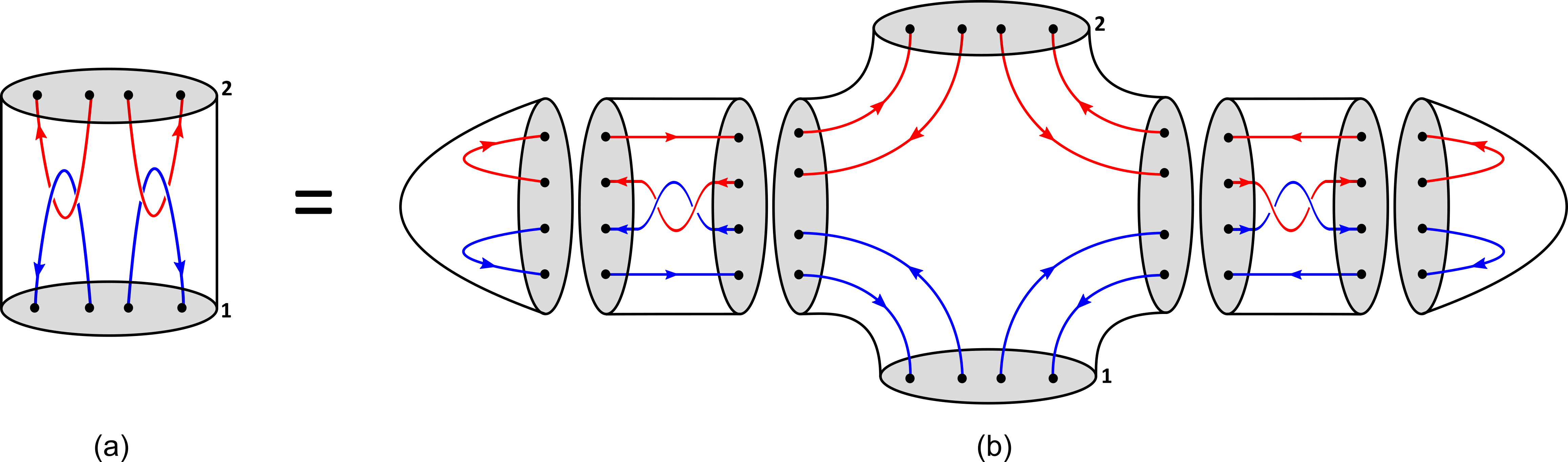}}
\caption[]{An example showing how the quantum state on $S^2 \cup S^2$ can be obtained by considering the state on multiple $S^2$ boundaries and gluing the extra $S^2$'s with the appropriate oppositely oriented boundaries of other quantum states.}
\label{Constructing-State}
\end{figure}

The braiding operators can be used to construct new quantum states. A typical example is shown in figure \ref{Alpha-States}. Consider the state $|{\alpha_1}\rangle$ in figure \ref{Alpha-States} where Wilson lines carrying representations $R_1, R_2, R_3, R_4$ stretch from boundary 1 to boundary 2 without any braiding. The Hilbert spaces associated with the two boundaries can be obtained as discussed previously and can be given as
\begin{align}
\text{basis}(\mathcal{H}_{1}) &= \{ |{\phi_{t_1,\,i_1j_1}^{(1)}}\rangle  \,\mid \, t_1 \in (\bar{R}_3 \otimes \bar{R}_4) \cap (\bar{R}_1 \otimes \bar{R}_2) \,,\, 1 \leq i_1 \leq N_{\bar{R}_3 \bar{R}_4}^{t_1}\,,\, 1 \leq j_1 \leq N_{\bar{R}_1 \bar{R}_2}^{t_1} \} \nonumber \\
\text{basis}(\mathcal{H}_{2}) &= \{ \langle{\phi_{t_2,\,i_2j_2}^{(2)}}|  \,\mid \, t_2 \in (R_3 \otimes R_4) \cap (R_1 \otimes R_2) \,,\, 1 \leq i_2 \leq N_{R_{3} R_{4}}^{t_2}\,,\, 1 \leq j_2 \leq N_{R_1 R_2}^{t_2} \}
 \nonumber ~.
\end{align} 
Here we used the `ket' and `bra' to denote the basis states since the Hilbert spaces $\mathcal{H}_{1}$ and $\mathcal{H}_{2}$ are dual to each other (the two $S^2$'s in this case are oppositely oriented). The state $|{\alpha_1}\rangle$ can be written as,
\begin{equation}
|{\alpha_1}\rangle = \sum_{t_1,\, i_1,\, j_1}\,\,\sum_{t_2,\, i_2,\, j_2} |{\phi_{t_1,\,i_1j_1}^{(1)}}\rangle \otimes \langle{\phi_{t_2,\,i_2j_2}}| \,\, \delta_{t_1 \bar{t}_2}\, \delta_{i_1 i_2}\, \delta_{j_1 j_2} ~,
\label{}
\end{equation}
where the delta functions have been used because this state can be seen as $S^2 \times I$ with four strands which is equivalent to the identity operation. We can also convert a `bra' to a `ket' which amounts to conjugation and thus will involve two 3j-phases (see ref. \cite{Nawata:2015xha}):
\begin{equation}
\langle{\phi_{t_2,\,i_2j_2}}| \quad \longrightarrow \quad \{R_1, R_2, \bar{t}_2, i_2\} \{R_3, R_4, \bar{t}_2, j_2\} \, |{\phi_{\bar{t}_2,\,j_2i_2}}\rangle
\end{equation}
Thus we can write the state as,
\begin{align}
|{\alpha_1}\rangle &= \sum_{t_1,\, i_1,\, j_1}\,\,\sum_{t_2,\, i_2,\, j_2} \{R_1, R_2, \bar{t}_2, i_2\} \{R_3, R_4, \bar{t}_2, j_2\} \, |{\phi_{t_1,\,i_1j_1}^{(1)}}\rangle \otimes |{\phi_{\bar{t}_2,\,j_2i_2}^{(2)}}\rangle \,\, \delta_{t_1 \bar{t}_2}\, \delta_{i_1 i_2}\, \delta_{j_1 j_2} \nonumber \\
&= \sum_{t} \sum_{i,\, j} \,\,\{R_1, R_2, t, i\} \{R_3, R_4, t, j\}\,\, |{\phi_{t,ij}^{(1)}}\rangle \otimes |{\phi_{t,ji}^{(2)}}\rangle ~.
\label{alpha-1}
\end{align}
 Now let us apply a series of braiding operators which results in the state $|{\alpha_2}\rangle$ as shown in figure \ref{Alpha-States}. This state can be given as,
\begin{equation}
|{\alpha_2}\rangle = \sum_{t} \sum_{i,\, j}\, \{R_1, R_2, t, i\} \{R_3, R_4, t, j\}\,\, |{\phi_{t,ij}^{(1)}}\rangle \otimes \left( b_3^{(-)} (b_2^{(-)})^{-1} b_1^{(+)} \right) |{\phi_{t,ij}^{(2)}}\rangle ~,
\end{equation}
which can be evaluated using eq.(\ref{B-eigenvalue}) and eq.(\ref{basis-transform}). However not all two boundary states can be obtained from $|{\alpha_1}\rangle$ by applying the braiding operators. For example, consider the state shown in figure \ref{Results}(d). In order to construct this, we need to take a four boundary state and then glue the extra two boundaries by taking its inner product with appropriate states as shown in figure \ref{Constructing-State}. The quantum states associated with the three-balls with one, two and four boundaries respectively as shown in figure \ref{Constructing-State}(b) serve as the basic building blocks \cite{Nawata:2013qpa}. In fact, the four boundary state can be generalized to $n$ boundaries which is the state $\ket{\Psi_2}$ given in section \ref{sec3}. Further the state shown in figure \ref{Constructing-State}(a) is the $n=2$ version of the state $\ket{\Psi_4}$ elaborated in section \ref{sec4} with one full twist each (i.e. $p_1 = p_2 = 1$ in figure \ref{vn-state}). 

So far, we have confined to the three-balls with two $S^2$ boundaries which could have local braiding or necklaces. Partial tracing of the Hilbert space associated with one of the $S^2$ boundaries is a straightforward exercise in determining the entanglement entropy. Note that the equivalence between the states in figure \ref{Constructing-State}(a) and  figure \ref{Constructing-State}(b) suggests that we could also consider the states on three-balls with more than two $S^2$ boundaries and study its entanglement structure from partial tracing some of the $S^2$ boundaries.
\subsection{Entanglement structure of the states on multiple $S^2$ boundaries}
Given a state $|{\Psi}\rangle$ on $n$ copies of $S^2$, we can compute the reduced density matrix by bi-partitioning the total Hilbert space as $(\mathcal{H}_A|\mathcal{H}_B)$ given in eq.(\ref{parts-HS}) and tracing out the $\mathcal{H}_B$ part. Let us take the example of $|{\alpha_1}\rangle \in \mathcal{H}_1 \otimes \mathcal{H}_2$ in eq.(\ref{alpha-1}). The total density matrix for this state can be evaluated as,
\begin{equation}
\rho_{\text{total}} = \frac{|{\alpha_1}\rangle \langle{\alpha_1} |}{\braket{\alpha_1}} ~.
\end{equation}
Tracing out $\mathcal{H}_2$ gives the following reduced density matrix:
\begin{equation}
\rho_{A} = \text{Tr}_{\mathcal{H}_B}(\rho_{\text{total}}) = \sum_{t'} \sum_{i', j'}\,\,  \langle \phi_{t',i'j'}^{(2)}  | \rho_{\text{total}} | \phi_{t',i'j'}^{(2)} \rangle ~.
\end{equation}
Using the orthonormal property of the conformal basis, this can be simplified to
\begin{equation}
\rho_{A} = \frac{1}{\text{dim}(\mathcal{H}_A)} \left(\sum_{t} \sum_{i, j} |{\phi_{t,ij}^{(1)}}\rangle \langle {\phi_{t,ij}^{(1)}}|\right)  ~.
\end{equation}
Thus we get the maximum entanglement entropy: $\text{EE} = \ln \text{dim}(\mathcal{H}_A)$, which is consistent with the result of \cite{Melnikov:2018zfn} that the quantum state $|{\alpha_1}\rangle$ is a maximally entangled or Bell state. 

This procedure of tracing out Hilbert space can be generalized for multiple $S^2$ boundaries. If the basis of the Hilbert space for $x^{\text{th}}$ $S^2$ boundary is denoted as $| \phi_{t_x,i_x j_x}^{(x)} \rangle $, then tracing out $\mathcal{H}_B$ can be achieved as,
\begin{equation}
\rho_{A} = \sum_{\substack{t_{m+1}, t_{m+2}, \ldots, t_n \\ i_{m+1}, i_{m+2}, \ldots, i_n  \\ j_{m+1}, j_{m+2}, \ldots, j_n }} \,\,  \langle \phi_{t_{m+1},\,i_{m+1} j_{m+1}}^{(m+1)}, \ldots, \phi_{t_{n},\,i_{n} j_{n}}^{(n)}   | \rho_{\text{total}} | \phi_{t_{m+1},\,i_{m+1} j_{m+1}}^{(m+1)}, \ldots, \phi_{t_{n},\,i_{n} j_{n}}^{(n)} \rangle ~,
\end{equation}
where $\rho_{\text{total}}$ is the total density matrix acting on $\mathcal{H}_A \otimes \mathcal{H}_B$.
In the next section, we will study the entanglement features of the states which are generalization of the state $|{\alpha_1}\rangle$ on $n$ copies of $S^2$.

\section{Quantum states $\ket{\Psi_1}$ and $\ket{\Psi_2}$}
\label{sec3}
\begin{figure}[h]
\centerline{\includegraphics[width=5.0in]{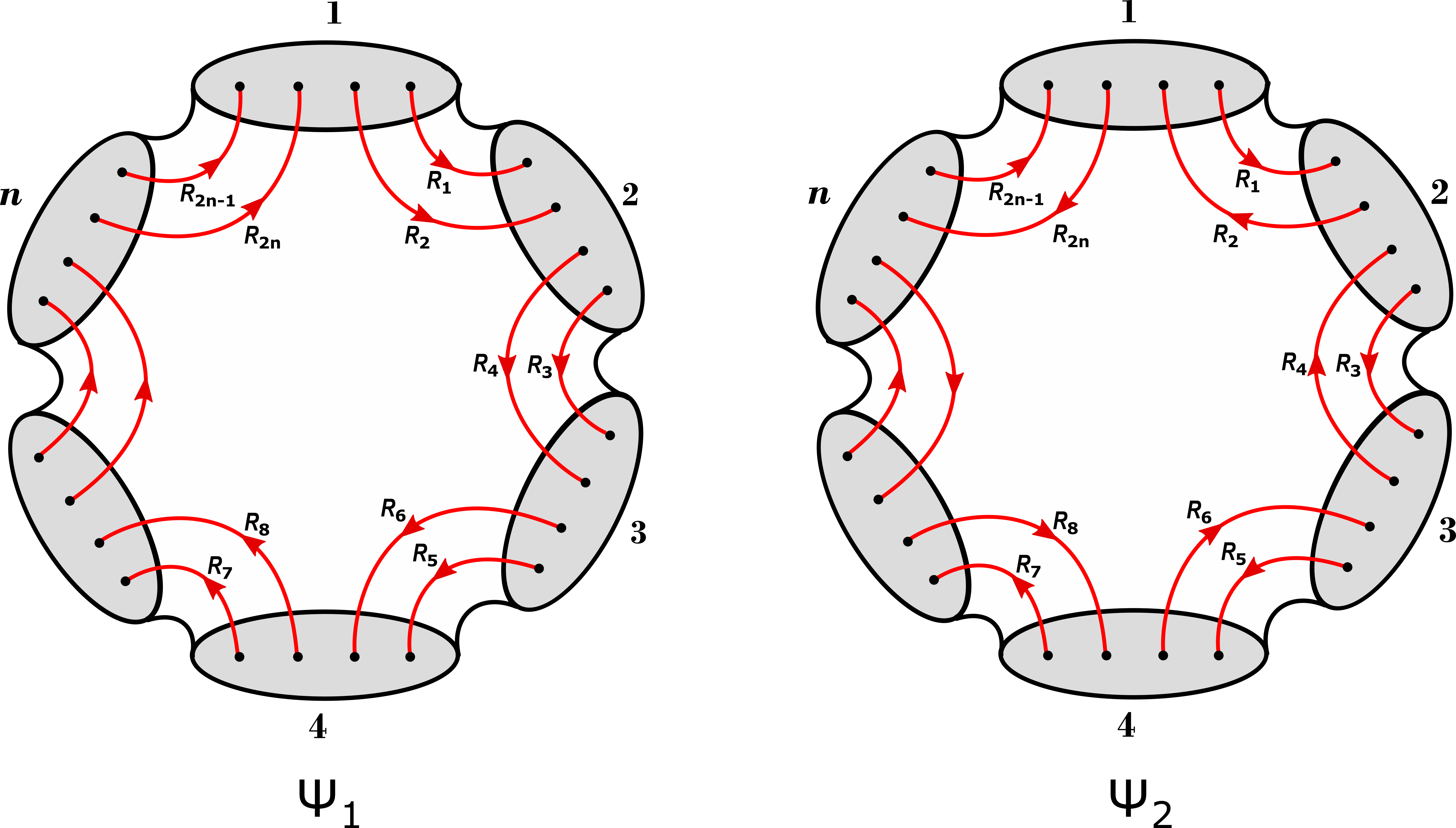}}
\caption[]{Manifold with $n$ number of $S^2$ boundaries. Each $S^2$ has four punctures. Wilson lines are threaded from one boundary to another without any braiding.}
\label{No-braid-WLine}
\end{figure}

In this section, we consider the states where a pair of Wilson lines stretch from $j^{\text{th}}$ boundary to $(j+1)^{\text{th}}$ boundary as shown in the figure \ref{No-braid-WLine}. Note that $|{\Psi_1}\rangle$ and $|{\Psi_2}\rangle$ in figure \ref{No-braid-WLine} differ in the relative orientation of the Wilson lines carrying representations $R_{2j-1}$ and $R_{2j}$ ($j=1,2,\ldots,n$). When $n=2$, $|{\Psi_1}\rangle$ and $|{\Psi_2}\rangle$ reduce to the state $|{\alpha_1}\rangle$ of figure \ref{Alpha-States} with appropriate orientation of Wilson lines.
These states have been obtained in \cite{Nawata:2015xha} which can be written in conformal basis as: 
\begin{align} 
|{\Psi_1}\rangle &= \sum_{\ell=1}^{\textsf{d}_1} \sum_{x_1, \ldots, x_n} \frac{\prod_{j=1}^n \left\{ R_{2j-1}, R_{2j},t_\ell, x_j \right\} }{(\text{dim}_q t_\ell)^{\frac{n-2}{2}}} |{\phi_{t_\ell,\, x_n x_1}^{(1)},\phi_{t_\ell,\, x_1 x_2}^{(2)},\ldots, \phi_{t_\ell,\, x_{n-1} x_{n}}^{(n)}}\rangle \nonumber \\
|{\Psi_2}\rangle &= \sum_{\ell=1}^{\textsf{d}_2} \sum_{y_1, \ldots, y_n} \frac{\prod_{j=1}^n \left\{ R_{2j-1}, \bar{R}_{2j},w_\ell, y_j \right\} }{(\text{dim}_q w_\ell)^{\frac{n-2}{2}}} |{\phi_{w_\ell,\, y_n y_1}^{(1)},\phi_{w_\ell,\, y_1 y_2}^{(2)}, \ldots, \phi_{w_\ell,\, y_{n-1} y_{n}}^{(n)}}\rangle ~,
\label{psi1}
\end{align}
where the intermediate states in the two cases are labeled by representations $t_\ell$ and $w_\ell$ such that $t_\ell \in \bigcap_{j=1}^n \left(R_{2j-1} \otimes R_{2j} \right)$ and $w_\ell \in \bigcap_{j=1}^n \left(R_{2j-1} \otimes \bar{R}_{2j} \right)$ respectively. The terms $\text{dim}_q t_\ell$ and $\text{dim}_q w_\ell$ are the quantum dimensions of the representations $t_\ell$ and $w_\ell$ respectively. Note that both $t_\ell$ and $w_\ell$ can occur multiple times in the corresponding tensor decompositions:
\begin{align}
R_{2j-1} \otimes R_{2j} = \bigoplus_{\ell=1}^{\textsf{d}_1} (N_{R_{2j-1}\, R_{2j}}^{t_\ell}) \,t_\ell \quad;\quad R_{2j-1} \otimes \bar{R}_{2j} = \bigoplus_{\ell=1}^{\textsf{d}_2} (N_{R_{2j-1}\,\bar{R}_{2j}}^{w_\ell}) \,w_\ell
\end{align}
and we need the extra labels $x_j$ and $y_j$ to keep track of multiple occurrences of $t_\ell$ and $w_\ell$ respectively. If we define the following:
\begin{equation}
N_{R_{2j-1}\, R_{2j}}^{t_\ell} \equiv a_j(t_\ell) \quad;\quad N_{R_{2j-1}\, \bar{R}_{2j}}^{w_\ell} \equiv b_j(w_\ell) ~,
\end{equation}
then $x_j = 1,2,\ldots,a_j(t_\ell)$ and $y_j = 1,2,\ldots,b_j(w_\ell)$ for a given $t_\ell$ and $w_\ell$ respectively.
The states $|{\Psi_1}\rangle$ and $|{\Psi_2}\rangle$ live in the total Hilbert spaces $\mathcal{H}(\Psi_1)$ and $\mathcal{H}(\Psi_2)$ respectively which are the product of $n$ Hilbert spaces:
\begin{alignat}{2}
\mathcal{H}({\Psi_1}) &= \bigotimes_{j=1}^n \mathcal{H}_j(\Psi_1)  &\quad;\quad \text{dim}(\mathcal{H}_j(\Psi_1)) &= \sum_{\ell=1}^{\textsf{d}_1} a_{j-1}(t_\ell) \, a_j(t_\ell) \nonumber \\
\mathcal{H}({\Psi_2}) &= \bigotimes_{j=1}^n \mathcal{H}_j(\Psi_2) &\quad;\quad \text{dim}(\mathcal{H}_j(\Psi_2)) &= \sum_{\ell=1}^{\textsf{d}_2} b_{j-1}(w_\ell) \, b_j(w_\ell) ~,
\end{alignat}
where $\mathcal{H}_j(\Psi_1)$ and $\mathcal{H}_j(\Psi_2)$ are Hilbert spaces associated with $S^2$ having four point conformal block given by representations $\left\{\bar{R}_{2j-3}, \bar{R}_{2j-2}, R_{2j}, R_{2j-1} \right\}$ and $\left\{\bar{R}_{2j-3}, R_{2j-2}, \bar{R}_{2j}, R_{2j-1} \right\}$ respectively which can be seen from the figure \ref{No-braid-WLine}.  Since the 3j-phases $\left\{ R_{2j-1}, R_{2j},t_\ell, x_j \right\}$ and $\left\{ R_{2j-1}, \bar{R}_{2j},w_\ell, y_j \right\}$ only take values $\pm 1$, they will not affect the entanglement structure of the quantum states $|{\Psi_1}\rangle$ and $|{\Psi_2}\rangle$ and can be removed by applying U(1) transformation (i.e. $\exp(i \pi)$ if phase is $-1$). Bi-partitioning the total Hilbert space as $(\mathcal{H}_A|\mathcal{H}_B) = (m|n-m)$ and tracing out $\mathcal{H}_B$ gives the reduced density matrix for the two states:
\begin{align} 
\rho_A(\Psi_1) &= \frac{1}{\braket{\Psi_1}} \sum_{\ell=1}^{\textsf{d}_1}  \sum_{\substack{x_0, x_1, \ldots,x_m \\ x_0', x_1', \ldots, x_{m}'}} f(t_\ell) \,\, \delta_{x_0 x_0'} \delta_{x_m x_m'} \bigotimes_{j=1}^m |{\phi_{t_\ell,\, x_{j-1} x_{j}}^{(j)}}\rangle \langle {\phi_{t_\ell,\, x'_{j-1} x'_{j}}^{(j)}}| \nonumber \\
\rho_A(\Psi_2) &= \frac{1}{\braket{\Psi_2}} \sum_{\ell=1}^{\textsf{d}_2}  \sum_{\substack{y_0, y_1, \ldots,y_m \\ y_0', y_1', \ldots, y_{m}'}} g(w_\ell)\,\, \delta_{y_0 y_0'} \delta_{y_m y_m'} \bigotimes_{j=1}^m |{\phi_{w_\ell,\, y_{j-1} y_{j}}^{(j)}}\rangle \langle{\phi_{w_\ell,\, y'_{j-1} y'_{j}}^{(j)}} | ~,
\label{reduced-rho}
\end{align}
where we have defined $x_0 \equiv x_n$ and $y_0 \equiv y_n$. The functions $f(t_\ell)$ and $g(w_\ell)$ are defined as,
\begin{equation}
f(t_\ell) = (\text{dim}_q t_\ell)^{2-n} \left(\prod_{j=m+1}^{n-1} a_j(t_\ell) \right) \quad;\quad g(w_\ell) = (\text{dim}_q w_\ell)^{2-n} \left(\prod_{j=m+1}^{n-1} b_j(w_\ell) \right)
\end{equation}
and the trace part is,
\begin{align} 
\braket{\Psi_1} = \sum_{\ell=1}^{\textsf{d}_1} \left( (\text{dim}_q t_\ell)^{2-n} \prod_{j=1}^{n} a_j(t_\ell) \right) ;\quad \braket{\Psi_2} = \sum_{\ell=1}^{\textsf{d}_2} \left( (\text{dim}_q w_\ell)^{2-n} \prod_{j=1}^{n} b_j(w_\ell) \right) ~.
\end{align}
These can be written as block diagonal matrices where each block correspond to $t_\ell$ and $w_\ell$ respectively as following: 
\begin{align}
\rho_A(\Psi_1) &= \frac{1}{\braket{\Psi_1}} \left(
\begin{array}{cccc}
 f(t_1)\, P(t_1) & & & \\
  & f(t_2)\, P(t_2) & & \\
&  & \ddots &  \\
  &  &  & f(t_{\textsf{d}_1})\, P(t_{\textsf{d}_1})
\end{array}
\right) \nonumber \\[10pt] 
\rho_A(\Psi_2) &= \frac{1}{\braket{\Psi_2}} \left(
\begin{array}{cccc}
 g(w_1)\, Q(w_1) &  &  &  \\
  & g(w_2)\, Q(w_2) &  &  \\
  &  & \ddots &  \\
 &  &  & g(w_{\textsf{d}_2})\, Q(w_{\textsf{d}_2})
\end{array}
\right)  ~.
\label{reduced-rho-matrix}
\end{align}
Let us pause for a moment and try to understand the structure of these density matrices. Take for example $\rho_A(\Psi_1)$. From eq.(\ref{reduced-rho}), we can see that its $\ell^{\text{th}}$ block will correspond to the row $|\underbrace{t_\ell, t_\ell, \ldots, t_\ell}_{m}\rangle$ and the column $\langle \underbrace{t_\ell, t_\ell, \ldots, t_\ell}_{m} |$ of the matrix $\rho_A(\Psi_1)$ and there are precisely $\textsf{d}_1$ number of non-zero blocks. There are other $(\textsf{d}_1^m - \textsf{d}_1)$ number of blocks for the states like $|{t_1, t_2, \ldots, t_m}\rangle$ (and its permutations) which are all 0 and have not been included in the eq.(\ref{reduced-rho-matrix}). Now let us see the structure of each block. The matrix $P(t_\ell)$ corresponding to the block $t_\ell$ is made of the following part of eq.(\ref{reduced-rho}):
\begin{equation}
P(t_\ell) = \sum_{\substack{x_0, x_1, \ldots,x_m \\ x_0', x_1' \ldots, x_{m}'}} \,\, \delta_{x_0 x_0'} \,\, \delta_{x_m x_m'} |{x_0 x_1, x_1 x_2, \ldots, x_{m-1} x_m}\rangle \langle{x_0' x_1', x_1' x_2', \ldots, x_{m-1}' x_m'} | ~,
\label{fun-Pt}
\end{equation} 
where $x_j$ takes values $1, 2, \ldots, a_j(t_\ell)$. Thus we can think of $P(t_\ell)$ to be acting on the following $m$ qudit Hilbert space:
\begin{equation}
\mathcal{H}_{P(t_\ell)} = \mathbb{C}^{a_0 a_1} \otimes \mathbb{C}^{a_1 a_2} \otimes \ldots \otimes \mathbb{C}^{a_{m-1} a_m} ~.
\end{equation}
A possible choice of the matrix form for $P(t_\ell)$ in this space can be given as the following Kronecker product: 
\begin{equation} 
P(t_\ell) =  \Theta_{1}(t_\ell) \otimes \Theta_{2}(t_\ell) \otimes \ldots \otimes \Theta_{m-1}(t_\ell) \otimes \mathbb{1}_{a_0} \otimes \mathbb{1}_{a_m} ~,
\label{P-matrix}
\end{equation}
where the two identity matrices appear because of the delta functions in eq.(\ref{fun-Pt}) and $\Theta_{j}(t_\ell)$ is a $a_j^2(t_\ell) \times a_j^2(t_\ell)$ matrix which can be written as the following block matrix:
\begin{equation} 
\Theta_{j}(t_\ell) = \left(
\begin{array}{cccc}
 J_{11} & J_{12} & \ldots & J_{1 a_j} \\
 J_{21} & J_{22} & \ldots & J_{2 a_j} \\
\vdots & \vdots & \ddots & \vdots \\
 J_{a_j 1} & J_{a_j2} & \ldots & J_{a_j a_j}
\end{array}
\right) ~,
\end{equation}
Each block $J_{pq}$ is a $a_j(t_\ell) \times a_j(t_\ell)$ single-entry matrix whose $(p,q)$ element is 1 and all the other elements are 0. Further $\Theta_{j}(t_\ell)$ has only one non-zero eigenvalue which is $a_j(t_\ell)$ and this means that the matrix $P(t_\ell)$ has only one non-zero eigenvalue: $a_1(t_\ell) a_2(t_\ell) \ldots a_{m-1}(t_\ell)$ which is repeated with a multiplicity $a_0(t_\ell) \, a_m(t_\ell)$. Thus the block $f(t_\ell) P(t_\ell)$ in eq.(\ref{reduced-rho-matrix}) will have the following eigenvalues (after substituting the value of $f(t_\ell)$),
\begin{equation}
\lambda_{t_\ell} = \frac{(\text{dim}_q t_\ell)^{2-n}}{\braket{\Psi_1}} \frac{\prod_{j=1}^{n} a_j(t_\ell)}{a_m(t_\ell)\, a_n(t_\ell)} \quad,\quad \text{mul}(\lambda_{t_\ell}) = a_m(t_\ell)\, a_n(t_\ell) ~,
\end{equation}
where $\lambda_{t_\ell}$ is repeated with multiplicity $\text{mul}(\lambda_{t_\ell})$ given above. Having found all the eigenvalues of the reduced density matrix, the entanglement entropy can be given as,
\begin{equation}
\boxed{\text{EE}(\Psi_1) = - \sum_{\ell=1}^{\textsf{d}_1} \text{mul}(\lambda_{t_\ell})\, \lambda_{t_\ell} \ln \lambda_{t_\ell}} ~.
\label{EE-Psi1}
\end{equation}
A similar analysis can be done for the quantum state $|{\Psi_2}\rangle$ and the eigenvalues can be obtained as,
\begin{equation}
\lambda_{w_\ell} = \frac{(\text{dim}_q w_\ell)^{2-n}}{\braket{\Psi_2}} \frac{\prod_{j=1}^{n} b_j(w_\ell)}{b_m(w_\ell)\, b_n(w_\ell)} \quad,\quad \text{mul}(\lambda_{w_\ell}) = b_m(w_\ell)\, b_n(w_\ell) ~,
\end{equation}
where $\lambda_{w_\ell}$ is repeated with multiplicity $\text{mul}(\lambda_{w_\ell})$ and thus the entanglement entropy is:
\begin{equation}
\boxed{\text{EE}(\Psi_2) = - \sum_{\ell=1}^{\textsf{d}_2} \text{mul}(\lambda_{w_\ell})\, \lambda_{w_\ell} \ln \lambda_{w_\ell}} ~.
\label{EE-Psi2}
\end{equation}
For $n=2$, all the eigenvalues of the reduced density matrix become equal. Thus the entanglement entropy is maximum and hence these states become Bell states.
\begin{alignat}{2}
   \text{EE}(\Psi_1^{(n=2)}) &= \ln \left( \sum_{\ell=1}^{\textsf{d}_1} a_1(t_\ell)\, a_2(t_\ell) \right) = \ln \text{dim}(\mathcal{H}_A) \quad &&\Longrightarrow \quad |{\Psi_1^{(n=2)}}\rangle = |\Psi_{\text{Bell}} \rangle \nonumber \\
   \text{EE}(\Psi_2^{(n=2)}) &= \ln \left( \sum_{\ell=1}^{\textsf{d}_2} b_1(w_\ell)\, b_2(w_\ell) \right) = \ln \text{dim}(\mathcal{H}_A) \quad &&\Longrightarrow \quad |{\Psi_2^{(n=2)}}\rangle = |\Psi_{\text{Bell}} \rangle ~.
	\label{}
\end{alignat}
Note that this entanglement entropy for two $S^2$ boundaries is the logarithm of the Chern-Simons partition function on $S^2 \times S^1$ in the presence of Wilson lines carrying representations $R_1$, $R_2$, $R_3$ and $R_4$ \cite{Melnikov:2018zfn}. More precisely, we have:
\begin{align}
   \text{EE}(\Psi_1^{(n=2)}) &= \ln Z(S^2 \times S^1\,;\, R_1,R_2,\bar{R}_3,\bar{R}_4)  \nonumber \\
   \text{EE}(\Psi_2^{(n=2)}) &= \ln Z(S^2 \times S^1\,;\, R_1, \bar{R}_2, \bar{R}_3, R_4) ~.
	\label{EE-Z-fn}
\end{align}

In the later sections, we will analyze the reduced density matrices for $n\geq3$ to get more insight into the structure of the states $|{\Psi_1}\rangle$ and $|{\Psi_2}\rangle$.
\subsection{Multiplicity free case}
When the representations $t_\ell$ and $w_\ell$ appear without any multiplicity in any of the tensor product (SU(2) gauge group for example), then the eigenvalues of the reduced density matrices labeled by $t_\ell$ and $w_\ell$ are given as:
\begin{equation}
\lambda_{t_\ell}(\Psi_1) = \frac{(\text{dim}_q t_\ell)^{2-n}}{\sum_{\ell'=1}^{\textsf{d}_1} (\text{dim}_q t_{\ell'})^{2-n}} \quad;\quad \lambda_{w_\ell}(\Psi_2) = \frac{(\text{dim}_q w_\ell)^{2-n}}{\sum_{\ell'=1}^{\textsf{d}_2} (\text{dim}_q w_{\ell'})^{2-n}}
\end{equation}
and the corresponding entanglement entropies will be,
\begin{equation}
\text{EE}(\Psi_1) = -\sum_{\ell=1}^{\textsf{d}_1} \lambda_{t_\ell} \ln \lambda_{t_\ell} \quad;\quad \text{EE}(\Psi_2) = -\sum_{\ell=1}^{\textsf{d}_2} \lambda_{w_\ell} \ln \lambda_{w_\ell} ~.
\end{equation}
\subsection{Separability criteria of the reduced density matrix}
In this section, we will study the reduced density matrices in eq.(\ref{reduced-rho-matrix}) acting on the Hilbert space $\mathcal{H}_A$ and find out whether they are separable or entangled. For this, let us first compute the entanglement negativity. Consider the bi-partition of $\mathcal{H}_A$ into two parts:
\begin{equation}
\mathcal{H}_A = \mathcal{H}_{A_1} \otimes \mathcal{H}_{A_2} = \left(\bigotimes_{i=1}^{r} \mathcal{H}_i \right)  \otimes \left(\bigotimes_{i=r+1}^{m} \mathcal{H}_i \right) ~.
\end{equation}
The partial transpose of the reduced density matrix $\rho_A(\Psi_1)$ with respect to $\mathcal{H}_{A_2}$ can be given as,
\begin{equation}
\rho^{\Gamma}(\Psi_1) = \frac{1}{\braket{\Psi_1}} \left(
\begin{array}{cccc}
 f(t_1) \, P^{\Gamma}(t_1) & 0 & \ldots & 0 \\
 0 & f(t_2) \, P^{\Gamma}(t_2) & \ldots & 0 \\
\vdots & \vdots & \ddots & \vdots \\
 0 & 0 & \ldots & f(t_{\textsf{d}_1}) \, P^{\Gamma}(t_{\textsf{d}_1})
\end{array}
\right) ~,
\end{equation}
where $P^{\Gamma}(t_\ell)$ denotes the partial transpose of $P(t_\ell)$. As given in eq.(\ref{fun-Pt}), $P(t_\ell)$ denotes the following part of the density matrix, where we have explicitly shown the breaking of the basis for $\mathcal{H}_{A_1}$ and $\mathcal{H}_{A_2}$:
\begin{align}
P(t_\ell) = \sum_{\substack{x_0, x_1, \ldots,x_m \\ x_0', x_1' \ldots, x_{m}'}} \,\, \delta_{x_0 x_0'} \,\, \delta_{x_m x_m'} & |\,\,{\underbrace{x_0\, x_1,\, x_1\, x_2,\, \ldots,\, x_{r-1}\, \textcolor{red}{x_{r}}}_{A_1},\,\, \underbrace{\textcolor{red}{x_{r}}\, x_{r+1},\, \ldots,\, x_{m-1}\, x_m}_{A_2}}\,\,\rangle \nonumber \\
& \langle \,\,{\underbrace{x_0'\, x_1',\, x_1'\, x_2',\, \ldots, x_{r-1}'\, \textcolor{blue}{x_{r}'}}_{A_1},\,\, \underbrace{\textcolor{blue}{x_{r}'}\, x_{r+1}',\, \ldots,\, x_{m-1}'\, x_m'}_{A_2}}\,\, | ~.
\label{Pt-fun-neg}
\end{align} 
As given previously, we can write $P(t_\ell)$ as the Kronecker product,
\begin{equation} 
P(t_\ell) =  \Theta_{1} \otimes \ldots \otimes \Theta_{r-1} \otimes \textcolor{red}{\Theta_{r}} \otimes \Theta_{r+1} \otimes \ldots \otimes \Theta_{m-1} \otimes \mathbb{1}_{a_0} \otimes \mathbb{1}_{a_m} ~.
\end{equation}
An important role is played by the matrix $\Theta_{r}$ here.  Notice that in eq.(\ref{Pt-fun-neg}), the last label of the basis set of $\mathcal{H}_{A_1}$ and the first label of the basis set of $\mathcal{H}_{A_2}$ are the same, namely $x_r$. Since $\Theta_{r}$ is the matrix with $x_{r}$ and $x_{r}'$ labeling its rows and columns, it will transform non-trivially under the operation of partial transpose. When we apply the partial transpose with respect to $A_2$, we get,
\begin{align}
P^{\Gamma}(t_\ell) = \sum_{\substack{x_0, x_1, \ldots,x_m \\ x_0', x_1' \ldots, x_{m}'}} \,\, \delta_{x_0 x_0'} \,\, \delta_{x_m x_m'} & |\,\,{\underbrace{x_0\, x_1,\, x_1\, x_2,\, \ldots,\, x_{r-1}\, \textcolor{red}{x_{r}}}_{A_1},\,\, \underbrace{\textcolor{blue}{x_{r}'}\, x_{r+1}',\, \ldots,\, x_{m-1}'\, x_m'}_{A_2}}\,\,\rangle \nonumber \\
& \langle \,\,{\underbrace{x_0'\, x_1',\, x_1'\, x_2',\, \ldots, x_{r-1}'\, \textcolor{blue}{x_{r}'}}_{A_1},\,\, \underbrace{\textcolor{red}{x_{r}}\, x_{r+1},\, \ldots,\, x_{m-1}\, x_m}_{A_2}}\,\, | ~
\label{}
\end{align} 
and hence $P^{\Gamma}(t_\ell)$ can be written as following,
\begin{equation} 
P^{\Gamma}(t_\ell) =  \Theta_{1} \otimes \ldots \otimes \Theta_{r-1} \otimes \textcolor{red}{\Theta_{r}^{\Gamma}} \otimes \Theta_{r+1}^T \otimes \ldots \otimes \Theta_{m-1}^T \otimes \mathbb{1}_{a_0} \otimes \mathbb{1}_{a_m}^T ~,
\end{equation}
where superscript $T$ denotes the usual transpose which does not affect the eigenvalues. The matrix $\Theta_{r}^{\Gamma}$ can be obtained from $\Theta_{r}$ as following:
\begin{equation} 
\Theta_{r}(t_\ell) = \left(
\begin{array}{cccc}
 J_{11} & J_{12} & \ldots & J_{1 a_{r}} \\
 J_{21} & J_{22} & \ldots & J_{2 a_r} \\
\vdots & \vdots & \ddots & \vdots \\
 J_{a_r 1} & J_{a_r 2} & \ldots & J_{a_r a_r}
\end{array}
\right) \quad;\quad \Theta_{r}^{\Gamma}(t_\ell) = \left(
\begin{array}{cccc}
 J_{11} & J_{21} & \ldots & J_{a_r 1} \\
 J_{12} & J_{22} & \ldots & J_{a_r 2} \\
\vdots & \vdots & \ddots & \vdots \\
 J_{1 a_r} & J_{2 a_r} & \ldots & J_{a_r a_r}
\end{array}
\right) ~,
\end{equation}
where $J_{pq}$ is the single-entry matrix defined earlier. The eigenvalues of $\Theta_{r}^{\Gamma}$ are $\pm 1$ with multiplicities given as following,
\begin{equation}
\lambda(\Theta_{r}^{\Gamma}) = \{-1, 1\} \quad;\quad \text{mul}(-1) = \frac{a_r(a_r-1)}{2} \quad,\quad \text{mul}(1) = \frac{a_r(a_r+1)}{2} ~.
\end{equation}
We only need to consider the negative eigenvalues to calculate entanglement negativity. The block of $\rho^{\Gamma}(\Psi_1)$ corresponding to representation $t_\ell$ which is $f(t_\ell) \, P^{\Gamma}(t_\ell)$ will have the following negative eigenvalues,
\begin{equation}
\lambda_{t_\ell}^{\Gamma} = - f_{t_\ell} \frac{a_1(t_\ell) a_2(t_\ell) \ldots a_{m-1}(t_\ell)}{a_r(t_\ell)} \quad,\quad \text{mul}(\lambda_{t_\ell}^{\Gamma}) = a_0(t_\ell) a_m(t_\ell) a_r(t_\ell) \left(\frac{a_r(t_\ell) - 1}{2}\right) ~.
\end{equation}
With all these negative eigenvalues, the entanglement negativity can be computed,
\begin{equation}
\boxed{\mathcal{N}(\Psi_1) = \frac{1}{\braket{\Psi_1}} \sum_{\ell=1}^{\textsf{d}_1} (\text{dim}_q t_\ell)^{2-n} \textcolor{red}{\left(\frac{a_r(t_\ell) - 1}{2}\right)} \prod_{j=1}^{n} a_j(t_\ell) } ~.
\label{neg-Psi1}
\end{equation}
Since the quantum dimension and $a_j(t_\ell)$ are always positive for any $t_\ell$, the negativity will vanish if and only if $a_r(t_\ell) = 1$ for all $t_\ell$, i.e. if and only if the fusion of $R_{2r-1}$ and $R_{2r}$ gives $t_\ell$ without multiplicity. A similar analysis can be done for the reduced density matrix  $\rho(\Psi_2)$ and the negativity can be obtained as,
\begin{equation}
\boxed{\mathcal{N}(\Psi_2) = \frac{1}{\braket{\Psi_2}} \sum_{\ell=1}^{\textsf{d}_2} (\text{dim}_q w_\ell)^{2-n} \left(\frac{b_r(w_\ell) - 1}{2}\right) \prod_{j=1}^{n} b_j(w_\ell) } ~.
\label{neg-Psi2}
\end{equation} 
Thus $a_r(t_\ell) = 1$ and $b_r(w_\ell) = 1$ for all values of $t_\ell$ and $w_\ell$, are both necessary and sufficient for the respective entanglement negativities $\mathcal{N}(\Psi_1)$ and $\mathcal{N}(\Psi_2)$ to vanish. Further, if the reduced density matrix is separable then negativity vanishes. This implies that $a_r(t_\ell)$ and $b_r(w_\ell)$ should be 1 for all the representations $t_\ell$ and $w_\ell$.

Now consider the case when all the multiplicities in the tensor decompositions of representations associated with the remaining $S^2$ boundaries $1,2,\ldots,m$ are unit, i.e. $a_0(t_\ell) = a_1(t_\ell) = \ldots = a_m(t_\ell) = 1$ and $b_0(t_\ell) = b_1(t_\ell) = \ldots = b_m(t_\ell) = 1$ for all the representations $t_\ell$ and $w_\ell$. In such a case the reduced density matrices can be written as,
\begin{align} 
\rho(\Psi_1) = \sum_{\ell=1}^{\textsf{d}_1} \frac{f(t_\ell)}{\braket{\Psi_1}} \,\, \rho_{t_\ell}^{A_1}  \otimes \rho_{t_\ell}^{A_2} \quad;\quad \rho(\Psi_2) = \sum_{\ell=1}^{\textsf{d}_2} \frac{g(w_\ell)}{\braket{\Psi_2}} \,\, \rho_{w_\ell}^{A_1}  \otimes \rho_{w_\ell}^{A_2} ~,
\label{sep-Psi12}
\end{align}
where $\rho_{X}^{A_1}$ and $\rho_{X}^{A_2}$ are the density matrices acting on Hilbert spaces $\mathcal{H}_{A_1}$ and $\mathcal{H}_{A_2}$ respectively and they correspond to the pure states:
\begin{align}
\rho_{X}^{A_1} &= |{{X}^{(1)}, {X}^{(2)}, \ldots, {X}^{(r)}}\rangle \langle{{X}^{(1)}, {X}^{(2)}, \ldots, {X}^{(r)}} | \nonumber \\
\rho_{X}^{A_2} &= |{{X}^{(r+1)}, {X}^{(r+2)}, \ldots, {X}^{(m)}}\rangle \langle{{X}^{(r+1)}, {X}^{(r+2)}, \ldots, {X}^{(m)}} | ~.
\end{align}
The coefficients are non-negative and they add up to 1:
\begin{equation}
\sum_{\ell=1}^{\textsf{d}_1} \frac{f(t_\ell)}{\braket{\Psi_1}} = 1 = \sum_{\ell=1}^{\textsf{d}_2} \frac{g(w_\ell)}{\braket{\Psi_2}} ~.
\end{equation}
Thus we see that the reduced density matrices are separable. From the above discussion, we can propose the following. \\[8pt]
\textbf{Proposition.} \emph{The necessary condition for the reduced density matrix $\rho$, corresponding to the states $|{\Psi_1}\rangle$ or $|{\Psi_2}\rangle$,  acting on $\left(\bigotimes_{i=1}^{r} \mathcal{H}_i \right)  \otimes \left(\bigotimes_{i=r+1}^{m} \mathcal{H}_i \right)$  to be separable is $R_{2r-1} \otimes R_{2r} = \bigoplus_{\ell} t_{\ell}$. On the other hand, the tensor product $R_{2i-1} \otimes  R_{2i}=\bigoplus_\ell t_\ell$ for each of the $m$ boundaries gives a sufficient condition for separability of $\rho$, i.e,}
\begin{empheq}[box=\fbox]{align}
   \rho = \text{separable} \quad &\Longrightarrow \quad a_r(t_\ell) = 1, \,\, \forall t_\ell \nonumber \\
a_0(t_\ell) = a_1(t_\ell) = \ldots = a_m(t_\ell) = 1, \,\, \forall t_\ell \quad &\Longrightarrow \quad \rho = \text{separable} ~.
	\label{prop-1}
\end{empheq}
For the group SU(2), the decomposition of the tensor product of any two representations is always multiplicity free. Thus one immediate result follows from the above discussion:\\[8pt]
\textbf{Corollary.} \emph{For SU(2) Chern-Simons theory, the states $\ket{\Psi_1}$ and $\ket{\Psi_2}$ are GHZ-like and the reduced density matrices are separable.}

Further if we consider all the representations (of SU($N$)) to be the same, then all the multiplicities will be equal (i.e. $a_i(t_\ell) = a_j(t_\ell)$ and $b_i(t_\ell) = b_j(t_\ell)$ for all values of $i$ and $j$). Thus $t_\ell$ and $w_\ell$ appearing without multiplicity in $R \otimes R$ and $R \otimes \bar{R}$ becomes a necessary and sufficient condition for reduced density matrices to be separable.  
\begin{equation}
\boxed{R \otimes R = \bigoplus_{\ell=1}^{\textsf{d}_1} t_\ell \Longleftrightarrow \rho(\Psi_1) = \text{separable} \quad;\quad R \otimes \bar{R} = \bigoplus_{\ell=1}^{\textsf{d}_2} w_\ell \Longleftrightarrow \rho(\Psi_2) = \text{separable}} ~. \nonumber
\end{equation}
From the analysis presented in this section, we can say that the multiplicity plays a crucial role in determining the entanglement structure of the quantum states $|{\Psi_1}\rangle$ and $|{\Psi_2}\rangle$. When there are no multiplicities, these states have a GHZ-like entanglement structure because the reduced density matrices are separable. However when the multiplicities are involved ($a_r > 1$ and $b_r>1$ to be precise), these states have a W-like entanglement structure and the reduced density matrices are non-separable.

So far in this section, we studied the entanglement properties of the states $\ket{\Psi_1}$ and $\ket{\Psi_2}$ and obtained the entropy values for a generic case which depends on the choice of representations carried by the Wilson lines. It also depends on the rank of the gauge group SU($N$) as well as the Chern-Simons level $k$. In the following subsection, we will consider explicit examples giving the analytic expressions of entropy as a function of $N$ and $k$. Further we will investigate how the entropy behaves in large $N$ or large $k$ limit and we provide various numerical plots supporting the results.  
\subsection{SU$(N)_k$ entanglement entropy and its large $k$ and large $N$ limit}
The spectrum of the reduced density matrix and hence the entanglement entropy depends on the quantum dimensions of various representations. For the SU$(N)$ group, these can be evaluated from the corresponding Young tableau using the $q$-numbers denoted as $[x]$. For example: 
\begin{equation}
\text{dim}_q\left(\,\text{$\tiny\yng(2,1)$}\,\right) = \frac{[N][N+1][N-1]}{[1][1][3]} ~.
\end{equation}
The $q$-numbers are written in terms of variable $q$ which is the primitive $(k+N)^{\text{th}}$ root of unity: $q=\exp\left(\tfrac{2\pi i}{k+N}\right)$. Thus the $q$-numbers for our computation will be given as,
\begin{equation}
[x] = \frac{q^{x/2} - q^{-x/2}}{q^{1/2} - q^{-1/2}} = \csc \left(\frac{\pi }{k+N}\right) \sin \left(\frac{\pi  x}{k+N}\right) ~.
\label{q-number}
\end{equation}
These numbers have the property that $[N+\alpha] = [k-\alpha]$. Thus the large $k$ limit or the large $N$ limit (with the other parameter fixed) of these numbers are given as following:
\begin{equation}
\lim_{k \to \infty}[\alpha] = \lim_{N \to \infty}[\alpha] = \alpha \quad;\quad \lim_{k \to \infty}[N\pm \alpha] = N\pm \alpha \quad;\quad \lim_{N \to \infty}[N \pm \alpha] = k \mp \alpha ~,
\end{equation}
where $\alpha$ is a finite positive integer. Hence in the $k \to \infty$ limit, $\text{dim}_q R \to \text{dim}(R)$ which is the usual dimension of the representation $R$. The corresponding $N \to \infty$ limit can be obtained by simply replacing each $(N\pm\alpha)$ factor in $\text{dim}(R)$ by $(k\mp\alpha)$. As an example, we will have:
\begin{equation}
\lim_{k \to \infty}\text{dim}_q\left(\,\text{$\tiny\yng(3)$}\,\right) = \frac{N(N+1)(N+2)}{6} \quad;\quad \lim_{N \to \infty}\text{dim}_q\left(\,\text{$\tiny\yng(3)$}\,\right) = \frac{k(k-1)(k-2)}{6} ~.
\end{equation}
Using this we can see that in the large $k$ or large $N$ limit, the eigenvalues of the reduced density matrix are rational numbers. The $k \to \infty$ limit of the entanglement entropy can be obtained by simply replacing $\text{dim}_q t \to \text{dim}(t)$ and $\text{dim}_q w \to \text{dim}(w)$ in the eq.(\ref{EE-Psi1}) and eq.(\ref{EE-Psi2}) respectively. The $N \to \infty$ limit can be computed as mentioned above where we first replace $\text{dim}_q t \to \text{dim}(t)$ and $\text{dim}_q w \to \text{dim}(w)$ and then replace all $(N\pm\alpha) \to (k\mp\alpha)$ in the expressions of $\text{dim}(t)$ and $\text{dim}(w)$ respectively.

In the following, we give some examples explicitly showing the computation of entanglement entropies and their large $k$ and large $N$ behavior.
\subsubsection{Wilson lines carrying symmetric representations of SU($N$)}
Let us compute the entropy when Wilson lines in figure \ref{No-braid-WLine} carry symmetric representations of SU$(N)$ as following:
\begin{equation}
R_{2j-1} = \sym{a} \quad,\quad R_{2j} = \sym{b} ~,
\end{equation}    
where we assume that $a \leq b$ without loss of generality. We will use the Young tableau notation in which a representation $R$ is specified by a set of non-negative integers: $R = (l_1, l_2,\ldots,l_{N-1})$ where $l_i$ denotes the number of boxes in the $i$-th row of Young tableau. From the group theory, we know the following tensor decompositions for $a \leq b$:
\begin{alignat}{3}
t &\in (R_{2j-1} \otimes R_{2j}) &&= \bigoplus_{\ell=0}^a \, (a{+}b{-}\ell, \ell) &&\equiv \bigoplus_{\ell=0}^a \,t_{\ell} \nonumber \\
w &\in (R_{2j-1} \otimes \bar{R}_{2j}) &&= \bigoplus_{\ell=0}^a \, (b{-}a{+}2\ell, b{-}a{+}\ell,\ldots,b{-}a{+}\ell) &&\equiv \bigoplus_{\ell=0}^a \,w_{\ell} ~.
\label{symm-decompose}
\end{alignat}
We should mention at this point that we want all the representations to be integrable (as already discussed in section 2) for which we must have $k \geq (a+b)$.\footnote{The maximum number of boxes in the first row of Young tableau happens for the representations $t_0$ and $w_a$ in the eq.(\ref{symm-decompose}) which is $(a+b)$. Thus $k \geq (a+b)$ will ensure that all the representations are integrable.}. The quantum dimensions needed for our computation are given as, 
\begin{align}
\text{dim}_q t_{\ell} &= \frac{[N+\ell-2]!\,[N+a+b-\ell-1]!\,[a+b-2\ell+1]}{[\ell]!\,[a+b-\ell+1]!\,[N-1]!\,[N-2]!} \nonumber \\[1pt]
\text{dim}_q w_{\ell} &= \frac{[N+\ell-2]!\,[N+b-a+\ell-2]!\,[N+b-a+2\ell-1]}{[\ell]!\,[b-a+\ell]!\,[N-1]!\,[N-2]!} ~,
\end{align}
where the factorial is defined as $[x]! = \prod_{i=1}^x [i]$ with $[0]!=1$. The 3j-phases for symmetric representations as given in \cite{Zodinmawia:2011osq} are:
\begin{equation}
\{R_{2j-1}, R_{2j}, t_{\ell}\} = (-1)^{\frac{a+b}{2}-\ell} \quad,\quad \{R_{2j-1}, \bar{R}_{2j}, w_{\ell}\} = (-1)^{\frac{b-a}{2}-\ell} ~.
\end{equation} 
Thus the quantum states for this example, up to overall phase, will be:
\begin{align} 
|{\Psi_1}\rangle = \sum_{\ell=0}^a  \frac{(-1)^{n\ell}}{(\text{dim}_q t_{\ell})^{\frac{n-2}{2}}} |{\phi_{\ell}^{(1)},\ldots, \phi_{\ell}^{(n)}}\rangle \,,\,\,\, |{\Psi_2}\rangle = \sum_{\ell=0}^a  \frac{(-1)^{n\ell}}{(\text{dim}_q w_{\ell})^{\frac{n-2}{2}}} |{\phi_{\ell}^{(1)},\ldots, \phi_{\ell}^{(n)}}\rangle~.
\label{psi1-symm}
\end{align}
The eigenvalues of the reduced density matrix for the bi-partition $(m|n-m)$ will be,
\begin{align} 
\lambda_{\rho_A}(\Psi_1) &= \frac{1}{\sum_{\ell=0}^a (\text{dim}_q t_{\ell})^{2-n}} \left\{ (\text{dim}_q t_{0})^{2-n}, (\text{dim}_q t_{1})^{2-n}, \ldots, (\text{dim}_q t_{a})^{2-n}  \right\} \nonumber \\ 
\lambda_{\rho_A}(\Psi_2) &= \frac{1}{\sum_{\ell=0}^a (\text{dim}_q w_{\ell})^{2-n}} \left\{ (\text{dim}_q w_{0})^{2-n}, (\text{dim}_q w_{1})^{2-n}, \ldots, (\text{dim}_q w_{a})^{2-n}  \right\} ~.
\end{align}
When $a=1$, i.e. $R_{2j-1}$ is a fundamental representation, each Hilbert space will be two dimensional and the eigenvalues are given as,
\begin{align} 
\lambda_{\rho_A}(\Psi_1) &= \left\{ \frac{\alpha_1^{n-2}}{\alpha_1^{n-2} + \beta_1^{n-2}}\,\,,\,\, \frac{\beta_1^{n-2}}{\alpha_1^{n-2} + \beta_1^{n-2}} \right\}  \nonumber \\ 
\lambda_{\rho_A}(\Psi_2) &= \left\{ \frac{\alpha_2^{n-2}}{\alpha_2^{n-2} + \beta_2^{n-2}}\,\,,\,\, \frac{\beta_2^{n-2}}{\alpha_2^{n-2} + \beta_2^{n-2}} \right\} ~,
\label{eigen-1-b}
\end{align} 
where we have defined,
\begin{alignat}{2} 
\alpha_1 &= \sin \left(\frac{\pi  (N-1)}{k+N}\right) \sin \left(\frac{\pi  b}{k+N}\right) \quad;\quad &&\beta_1 = \sin \left(\frac{\pi }{k+N}\right) \sin \left(\frac{\pi  (b+N)}{k+N}\right)  \nonumber \\ 
\alpha_2 &= \sin \left(\frac{\pi }{k+N}\right) \sin \left(\frac{\pi  b}{k+N}\right)  \quad;\quad &&\beta_2 = \sin \left(\frac{\pi  (N-1)}{k+N}\right) \sin \left(\frac{\pi  (b+N)}{k+N}\right) ~.
\end{alignat} 
Note that the two eigenvalues may become equal at certain values of $N$ and $k$, i.e. $\alpha_1 = \beta_1$ and $\alpha_2 = \beta_2$ respectively. In such a case, the entanglement entropy will be maximum ($\ln 2$) and the two states $|{\Psi_1}\rangle$ and $|{\Psi_2}\rangle$ will be the maximally entangled GHZ states. The values at which this may happen are:
\begin{equation}
\boxed{|{\Psi_1}\rangle = |{\text{GHZ}}\rangle: \begin{cases}
\text{SU$(2)_{k=2b}$} \\
\text{SU$(3)_{k=9}$ for $b=2$} \\
\text{SU$(N)_{k=N}$ for $b=1$}
\end{cases}}
 \boxed{|{\Psi_2}\rangle = |{\text{GHZ}}\rangle: \begin{cases}
\text{SU$(2)_{k=2b}$} \\
\text{SU$(3)_{k=9}$ for $b=7$} \\
\text{SU$(b+1)_{k=b+1}$}
\end{cases}}  ~.
\label{GHZ-fund}
\end{equation} 
Thus if we consider the case where all the Wilson lines carry the fundamental representation of SU($N$), then the state $|{\Psi_1}\rangle$ will be a GHZ state for the group SU$(N)_{k=N}$ and the state $|{\Psi_2}\rangle$ will be a GHZ state for the group SU$(2)_{k=2}$.

\begin{figure}[t]
\centerline{\includegraphics[width=6.0in]{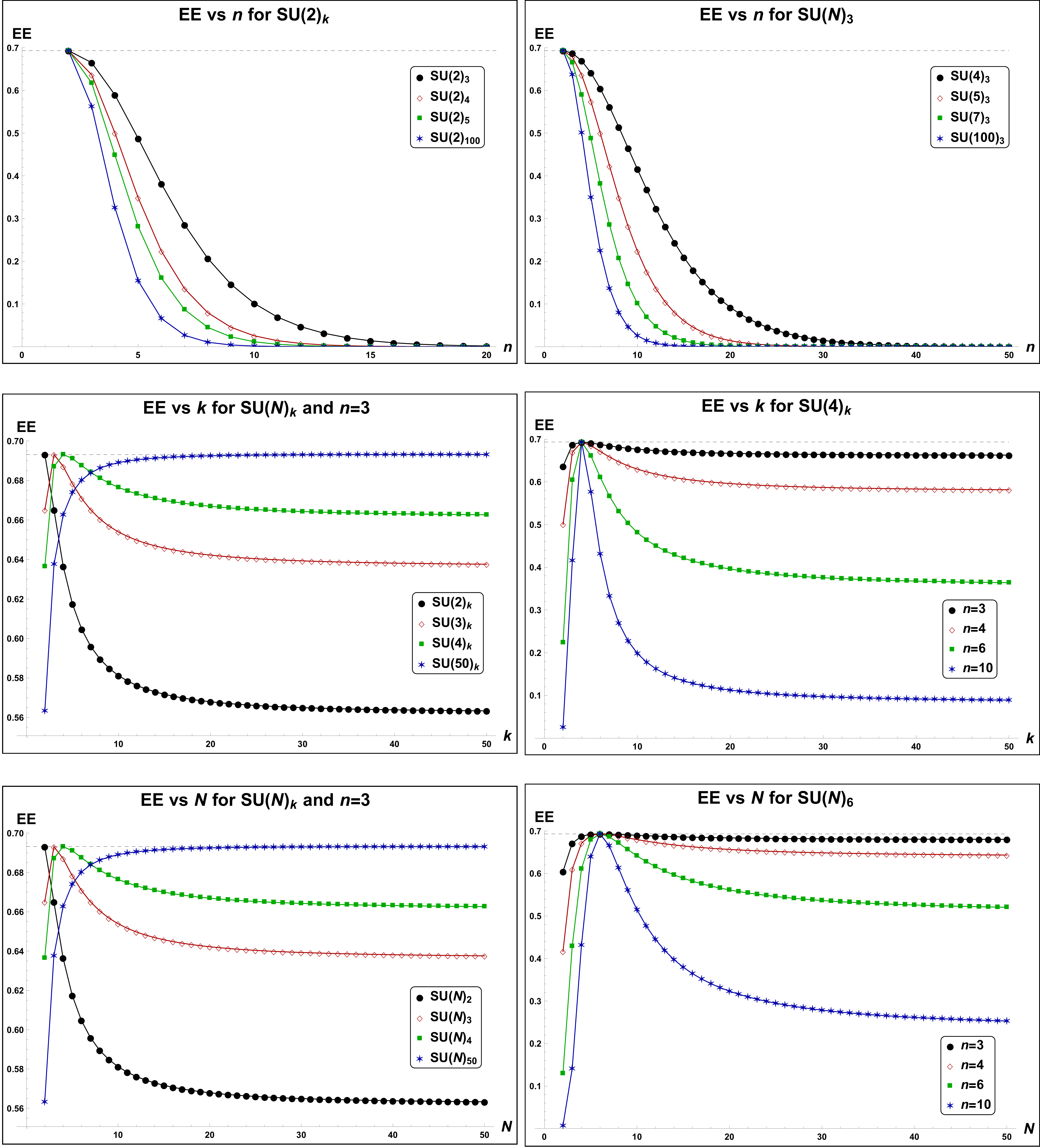}}
\caption[]{The variation of SU$(N)_k$ entanglement entropy for the state $|{\Psi_1}\rangle$ when all the Wilson lines carry fundamental representation. The horizontal dotted line corresponds to $\text{EE} = \ln2$ which is the maximum possible value for entropy.}
\label{EEPlotsPsi1-fund}
\end{figure}

\begin{figure}[t]
\centerline{\includegraphics[width=6.0in]{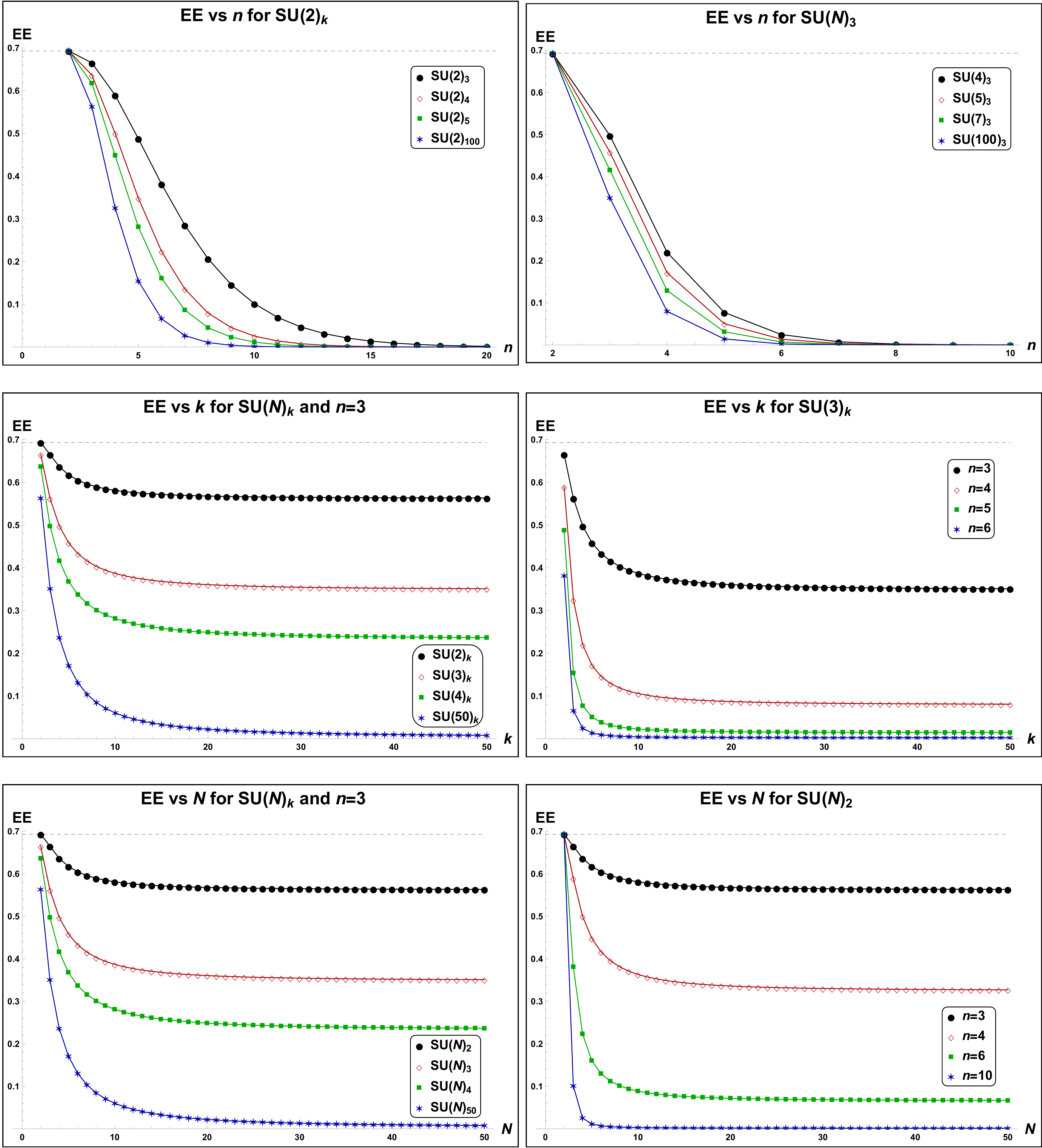}}
\caption[]{The variation of SU$(N)_k$ entanglement entropy for the state $|{\Psi_2}\rangle$ when all the Wilson lines carry fundamental representation. The horizontal dotted line corresponds to $\text{EE} = \ln2$ which is the maximum possible value for entropy.}
\label{EEPlotsPsi2-fund}
\end{figure}

In figure \ref{EEPlotsPsi1-fund}, we have plotted the behavior of entanglement entropy for the state $|{\Psi_1}\rangle$ when $b=1$, i.e. when all the representations $R_i$ are fundamental.   
We can see that entropy decreases and eventually goes to 0 as we increase the number of boundaries $n$ (for a fixed $k$ and $N$). However, when either $k$ or $N$ increases (with other parameters fixed), the entropy first increases, becomes maximum at $k=N$ (which is consistent from eq.(\ref{GHZ-fund})) and then gradually decreases and converges to a finite value as $k \to \infty$ or $N \to \infty$. The third and the fifth plots in figure \ref{EEPlotsPsi1-fund} also show that when both $k$ as well as $N$ become large, the entanglement entropy converges to its maximum value $\ln 2$. This behavior can be seen from eqn.(\ref{eigen-1-b}). When either $k \to \infty$ or $N \to \infty$ (and the other parameter is fixed), the entropy converges as following:
\begin{align}
\lim_{k \to \infty}\text{EE}(\Psi_1) &= \tfrac{x_1^n \log \left(x_1^{2-n}+1\right)+x_1^2 \log \left(x_1^{n-2}+1\right)}{x_1^n+x_1^2},\, \lim_{N \to \infty}\text{EE}(\Psi_1) = \tfrac{y_1^n \log \left(y_1^{2-n}+1\right)+y_1^2 \log \left(y_1^{n-2}+1\right)}{y_1^n+y_1^2} \nonumber \\
\lim_{k \to \infty}\text{EE}(\Psi_2) &= \tfrac{x_2^n \log \left(x_2^{2-n}+1\right)+x_2^2 \log \left(x_2^{n-2}+1\right)}{x_2^n+x_2^2},\, \lim_{N \to \infty}\text{EE}(\Psi_2) = \tfrac{y_2^n \log \left(y_2^{2-n}+1\right)+y_2^2 \log \left(y_2^{n-2}+1\right)}{y_2^n+y_2^2} \nonumber,
\end{align}
where we have,
\begin{equation}
x_1 = \frac{N+b}{b(N-1)},\quad y_1 = \frac{k-b}{b(k+1)},\quad x_2 = \frac{(N+b)(N-1)}{b},\quad y_2 = \frac{(k-b)(k+1)}{b} ~.
\end{equation}
When both $k \to \infty$ and $N \to \infty$ (in no particular order), the entropy converges to:
\begin{equation}
\lim_{\substack{k \to \infty \\ N \to \infty}}\text{EE}(\Psi_1) = \frac{b^n \log \left(b^{2-n}+1\right)+b^2 \log \left(b^{n-2}+1\right)}{b^n+b^2} \quad;\quad \lim_{\substack{k \to \infty \\ N \to \infty}}\text{EE}(\Psi_2) = \ln(2)\, \delta_{n,2} ~.
\label{EElargekN-con}
\end{equation}
The first term for $b=1$ becomes $\ln 2$. Thus the state $|{\Psi_1}\rangle$ becomes a GHZ state and  $|{\Psi_2}\rangle$ becomes a non-entangled state in the large level and large rank limit:
\begin{equation}
\boxed{\text{Wilson lines in fundamental representation of SU$(N)$: } \lim_{N \to \infty} \lim_{k \to \infty} |{\Psi_1}\rangle = |{\text{GHZ}}\rangle} ~.
\end{equation}
For completeness, we have also given various plots in figure \ref{EEPlotsPsi2-fund}, showing the behavior of entanglement entropy for the state $|{\Psi_2}\rangle$ for fundamental representations of SU$(N)$.   

Note that using the level-rank duality \cite{Naculich:2007nc, naculich1990group, mlawer1991group, naculich1990duality}, our entropy results for representations $R$ can be extended to the representations $R^T$ by simply exchanging $N \leftrightarrow k$.\footnote{We thank Howard Schnitzer for pointing this out.} Here $R^T$ represents a transposed representation where the Young tableau for $R^T$ is obtained by transposing the Young tableau of $R$. As an example, 
\begin{equation}
R = \text{$\tiny\yng(3,2)$} \quad \Longrightarrow \quad  R^T = \text{$\tiny\yng(2,2,1)$} ~.
\end{equation}
Now consider the decomposition of tensor product of symmetric representations in eq.(\ref{symm-decompose}). For the transposed representations, this decomposition can be given as,
\begin{align}
(R_{2j-1} \otimes R_{2j}) = \bigoplus_{\ell=0}^a \,t_{\ell} \quad &\Longrightarrow \quad (R_{2j-1}^T \otimes R_{2j}^T) = \bigoplus_{\ell=0}^a \,t_{\ell}^T \nonumber \\
(R_{2j-1} \otimes \bar{R}_{2j}) = \bigoplus_{\ell=0}^a \, w_{\ell} \quad &\Longrightarrow \quad (R_{2j-1}^T \otimes \bar{R}_{2j}^T) = \bigoplus_{\ell=0}^a \, w_{\ell}^T ~.
\label{}
\end{align} 
Moreover, using the property $[N+\alpha] = [k-\alpha]$ of the $q$-number for any integer $\alpha$, it is not difficult to see that the quantum dimensions of $t_{\ell}$ and $w_{\ell}$ in SU$(N)_k$ will be the same as the quantum dimensions of $t_{\ell}^T$ and $w_{\ell}^T$ respectively in SU$(k)_N$. Thus the SU$(N)_k$ entropy obtained for the representations $R_{2j-1}$ and $R_{2j}$ will be equal to the SU$(k)_N$ entropy for the representations $R_{2j-1}^T$ and $R_{2j}^T$.

The tensor decomposition of two symmetric representations of SU($N$) is multiplicity free. Thus the reduced density matrices for this example will be separable with vanishing entanglement negativity. However while taking the tensor product of mixed representations of SU($N$), we will encounter multiplicity. In the following, we will consider two examples of mixed representations and give the results for entropy and negativity. 
\subsubsection{Wilson lines carrying mixed representations of SU($N$)}
\begin{figure}[t]
\centerline{\includegraphics[width=6.0in]{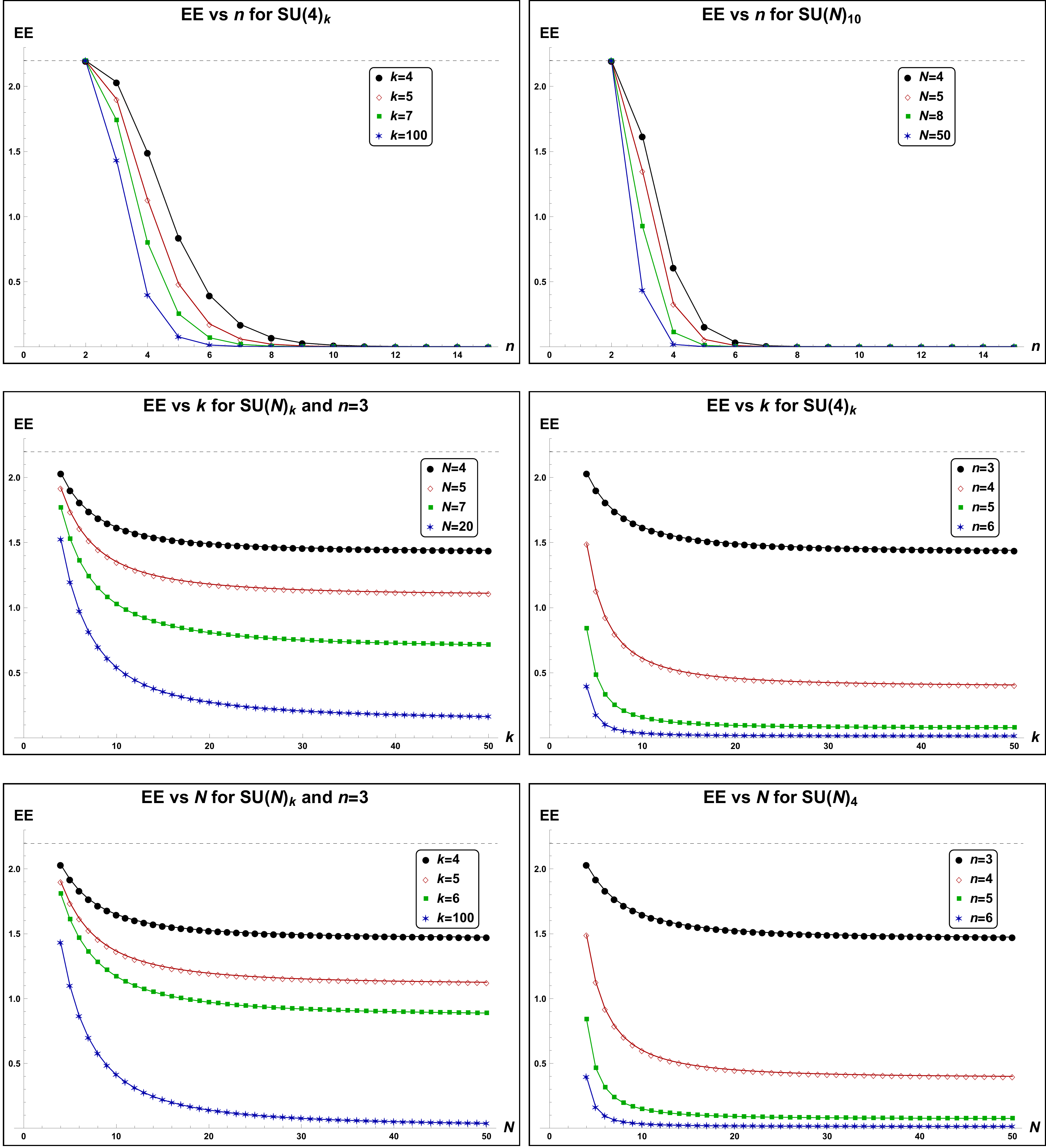}}
\caption[]{The variation of SU$(N)_k$ entanglement entropy when the Wilson lines carry adjoint representation. The first two plots show that $\text{EE} \to 0$ for $n \gg 1$. The other four plots show that entropy decreases and converges to a fixed value as $k \to \infty$ or $N \to \infty$ respectively.}
\label{EEPlotsPsi1}
\end{figure}

Let us first consider the case when all the Wilson lines carry the adjoint representation of SU($N$). Since adjoint representation is self conjugate, the two states $|{\Psi_1}\rangle$ and $|{\Psi_2}\rangle$ will be the same. We will have the following decomposition for $N \geq 4$:\footnote{For SU(2), $\text{adj} \otimes \text{adj} = \bullet \oplus \text{$\tiny\yng(2)$} \oplus \text{$\tiny\yng(4)$}$ which has already been covered in previous section (see eq.(\ref{symm-decompose})). For SU(3), we have to discard representation $t_2$ in eq.(\ref{adj-decompose}).}
\begin{equation}
\text{adj} \otimes \text{adj} = t_1 \oplus t_2 \oplus t_3 \oplus t_4 \oplus t_5 \oplus 2\,t_6 ~,
\label{adj-decompose}
\end{equation}
where various representations are given as following:
\begin{equation}
\{t_1, t_2, t_3, t_4, t_5, t_6 \} = \left\{ \bullet,\, (2^2, 1^{N-4}),\, (3, 1^{N-3}),\, (3^2, 2^{N-3}),\, (4, 2^{N-2}),\, (2, 1^{N-2}) \right\} ~.
\end{equation}
In writing these representations, we have used a compact notation to denote the number of boxes in various rows of the Young tableau, where $(l_i^x)$ means that $l_i$ is repeated $x$ times. For example, $(2^2, 1^2)$ denotes the representation $(2,2,1,1)$ in which the first two rows of the Young tableau have two boxes each and the third and fourth rows have one box each. 
The quantum dimensions of these representations will be,
\begin{align}
&\text{dim}_q t_2 = \frac{[N+1][N-3][N]^2}{[2]^2[N-2]},\quad \text{dim}_q t_3 = \text{dim}_q t_4 = \frac{[N+2][N-2][N+1][N-1]}{[2]^2} \nonumber \\
&\text{dim}_q t_5 = \frac{[N+3][N-1][N]^2}{[2]^2},\quad \text{dim}_q t_6 = [N+1][N-1],\quad \text{dim}_q t_1 = 1 ~.
\end{align}
Thus the quantum state $|{\Psi_1}\rangle = |{\Psi_2}\rangle \equiv |{\Psi}\rangle$ for this case can be written as,
\begin{equation}
|{\Psi}\rangle = \sum_{\ell=1}^5 \frac{ \left\{\text{adj}, \text{adj},t_\ell \right\}^n }{(\text{dim}_q t_\ell)^{\frac{n-2}{2}}}  \bigotimes_{i=1}^n |{\phi_{\ell}^{(i)}}\rangle + \sum_{x_1, \ldots, x_n}\frac{\prod_{j=1}^n \left\{ \text{adj}, \text{adj},t_6, x_j \right\} }{(\text{dim}_q t_6)^{\frac{n-2}{2}}} \bigotimes_{i=1}^n |{\phi_{6,\, x_{i-1} x_i}^{(i)}}\rangle ,
\end{equation}
where each variable $x_i$ takes value 1 and 2. The non-vanishing eigenvalues of the reduced density matrix will be,
\begin{equation} 
\lambda_{\rho_A} = \frac{1}{\braket{\Psi}} \left\{ \lambda_{t_1}, \lambda_{t_2}, \lambda_{t_3}, \lambda_{t_4}, \lambda_{t_5}, \lambda_{t_6}, \lambda_{t_6}, \lambda_{t_6}, \lambda_{t_6} \right\} ~,
\end{equation}
where $\braket{\Psi} = \lambda_{t_1}+\lambda_{t_2}+\lambda_{t_3}+\lambda_{t_4}+\lambda_{t_5}+4\lambda_{t_6}$ and,
\begin{equation}
\lambda_{t_\ell} = \frac{1}{(\text{dim}_q t_\ell)^{n-2}} \quad (\text{for } \ell=1,2,3,4,5) \quad;\quad \lambda_{t_6} = \frac{2^{n-2}}{(\text{dim}_q t_6)^{n-2}} ~.
\end{equation}
The entanglement entropy will be maximum for $n=2$ which is $\ln 9$ (for SU(3), it is $\ln 8$) and decreases and goes to 0 as $n$ increases as shown in the first two plots of figure \ref{EEPlotsPsi1}. Similarly for a given $n$, the entropy decreases as $k$ or $N$ increases (with other parameter fixed) but converges to a fixed value as $k \to \infty$ or $N \to \infty$ respectively which is clear from the plots in figure \ref{EEPlotsPsi1}.
These limiting values can be obtained from the eigenvalues of the reduced density matrix which converge to the following when $k \to \infty$ for a fixed $N$:
\begin{align} 
\lim_{k \to \infty}\lambda_{\rho_A} &= \frac{1}{\text{trace}_1} \left\{ 1, \tfrac{4^{n-2} (N-2)^{n-2}}{\left(N^4-2 N^3-3 N^2\right)^{n-2}}, \tfrac{4^{n-2}}{\left(N^4-5 N^2+4\right)^{n-2}}, \tfrac{4^{n-2}}{\left(N^4-5 N^2+4\right)^{n-2}}, \right. \nonumber \\
& \left. \tfrac{4^{n-2}}{\left(N^4+2 N^3-3 N^2\right)^{n-2}}, \tfrac{2^{n-2}}{\left(N^2-1\right)^{n-2}}, \tfrac{2^{n-2}}{\left(N^2-1\right)^{n-2}}, \tfrac{2^{n-2}}{\left(N^2-1\right)^{n-2}}, \tfrac{2^{n-2}}{\left(N^2-1\right)^{n-2}} \right\} ~.
\end{align}
Similarly when $N \to \infty$ for a fixed $k$, the eigenvalues converge to:
\begin{align} 
\lim_{N \to \infty}\lambda_{\rho_A} &= \frac{1}{\text{trace}_2} \left\{1,\tfrac{4^{n-2} (k+2)^{n-2}}{\left(k^4+2 k^3-3 k^2\right)^{n-2}},\tfrac{4^{n-2}}{\left(k^4-5 k^2+4\right)^{n-2}},\tfrac{4^{n-2}}{\left(k^4-5 k^2+4\right)^{n-2}}, \right. \nonumber \\
& \left. \tfrac{4^{n-2}}{\left(k^4-2 k^3-3 k^2\right)^{n-2}}, \tfrac{2^{n-2}}{\left(k^2-1\right)^{n-2}},\tfrac{2^{n-2}}{\left(k^2-1\right)^{n-2}},\tfrac{2^{n-2}}{\left(k^2-1\right)^{n-2}},\tfrac{2^{n-2}}{\left(k^2-1\right)^{n-2}}\right\} ~,
\end{align}
where $\text{trace}_1$ and $\text{trace}_2$ are appropriate factors which make sure that the reduced density matrix has unit trace. When both $k \to \infty$ and $N \to \infty$ (in no particular order), then $\text{EE} \to 0$ and the state $|{\Psi}\rangle$ tends to an unentangled state:
\begin{equation}
\lim_{\substack{k \to \infty \\ N \to \infty}} \text{EE}(\Psi) = 0 ~.
\label{EEAdj-con}
\end{equation}
Moreover, using the results of the earlier sections, we can also compute the entanglement negativity which comes out to be:
\begin{equation}
\mathcal{N} = \frac{2^{n-1}\, (\text{dim}_q t_6)^{2-n}}{\braket{\Psi}} = \frac{2^{n-1}\, [N+1]^{2-n}\, [N-1]^{2-n}}{\braket{\Psi}} ~.
\end{equation}
We can see that $\mathcal{N} \neq 0$ and hence the state $|{\Psi}\rangle$ has a W-like entanglement structure with non-separable reduced density matrix.

Next consider an example when all the Wilson lines carry the mixed representation $(2,1) = \text{$\tiny\yng(2,1)$}$ of SU($N$). In this case, we will have:
\begin{align}
R_{2j-1} \otimes R_{2j} &= t_1 \oplus t_2 \oplus t_3 \oplus t_4 \oplus t_5 \oplus t_6 \oplus 2\,t_7 \nonumber \\
R_{2j-1} \otimes \bar{R}_{2j} &= w_1 \oplus w_2 \oplus w_3 \oplus w_4 \oplus w_5 \oplus w_6 \oplus 2\,w_7 ~.
\end{align}
where various representations $t_\ell \in (R_{2j-1} \otimes R_{2j})$ and $w_\ell \in (R_{2j-1} \otimes \bar{R}_{2j})$ are given as following:\footnote{For SU(3), representations $t_2, t_3$ and $w_2, w_6$ do not appear in the tensor decomposition.}
\begin{align}
\{t_1 \ldots t_7 \} &= \left\{ (2^3),\, (2^2, 1^2),\, (3, 1^3),\, (3^2),\, (4, 1^2),\, (4, 2),\, (3, 2, 1) \right\} \nonumber \\
\{w_1 \ldots w_7 \} &= \left\{ \bullet, (2^2, 1^{N-4}), (3, 1^{N-3}), (3^2, 2^{N-3}), (4, 2^{N-2}), (4, 3, 2^{N-4}, 1), (2, 1^{N-2}) \right\} \nonumber ~.
\end{align}
The non-vanishing eigenvalues of the reduced density matrix are:
\begin{align} 
\lambda(\Psi_1) &= \frac{1}{\braket{\Psi_1}} \left\{ \lambda_{t_1}, \lambda_{t_2}, \lambda_{t_3}, \lambda_{t_4}, \lambda_{t_5}, \lambda_{t_6}, \lambda_{t_7}, \lambda_{t_7}, \lambda_{t_7}, \lambda_{t_7} \right\} \nonumber \\
\lambda(\Psi_2) &= \frac{1}{\braket{\Psi_2}} \left\{ \lambda_{w_1}, \lambda_{w_2}, \lambda_{w_3}, \lambda_{w_4}, \lambda_{w_5}, \lambda_{w_6}, \lambda_{w_7}, \lambda_{w_7}, \lambda_{w_7}, \lambda_{w_7} \right\} ~,
\end{align}
where $\braket{\Psi_1} = \lambda_{t_1} + \lambda_{t_2} + \lambda_{t_3} + \lambda_{t_4} + \lambda_{t_5} + \lambda_{t_6} + 4\lambda_{t_7}$ and $\braket{\Psi_2} = \lambda_{w_1} + \lambda_{w_2} + \lambda_{w_3} + \lambda_{w_4} + \lambda_{w_5} + \lambda_{w_6} + 4\lambda_{w_7}$ and we have:
\begin{alignat}{2}
\lambda_{t_\ell} &= \frac{1}{(\text{dim}_q t_\ell)^{n-2}} \quad (\text{for } \ell=1,2,3,4,5,6) &\quad;\quad \lambda_{t_7} &= \frac{2^{n-2}}{(\text{dim}_q t_7)^{n-2}} \nonumber \\
\lambda_{w_\ell} &= \frac{1}{(\text{dim}_q w_\ell)^{n-2}} \quad (\text{for } \ell=1,2,3,4,5,6) &\quad;\quad \lambda_{w_7} &= \frac{2^{n-2}}{(\text{dim}_q w_7)^{n-2}} ~.
\end{alignat}
When $k \to \infty$, the quantum dimensions become the usual dimension which for the SU($N$) group are given as:
\begin{align}
& \text{dim}(t_1) = \tfrac{(N^2-N) (N+1)!}{144 (N-3)!} \, , \, \text{dim}(t_2) = \tfrac{N (N+1)!}{80 (N-4)!} \, , \, \text{dim}(t_3) = \tfrac{(N+2)!}{72 (N-4)!} \, , \, \text{dim}(t_4) = \tfrac{(N^2+N) (N+2)!}{144 (N-2)!} \nonumber \\
& \text{dim}(t_5) = \tfrac{(N+3)!}{72 (N-3)!} \, , \, \text{dim}(t_6) = \tfrac{N (N+3)!}{80 (N-2)!} \, , \, \text{dim}(t_7) = \tfrac{N (N+2)!}{45 (N-3)!} \, , \, \text{dim}(w_1) = 1 \, , \, \text{dim}(w_2) = \tfrac{N^2(N-3)(N+1)}{4} \nonumber \\
& \text{dim}(w_3) = \text{dim}(w_4) = \tfrac{(N-2)(N-1)(N+1)(N+2)}{4} \, , \, \text{dim}(w_5) = \tfrac{N^2(N-1)(N+3)}{4} \nonumber \\ 
& \text{dim}(w_6) = \tfrac{(N-3)(N-1)^2(N+1)^2(N+3)}{9} \, , \, \text{dim}(w_7) = (N-1)(N+1) ~.
\label{dim(2,1)}
\end{align}
The entanglement entropy can be computed for $k \to \infty$ by simply replacing $\text{dim}_q t_\ell$ and $\text{dim}_q w_\ell$ by $\text{dim}(t_\ell)$  and $\text{dim}(w_\ell)$ respectively as given in eq.(\ref{dim(2,1)}). Similarly the $N \to \infty$ limit of the entropy can be obtained by replacing the quantum dimensions by usual dimensions and then replacing each $(N \pm x)$ factor in eq.(\ref{dim(2,1)}) by $(k \mp x)$. When both $k$ and $N$ goes to infinity (in no particular order), the eigenvalues of the reduced density matrix converge to the following:
\begin{align} 
\lim_{\substack{k \to \infty \\ N \to \infty}}\lambda(\Psi_1^{(n \geq 2)}) &= \left\{ \tfrac{5^2 9^n}{\alpha}, \tfrac{9^2 5^n}{\alpha}, \tfrac{5^2 2^{2-n} 9^n}{\alpha}, \tfrac{5^2 9^n}{\alpha}, \tfrac{5^2 2^{2-n} 9^n}{\alpha}, \tfrac{9^2 5^n}{\alpha}, \tfrac{8^{2-n} 45^n}{\alpha}, \tfrac{8^{2-n} 45^n}{\alpha}, \tfrac{8^{2-n} 45^n}{\alpha}, \tfrac{8^{2-n} 45^n}{\alpha} \right\} \nonumber \\
\lim_{\substack{k \to \infty \\ N \to \infty}}\lambda(\Psi_2^{(n \geq 3)}) &= \left\{1,0,0,0,0,0,0,0,0,0 \right\};\quad \lambda(\Psi_2^{(n = 2)}) = \left\{\tfrac{1}{10}, \tfrac{1}{10}, \tfrac{1}{10}, \tfrac{1}{10}, \tfrac{1}{10}, \tfrac{1}{10}, \tfrac{1}{10}, \tfrac{1}{10}, \tfrac{1}{10}, \tfrac{1}{10} \right\} ~,
\label{EE-mixed-con}
\end{align}
where $ \alpha \equiv 162 \times 5^n+50 \times 9^n+25 \times 2^{3-n} 9^n+2^{8-3 n} 45^n$. Thus in this limit, the state $|{\Psi_2^{(n \geq 3)}}\rangle$ becomes an unentangled state. The state $|{\Psi_1}\rangle$ however remains entangled and the entropy decreases monotonically as $n$ increases and converges to $\ln 2$ as $n$ becomes very large as shown in figure \ref{EEkNinfvsn}:
\begin{figure}[t]
\centerline{\includegraphics[width=4.0in]{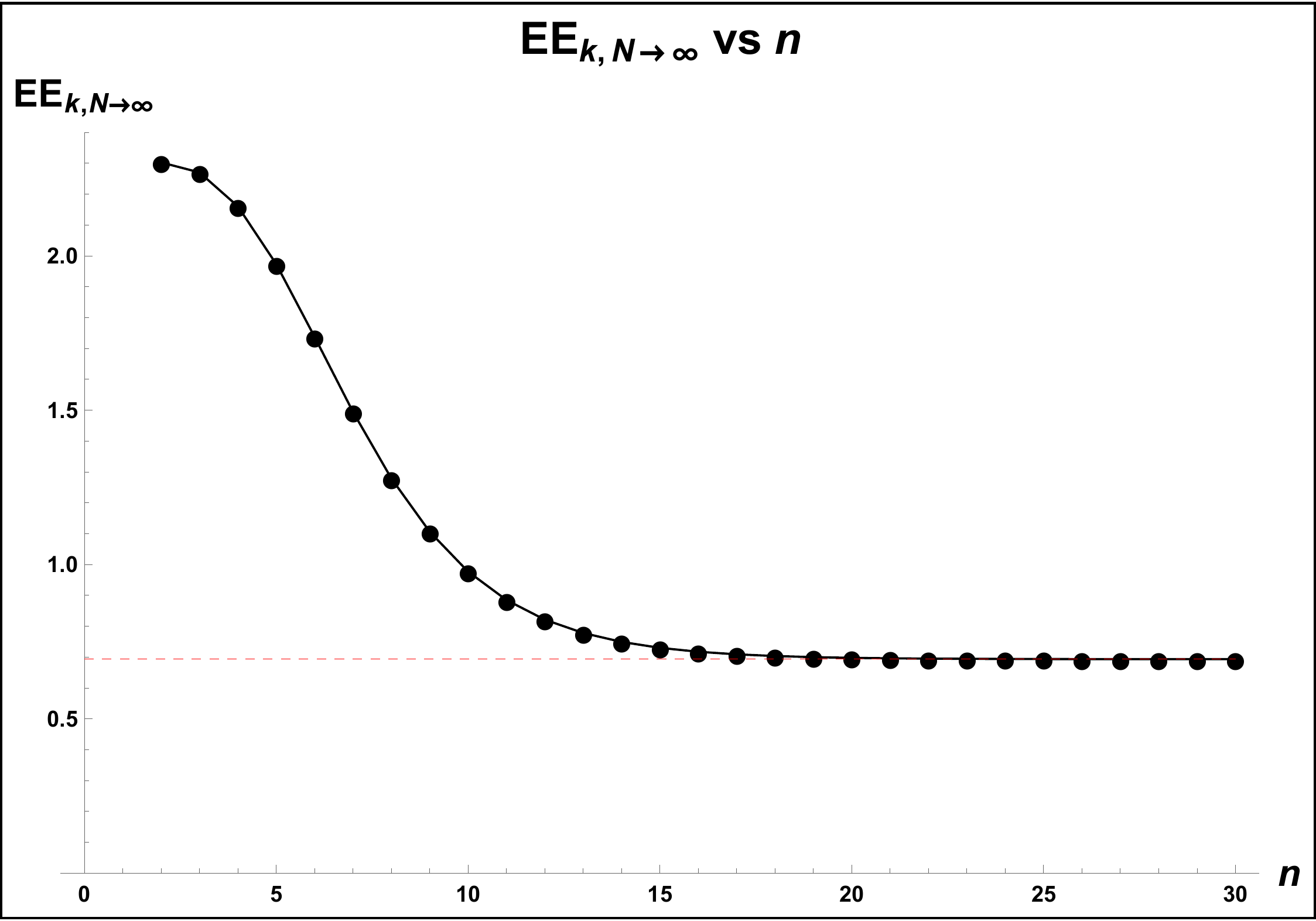}}
\caption[]{The plot showing the variation of $\lim(\substack{k \to \infty \\ N \to \infty})\text{EE}(\Psi_1)$ as a function of $n$. The entropy is maximum ($\ln 10$) at $n=2$ and converges to $\ln 2$ as $n \to \infty$.}
\label{EEkNinfvsn}
\end{figure}
\begin{equation}
\left(\lim_{\substack{k \to \infty \\ N \to \infty}} \text{EE}(\Psi_1)\right)_{n \to \infty} = \ln(2) ~. 
\label{EEminforrep=(2,1)}
\end{equation}
The entanglement negativity in this example can be evaluated as:
\begin{equation}
\mathcal{N}(\Psi_1) = \frac{2^{n-1}\, (\text{dim}_q t_7)^{2-n}}{\text{trace}_1} \quad;\quad \mathcal{N}(\Psi_2) = \frac{2^{n-1}\, (\text{dim}_q w_7)^{2-n}}{\text{trace}_2} ~.
\end{equation}
Clearly $\mathcal{N} \neq 0$ and the states have a W-like entanglement structure with non-separable reduced density matrices.
\subsection{Braiding between the Wilson lines for states $\ket{\Psi_1}$ and $\ket{\Psi_2}$}
The braiding between the Wilson lines for the quantum states $|{\Psi_1}\rangle$ and $|{\Psi_2}\rangle$ can be achieved by acting the braiding operators ($b_1$, $b_2$ or $b_3$) on the basis states.\footnote{One such example was shown in the figure \ref{Alpha-States} for the state $\ket{\alpha_1}$ which is the $n=2$ case of $|{\Psi_1}\rangle$.} Let us consider $|{\Psi_1}\rangle$ first. Any braiding between the Wilson lines carrying representations $R_{2j-1}$ and $R_{2j}$ can be seen as the action of $b_1^{(+)}$ operator on the $j^{\text{th}}$ boundary. Similarly the braiding between the Wilson lines $R_{2j-2}$ and $R_{2j-3}$ can be seen as the action of $b_3^{(+)}$ operator on the $j^{\text{th}}$ boundary. The action of these operators on the $j^{\text{th}}$ basis is given as:
\begin{align}
b_1^{(+)} |{\phi_{t,\, x_{j-1} x_j}^{(j)}}\rangle &= \{R_{2j-1}, R_{2j}, t, x_{j-1}\} \exp\left(\pm\frac{i\pi (C_{R_{2j-1}}+C_{R_{2j}}-C_{t})}{k+N}\right) |{\phi_{t,\, x_{j-1} x_j}^{(j)}}\rangle \nonumber \\
b_3^{(+)} |{\phi_{t,\, x_{j-1} x_j}^{(j)}}\rangle &= \{R_{2j-2}, R_{2j-3}, t, x_{j}\} \exp\left(\pm\frac{i\pi (C_{R_{2j-2}}+C_{R_{2j-3}}-C_{t})}{k+N}\right) |{\phi_{t,\, x_{j-1} x_j}^{(j)}}\rangle \nonumber ~.
\end{align}
Thus the new basis state is related to the old basis by U(1) transformation, which does not affect the entanglement structure. It is also possible to have braiding between the Wilson lines $R_{2j}$ and $R_{2j-2}$ which can be obtained by applying $b_2^{(-)}$ on the $j^{\text{th}}$ basis state. Since the operator $b_2$ does not act directly on the `$t$' basis, we need to change it to `$s$' basis using the transformation rules given in eq.(\ref{basis-transform}) and then apply $b_2$. Then we can transform back to the `$t$' using eq.(\ref{basis-transform}). The end result will be the following basis state:
\begin{align}
b_2^{(-)} |{\phi_{t,\, x_{j-1} x_j}^{(j)}}\rangle &= \sum_{\substack{s,u,v\\t',\,x_{j-1}',\, x_j'}} a_{s,uv}^{t,\,x_{j-1} x_j}\left[
\begin{array}{cc}
R_{2j-1} & R_{2j} \\
\bar{R}_{2j-2} & \bar{R}_{2j-3}
\end{array}
\right] \, a_{s,uv}^{t',\,x_{j-1}' x_j'}\left[
\begin{array}{cc}
R_{2j-1} & R_{2j} \\
\bar{R}_{2j-2} & \bar{R}_{2j-3}
\end{array}
\right]^{*} \nonumber \\
&\times \{R_{2j}, \bar{R}_{2j-2}, t, x_{j}\} \exp\left(\pm\frac{i\pi (C_{{R}_{2j}}+C_{\bar{R}_{2j-2}}-C_{s})}{k+N}\right) |{\phi_{t',\, x_{j-1}' x_j'}^{(j)}}\rangle   ~.
\end{align}
If we denote the Racah matrix $a_{s,uv}^{t,ij}$ as $\mathcal{R}$ and the matrix consisting of the eigenvalues of the braiding operator $b_2$ as $\Lambda$, we can write the above equation as,
\begin{equation}
b_2^{(-)} |\phi_{t,\, x_{j-1} x_j}^{(j)}\rangle = \sum_{t',\, x_{j-1}',\, x_j'} \left({\mathcal{R}}^{\dagger} \Lambda \mathcal{R} \right)_{t',\,x_{j-1}' x_j'}^{t,\,x_{j-1} x_j} |{\phi_{t',\, x_{j-1}' x_j'}^{(j)}}\rangle ~.
\end{equation}
where the subscript and superscript in $({\mathcal{R}}^{\dagger} \Lambda \mathcal{R})$ label the row and column respectively of the matrix ${\mathcal{R}}^{\dagger} \Lambda \mathcal{R}$. Since $\mathcal{R}$ and $\Lambda$ are the unitary matrices, so is ${\mathcal{R}}^{\dagger} \Lambda \mathcal{R}$ and hence the new basis is related to old one by unitary transformation. A similar analysis can also be done for the state $|{\Psi_2}\rangle$. This means that any braiding between the Wilson lines for the states $|{\Psi_1}\rangle$ and $|{\Psi_2}\rangle$ will not affect the entanglement structure of these states. This is in accordance with \cite{Melnikov:2018zfn} where, using the replica trick, the authors show that the entanglement structure of the state $|{\alpha_1}\rangle$ of figure \ref{Alpha-States} does not change under the braiding between the Wilson lines.  

Thus we see that if the Wilson lines only braid locally, the entanglement structure of the quantum states remains same. In order to see the affect of the braiding on the entanglement entropy, we must consider the states where Wilson lines connecting punctures on one of the $S^2$ braid with the Wilson lines connecting the punctures on a different $S^2$, which we call as horizontal braiding (necklaces). In the next section, we study the states having horizontal braiding between the Wilson lines and study its affect on the entanglement structure.
\section{Horizontal braiding and its affect on entanglement entropy}
\label{sec4}
In this section, we will consider the states with horizontal braiding and study their entanglement structure by explicitly evaluating the entropy and negativity. Particularly, we will work out the states corresponding to the necklaces in the following subsections.
\subsection{Quantum state $\ket{\Psi_3}$ on $S^2 \cup S^2$}
Consider the state $\ket{\Psi_3}$ shown in figure \ref{v1-state}. The Wilson lines carrying representations $R_1$ and $R_2$ undergo horizontal braiding $2p$ number of times (i.e. twisted $p$ number of times) while the other two Wilson lines with representations $R_3$ and $R_4$ go straight from one boundary to the other. This state can be constructed as given in Appendix A and is given as:
\begin{figure}[h]
\centerline{\includegraphics[width=2.0in]{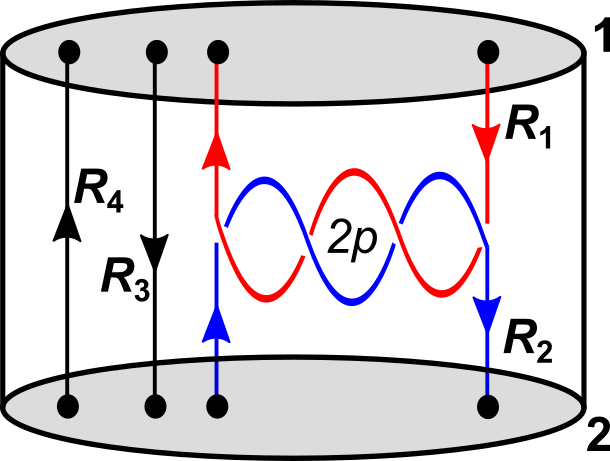}}
\caption[]{The quantum state $\ket{\Psi_3}$ on two $S^2$ boundaries labeled by 1 and 2. The Wilson lines carrying rep $R_1$ and $R_2$ are braided even number of times.}
\label{v1-state}
\end{figure}
\begin{align}
\ket{\Psi_3} & = \sum_{\ell=1}^{\textsf{d}} \sum_{r_1, r_2, r_3} \frac{ \left\{ R_{1}, \bar{R}_{1}, t_\ell, r_1 \right\} \left\{ R_{2}, \bar{R}_{2},t_\ell, r_2 \right\} \left\{ R_{4}, \bar{R}_{3}, t_\ell, r_3 \right\} \left\{ R_{2} \right\} \left\{ R_{3} \right\} }{\sqrt{\text{dim}_q t_\ell}\, \sqrt{(\text{dim}_q R_1)(\text{dim}_q R_2)(\text{dim}_q R_3)(\text{dim}_q R_4)}}  \nonumber \\
& \times F(t_\ell, r_1, r_2, p) \,\, | \phi_{t_\ell,\, r_3  r_1}^{(1)},\, \phi_{t_\ell,\, r_2 r_3}^{(2)} \rangle ~.
\label{Psi3state}
\end{align}
Here the representation $t_\ell \in (R_{1} \otimes \bar{R}_{1}) \cap (R_{2} \otimes \bar{R}_{2}) \cap (R_{4} \otimes \bar{R}_{3})$ which may appear in each of this decomposition multiple times. We denote the corresponding multiplicities as $a_1(t_\ell) \equiv N_{R_1 \bar{R}_1}^{t_\ell}$, $a_2(t_\ell) \equiv N_{R_2 \bar{R}_2}^{t_\ell} $ and $a_3(t_\ell) \equiv N_{R_4 \bar{R}_3}^{t_\ell}$ respectively. The labels $r_1, r_2, r_3$ are used to keep track of these multiple occurrences where $r_i$ takes values from 1 to $a_i(t_\ell)$. The notation $\{ R \} \equiv \{ R, \bar{R}, 0\}$ is used for the 2j-phase and the function $F$ is defined as, 
\begin{align} 
F(t_\ell, r_1, r_2, p) = \sum_{s,\, u} \left\{ \bar{R}_{1}, R_{2}, s, u \right\} \sqrt{\text{dim}_q s}\,\, a_{s,\, uu}^{t_\ell,\, r_1 r_2}\left[
\begin{array}{cc}
 R_{1} & \bar{R}_{1} \\
 R_{2} & \bar{R}_{2} \\
\end{array}
\right] \exp\left[\frac{2 \pi i \, p \left( C_{s} - C_{\bar{R}_{1}} - C_{R_{2}} \right)}{k+N} \right] ~,
\label{F-function-Psi3}
\end{align}
where the summation is over representation $s \in (\bar{R}_{1} \otimes R_{2})$ and $u$ is the corresponding multiplicity label. Note that the factors like $\{R_{2}\}$, $\{R_{3}\}$, $\text{dim}_q R_i$ are constant and will be canceled out while normalizing $\ket{\Psi_3}$. Also since the 3j-phases in eq.(\ref{Psi3state}) are overall phases, they do not affect the entanglement structure. The state $\ket{\Psi_3} \in \mathcal{H}_1 \otimes \mathcal{H}_2$, where $\mathcal{H}_1$ and $\mathcal{H}_2$ are the Hilbert spaces associated with the two $S^2$ boundaries. After tracing out $\mathcal{H}_2$, we get the following reduced density matrix:
\begin{equation} 
\rho_A = \sum_{\ell=1}^{\textsf{d}} \sum_{r_1,r_1', r_2, r_3} \frac{ F(t_\ell, r_1, r_2, p) \, F^{*}(t_\ell, r_1', r_2,p)}{(\text{dim}_q t_\ell) \, \braket{\Psi_3}} \ket{\phi_{t_\ell,\, r_3 r_1}^{(1)}} \bra{\phi_{t_\ell,\, r_3 r_1'}^{(1)}} ~,
\label{redrhoPsi3}
\end{equation}
where the trace is given as,
\begin{equation} 
\braket{\Psi_3} = \sum_{\ell=1}^{\textsf{d}} \sum_{r_1, r_2, r_3} \frac{|F(t_\ell, r_1, r_2, p)|^2}{\text{dim}_q t_\ell} = \sum_{\ell=1}^{\textsf{d}} \sum_{r_1, r_2} \frac{a_3(t_\ell)\,|F(t_\ell, r_1, r_2, p)|^2}{\text{dim}_q t_\ell} ~.
\label{}
\end{equation}
We can write the result in a matrix form as following. First let us define $A(t_\ell)$ whose rows and columns are labeled by $r_1$ and $r_1'$ respectively. So this will be a square matrix of order $a_1(t_\ell)$ whose elements are given as,
\begin{equation} 
A(t_\ell)_{r_1, r_1'} = \sum_{r_2} F(t_\ell, r_1, r_2, p) \, F^{*}(t_\ell, r_1', r_2, p) ~.
\label{}
\end{equation}
The summation over $r_3$ in eq.(\ref{redrhoPsi3}) gives the identity matrix $\mathbb{1}$ of order $a_3(t_\ell)$. Thus the summation over various multiplicity labels in eq.(\ref{redrhoPsi3}) can be written as the Kronecker product $A(t_\ell) \otimes \mathbb{1} \equiv M(t_\ell)$. Thus the final matrix can be given in a block diagonal form as,
\begin{equation} 
\rho_A = \frac{1}{\braket{\Psi_3}} \left(
\begin{array}{cccc}
 \frac{M(t_1)}{\text{dim}_q t_1} & & & \\
  & \frac{M(t_2)}{\text{dim}_q t_2} & & \\
 & & \ddots &  \\
  &  &  & \frac{M(t_{\textsf{d}})}{\text{dim}_q t_{\textsf{d}}} \\
\end{array}
\right) ~.
\label{}
\end{equation} 
If $\lambda_i(A(t_\ell))$ denotes the eigenvalue of $A(t_\ell)$, then matrix $M(t_\ell)$ will also have same eigenvalues where each eigenvalue will have multiplicity $a_3(t_\ell)$. Thus the eigenvalue corresponding to the block $t_\ell$ in the reduced density matrix can be given as,
\begin{equation}
\lambda_i(t_\ell) = \frac{\lambda_i(A(t_\ell))}{\braket{\Psi_3}(\text{dim}_q t_\ell)} \quad;\quad \text{mul}(\lambda_i(t_\ell)) = a_3(t_\ell) ~.
\end{equation}
Thus the entanglement entropy is given as,
\begin{equation}
\boxed{\text{EE} = -\sum_{\ell=1}^{\textsf{d}} \sum_i a_3(t_\ell) \lambda_i(t_\ell) \ln \lambda_i(t_\ell)} ~.
\end{equation}
When there is no multiplicity, i.e. when $a_1(t_\ell) = a_2(t_\ell) = a_3(t_\ell) = 1$, the entropy will become:
\begin{equation}
\text{EE} = -\sum_{\ell=1}^{\textsf{d}} \frac{|F(t_\ell, p)|^2}{\braket{\Psi_3}(\text{dim}_q t_\ell)} \ln\left( \frac{|F(t_\ell, p)|^2}{\braket{\Psi_3}(\text{dim}_q t_\ell)} \right);\quad \braket{\Psi_3} = \sum_{\ell'=1}^{\textsf{d}} \frac{|F(t_{\ell'}, p)|^2}{\text{dim}_q t_{\ell'}} ~.
\end{equation}
\subsubsection{Periodic entanglement structure}
The entanglement structure for the state $|{\Psi_3}\rangle$ is periodic in the twist number $p$. Consider the function $F$ given in eq.(\ref{F-function-Psi3}). When $p$ is 0, the exponential factor becomes unit. In such a case, using the identity of the Racah matrix given in eq.(\ref{Racah-prop}) and its unitary property in eq.(\ref{Racah-unitary}),  we get:
\begin{equation} 
F(t_\ell, r_1, r_2, 0) = \frac{\sqrt{\text{dim}_q R_{1}} \sqrt{\text{dim}_q R_{2}}}{\left\{ R_{2} \right\}} \, \delta_{t_\ell,\,0} \, \delta_{r_{1},\,1} \, \delta_{r_{2},\,1} ~.
\end{equation}
This sets $t_\ell=0$. Since the trivial representation always occurs once in the tensor decomposition of $R \otimes \bar{R}$ for any $R$, all the multiplicities become one and the quantum state of eq.(\ref{Psi3state}), after normalization, becomes a product state (which is consistent with the arguments presented in \cite{Melnikov:2018zfn}):
\begin{equation}
\begin{array}{c}
\includegraphics[width=0.18\linewidth]{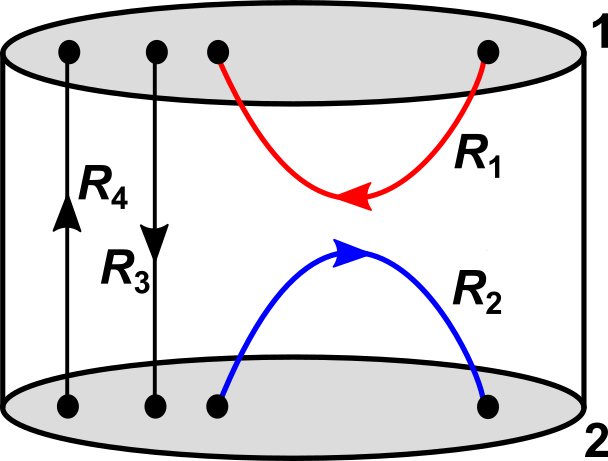}
\end{array} = |\phi_{0}^{(1)}\rangle \otimes |\phi_{0}^{(2)}\rangle ~.
\end{equation}
Further, we show that the state becomes product state not only at $p = 0$ but also for non-zero values of $p$. For example, the exponential factor in eq.(\ref{F-function-Psi3}) becomes unit when $p$ is a multiple of $2N(k+N)$ and $\ket{\Psi_3}$ will be non-entangled.\footnote{The quadratic Casimirs in general have a denominator $2N$, see for example eq.(\ref{quadCas}).} In general, this happens whenever $p$ is a multiple of $\textsf{p}$, where $\textsf{p}$ is an integer which depends on the values of $N$ and $k$ (and also the choice of representations $R_1, R_2, R_3, R_4$) such that $\textsf{p}$ is a divisor of $2N(k+N)$. In fact, the entanglement structure repeats itself in a periodic way when we increase $p$ as following: 
\begin{equation}
\text{EE}(p) = \text{EE}\left(p+\textsf{p}\mathbb{Z}\right) \quad;\quad \text{EE}\left(p=0\right) = 0 ~.
\label{}
\end{equation}
Moreover, there is a symmetry around $p=\textsf{p}$ in the sense that:
\begin{equation} 
\text{EE}(p) = \text{EE}\left(\textsf{p} - p \right) ~.
\end{equation}
We will denote \textsf{p} as `periodicity' which is the fundamental period of the periodic behavior of the entanglement structure. Its value can be obtained on a case by case basis. In the following we compute the entanglement entropy for the fundamental representations of SU($N$) showing this behavior and will give the explicit values of \textsf{p}.
\subsubsection{Fundamental representation of SU($N$)}
\begin{figure}[t]
\centerline{\includegraphics[width=6.0in]{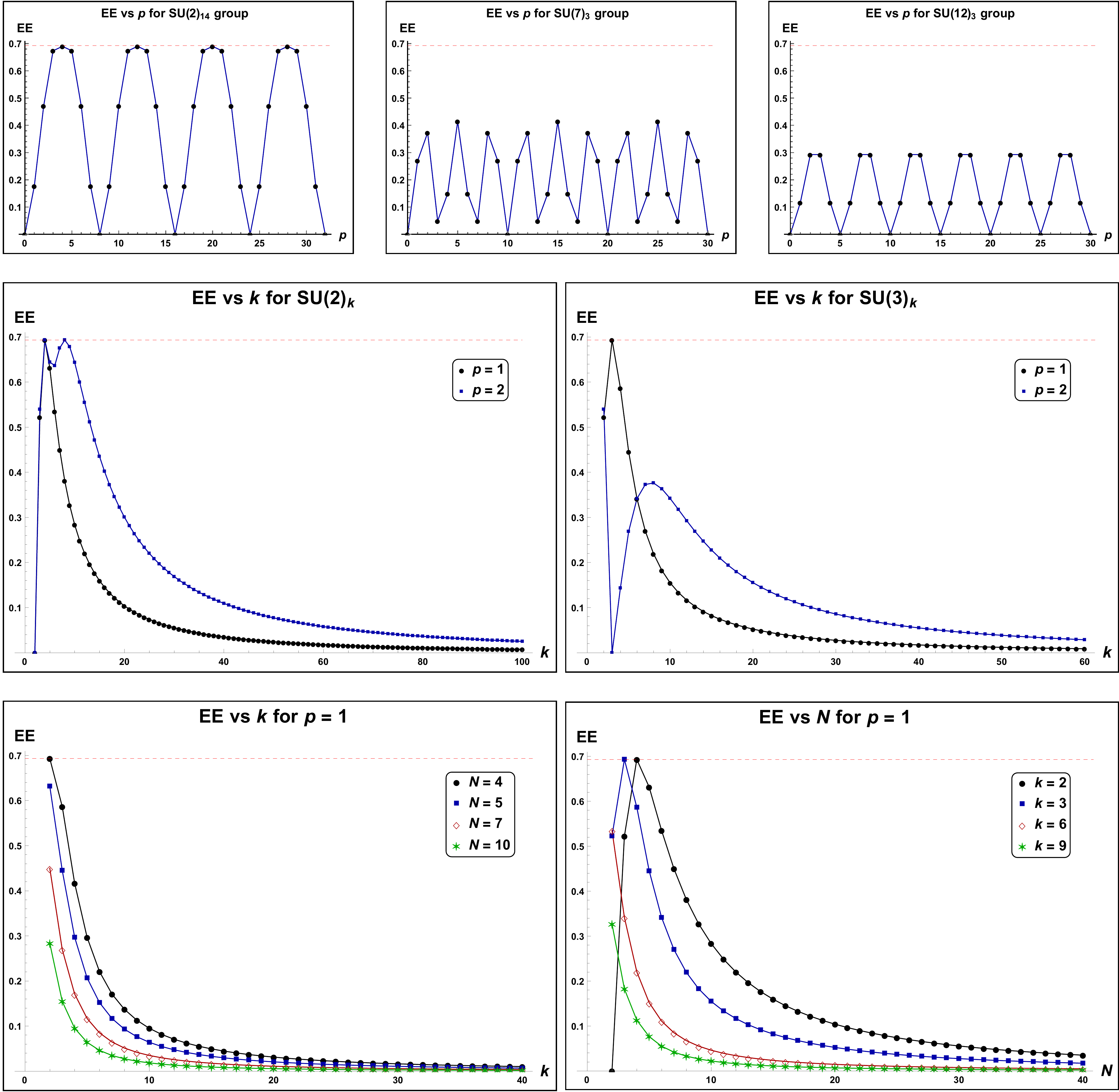}}
\caption[]{Variation of SU$(N)_k$ entanglement entropy for the state $\ket{\Psi_3}$ as a function of $k$, $N$ and $p$. The horizontal dashed line represents the maximum possible value ($\ln 2$) of entropy.}
\label{EEPlots1Necklace}
\end{figure}

When all the Wilson lines carry the fundamental representation, each Hilbert space is two dimensional. The representation $t_\ell \in \text{$\tiny\yng(1)$} \otimes \bar{\text{$\tiny\yng(1)$}} = \bullet \oplus \text{adj}$. We assume $k \geq 2$ so that both the representations are integrable. The quantum dimensions and quadratic Casimirs are given as,
\begin{equation}
\text{dim}_q (\bullet) = 1,\quad \text{dim}_q (\text{adj}) = [N+1][N-1] \quad;\quad C_{\bullet} = 0,\quad C_{\text{adj}} = N  ~.
\end{equation}
We also need the Racah matrix which can be taken from \cite{Zodinmawia:2011osq}:
\begin{equation}
a_{ts}\left[
\begin{array}{cc}
 \text{$\tiny\yng(1)$} & \bar{\text{$\tiny\yng(1)$}} \\
 \text{$\tiny\yng(1)$} & \bar{\text{$\tiny\yng(1)$}}
\end{array}
\right] = \left(
\begin{array}{c|cc}
  & \bullet & \text{adj} \\ \hline
\bullet & -\frac{1}{[N]} & \frac{\sqrt{[N+1]\,[N-1]}}{[N]} \\[8pt]
\text{adj} & \frac{\sqrt{[N+1]\,[N-1]}}{[N]} & \frac{1}{[N]}
\end{array}
\right) ~.
\label{Rac-fund}
\end{equation}
The phase factors are given as $\{ \text{$\tiny\yng(1)$},\, \bar{\text{$\tiny\yng(1)$}},\, \bullet \} = 1$ and $\{ \text{$\tiny\yng(1)$},\, \bar{\text{$\tiny\yng(1)$}},\, \text{adj} \} = -1$. Using various information, we can compute the function $F$ as (ignoring the constant factors $C_{\bar{R}_1}$ and $C_{R_2}$ in the exponential):
\begin{align}
F(\bullet,\, p) &= -\left(1 + [N+1][N-1] \exp[\frac{2\pi i\, pN}{k+N}]\right) \frac{1}{[N]} \nonumber \\
F(\text{adj},\, p) &= \left(1 - \exp[\frac{2\pi i\, pN}{k+N}]\right) \frac{\sqrt{[N+1]\,[N-1]}}{[N]} ~.
\end{align} 
The two eigenvalues of the reduced density matrix will be given as,
\begin{equation}
\lambda(\rho_A) = \{1-x,\, x\}, \quad x = \frac{4 \sin^2\left(\frac{\pi p N}{k+N}\right)}{2 ([N-1] [N+1]-1) \cos \left(\frac{2 \pi p N}{k+N}\right)+[N-1]^2 [N+1]^2+3}
\end{equation}
and the entropy will be:
\begin{equation}
\text{EE} = -(1-x)\ln(1-x) - x\ln x ~.
\end{equation}
Interestingly, the entropy is invariant under the exchange $N \longleftrightarrow k$. Moreover, we can see that the spectrum of $\rho_A$ and hence the entropy has a periodic behavior in twist number and the fundamental period is given as\footnote{We find that for SU(2), the entanglement entropy vanishes at $k=2$.}
\begin{equation}
\textsf{p} = \frac{k+N}{\text{gcd}(k,N)} ~.
\label{}
\end{equation}
Since we are working in a two dimensional Hilbert space, $\text{EE} \leq \ln 2$. Further, whenever $\text{EE} = \ln 2$ (which happens when $x=1/2$), the state $|{\Psi_3}\rangle$ will be a maximally entangled Bell state. Doing a numerical check, we find the following values at which this can happen:
\begin{equation}
\begin{rcases}
    \text{SU(2)}_{k=4K} & \text{with twists $p=K, K+1$ (mod $2K+1$)} \\
		\text{SU(3)}_{k=3} & \text{with twists $p=1$ (mod 2)} \\
		\text{SU($4M$)}_{k=2} & \text{with twists $p=M, M+1$ (mod $2M+1$)} 
\end{rcases} \quad \Longrightarrow |{\Psi_3}\rangle = |{\Psi_{\text{Bell}}}\rangle ~,
\label{Bell-state-Psi3}
\end{equation}
where $K\geq 1$ and $M \geq 1$ are integers. In figure \ref{EEPlots1Necklace}, we have plotted the variation of entropy for various values of $k$, $N$ and $p$. The first three plots show the periodic behavior in entropy as $p$ increases (for fixed values of $k$ and $N$). We find that whenever the periodicity $\textsf{p}$ is even, the entropy is maximum for $p = \textsf{p}/2$ which is evident from the plots.

We also find the asymptotic behavior of the entanglement entropy for large $k$ or large $N$ (with other parameters fixed) values which is given as:
\begin{equation} 
\text{EE}(k \gg 1) \sim \frac{4 p^2 \pi^2}{k^2 N^2} \left(1 - \ln\left[ \frac{4 p^2 \pi^2}{k^2 N^2} \right]  \right) \,\,;\quad \text{EE}(N \gg 1) \sim \frac{4 p^2 \pi^2}{k^2 N^2} \left(1 - \ln\left[ \frac{4 p^2 \pi^2}{k^2 N^2} \right]  \right) ~.
\label{largekN-1neck}
\end{equation}
From this, we can see that the state is no longer entangled ($\text{EE} \to 0$) as $k \to \infty$ or $N \to \infty$. The last four plots in figure \ref{EEPlots1Necklace} show that entropy decreases as $k$ or $N$ increases which is in accordance with the eq.(\ref{largekN-1neck}).

So far we have considered the braiding between the Wilson lines $R_1$ and $R_2$. It is also possible to have horizontal braiding between the Wilson lines $R_3$ and $R_4$ (similar to the figure \ref{Constructing-State}(a)). We will study the entanglement structure of this kind of state in the next subsection.
\subsection{Quantum state $\ket{\Psi_4}$ on $n$ copies of $S^2$}
\begin{figure}[h]
\centerline{\includegraphics[width=6.0in]{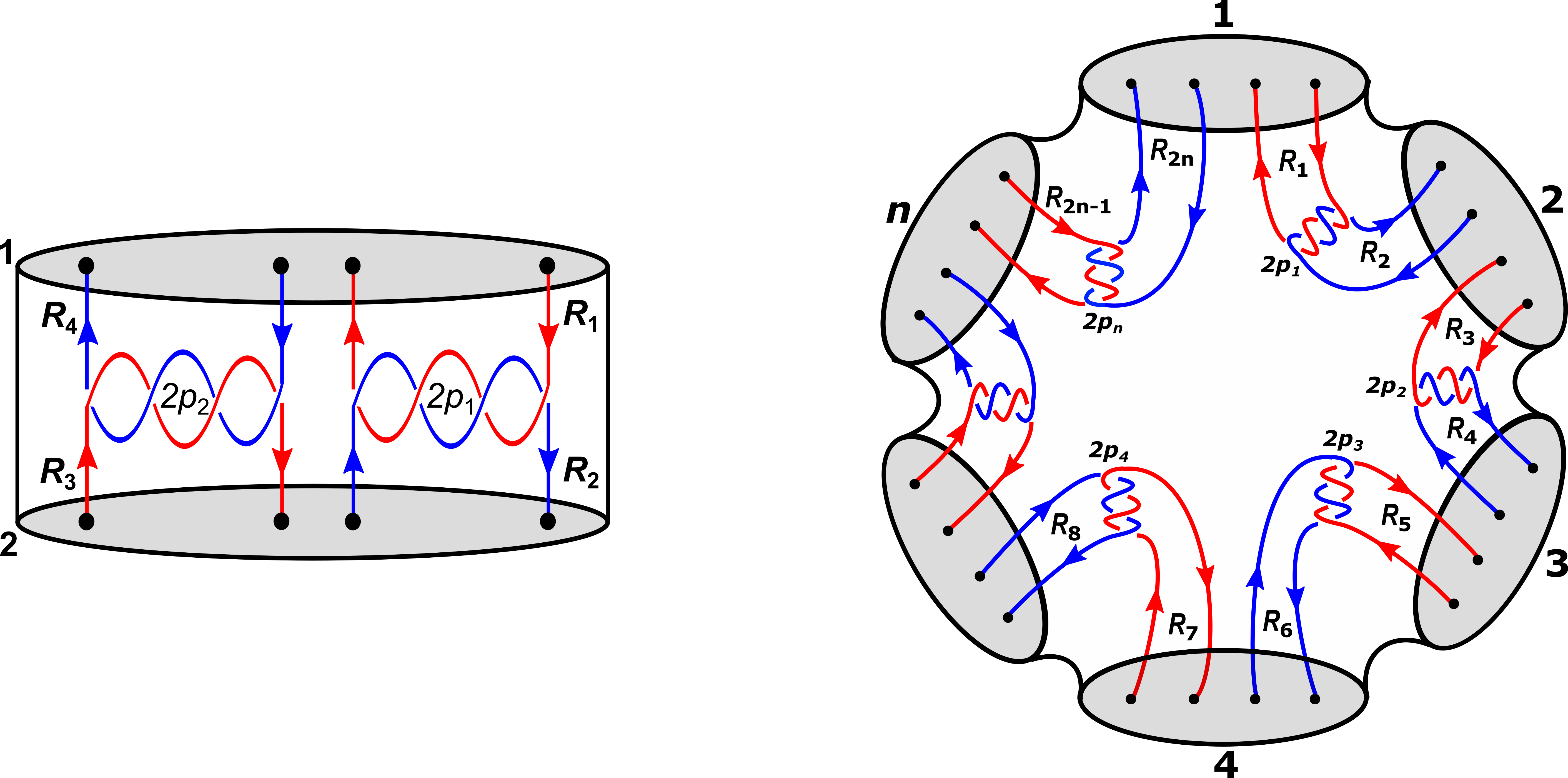}}
\caption[]{The quantum state $|{\Psi_4}\rangle$ on $n$ copies of $S^2$. The Wilson lines carrying representations $R_{2j-1}$ and $R_{2j}$ braid $2p_j$ number of times ($j=1,2,\ldots,n$). The figure on the left is the state for $n=2$.}
\label{vn-state}
\end{figure}

In this section, we will analyze the entanglement structure of the state $|{\Psi_4}\rangle$ on $n$ copies of $S^2$ as shown in figure \ref{vn-state}, where the Wilson lines connecting the punctures on one boundary undergo non-trivial braiding with the Wilson lines connecting the punctures on other boundary. We have also shown the state corresponding to $n=2$ in figure \ref{vn-state}. The Wilson lines carrying representations $R_{2j-1}$ and $R_{2j}$ are braided $2p_j$ number of times.
This state can be constructed as given in the appendix A and can be written as:
\begin{align}
|{\Psi_4}\rangle & = \sum_{\ell=1}^{\textsf{d}} \sum_{x_1, x_2, \ldots, x_{2n}} \frac{ \prod_{j=1}^{2n} \left\{R_{j}, \bar{R}_{j}, t_\ell, x_j \right\}  }{(\text{dim}_q t_\ell)^{n-1}} \left(\prod_{j=1}^n \left\{ R_{2j} \right\} \right) \nonumber \\ 
& \times \left(\prod_{j=1}^n F_j(t_\ell, x_{2j-1}, x_{2j}, p_j) \right) |{\phi_{t_\ell;\, x_{2n}, x_1}^{(1)},\, \phi_{t_\ell;\, x_2, x_3}^{(2)},\,\ldots,\, \phi_{t_\ell;\, x_{2n-2}, x_{2n-1}}^{(n)}}\rangle ~.
\label{Psi3-nbdy}
\end{align}
Here the representation $t_\ell \in \bigcap_{j=1}^{2n} (R_{j} \otimes \bar{R}_{j})$ which may appear in the decomposition of $(R_{j} \otimes \bar{R}_{j})$ multiple times. We denote this multiplicity as $a_j(t_\ell) = N_{R_j \bar{R}_j}^{t_\ell}$. To keep track of this, we have used the label $x_j$ which runs from 1 to $a_j(t_\ell)$ for a particular representation $t_\ell$. The notation $\left\{ R \right\} \equiv \left\{ R, \bar{R}, 0\right\}$ is used for the 2j-phase and the function $F$ is defined as,
\begin{align} 
F_j(t_\ell, x_{2j-1}, x_{2j}, p_j) &= \sum_{s_j,\, y_j} \left\{ \bar{R}_{2j-1}, R_{2j}, s_j, y_j \right\} \sqrt{\text{dim}_q s_j}\,\, a_{s_j,\, y_j,y_j}^{t_\ell,\, x_{2j-1}, x_{2j}}\left[
\begin{array}{cc}
 R_{2j-1} & \bar{R}_{2j-1} \\
 R_{2j} & \bar{R}_{2j} \\
\end{array}
\right] \nonumber \\
&\times \exp\left[\frac{2i \pi \, p_j \left( C_{s_j} - C_{\bar{R}_{2j-1}} - C_{R_{2j}} \right)}{k+N} \right] ~,
\label{F-function}
\end{align}
where the summation is over representation $s_j \in (\bar{R}_{2j-1} \otimes R_{2j})$ and $y_j$ is the corresponding multiplicity label which keeps track of multiple occurrences of $s_j$ in $(\bar{R}_{2j-1} \otimes R_{2j})$. 
The state $\ket{\Psi_4}$ lives in the total Hilbert space $\mathcal{H}$ which can be given as the tensor product $\mathcal{H} = \bigotimes_{j=1}^n \mathcal{H}_j$,  with $\mathcal{H}_j$ denoting the Hilbert space associated with the $j^{\text{th}}$ $S^2$ with four point conformal block given by representations $\left\{\bar{R}_{2j-2}, R_{2j-2}, \bar{R}_{2j-1}, R_{2j-1} \right\}$. The dimension of each Hilbert space will be: $\text{dim}(\mathcal{H}_j) = \sum_{\ell=1}^{\textsf{d}} a_{2j-2}(t_\ell) \, a_{2j-1}(t_\ell)$. Similar to the previous examples, we bi-partition the total Hilbert space as $(\mathcal{H}_A|\mathcal{H}_B)$ or $(m|n-m)$ and trace out $\mathcal{H}_B$ to get the following reduced density matrix:
\begin{align} 
\rho_A & = \sum_{\ell=1}^{\textsf{d}} \sum_{\substack{x_1,\, x_2,\, \ldots,\, x_{2n} \\ x_1',\, x_{2}',\,\ldots,\, x_{2m-1}',\,x_{2n}'}} \frac{(\text{dim}_q t_\ell)^{2-2n}}{\braket{\Psi_4}} \prod_{j=0}^{m} F_j(t_\ell, x_{2j-1}, x_{2j}, p_j) \, F_j^{*}(t_\ell, x_{2j-1}', x_{2j}', p_j) \nonumber \\ 
& \times \prod_{j=m+1}^{n-1} F_j(t_\ell, x_{2j-1}, x_{2j}, p_j) \, F_j^{*}(t_\ell, x_{2j-1}, x_{2j}, p_j) \, \delta_{x_{2m},x_{2m}'} \, \delta_{x_{2n-1},x_{2n-1}'} \nonumber \\ 
& \times |{\phi_{t_\ell;\, x_{0}, x_1}^{(1)},\, \phi_{t_\ell;\, x_2, x_3}^{(2)},\,\ldots,\, \phi_{t_\ell;\, x_{2m-2}, x_{2m-1}}^{(m)}}\rangle \langle{\phi_{t_\ell;\, x_{0}', x_1'}^{(1)},\, \phi_{t_\ell;\, x_2', x_3'}^{(2)},\,\ldots,\, \phi_{t_\ell;\, x_{2m-2}', x_{2m-1}'}^{(m)}} | ~,
\label{rho-red-Psi3-op}
\end{align}
where we have used the notations $F_0,\, x_{-1},\, x_0$ and $p_0$ to denote $F_n,\, x_{2n-1},\, x_{2n}$ and $p_n$ respectively and we have defined trace as:
\begin{equation}
\braket{\Psi_4} = \sum_{\ell=1}^{\textsf{d}}\,\, \sum_{x_1,\, x_2,\, \ldots,\, x_{2n}} \frac{\prod_{j=1}^n |F_j(t_\ell, x_{2j-1}, x_{2j}, p_j) |^2}{(\text{dim}_q t_\ell)^{2n-2}} ~.
\end{equation}
We can write $\rho_A$ into a block diagonal form, where each block corresponds to a particular representation $t_\ell$:\footnote{We should remind again that in the eq.(\ref{rho-red-Psi3}), we have only included the blocks which are non-zero. The actual matrix is much bigger having order $= \text{dim}(\mathcal{H}_A)$.}
\begin{align} 
\rho_A = \frac{1}{\braket{\Psi_4}} \left(
\begin{array}{cccc}
 \dfrac{\prod_{j=m+1}^{n-1} \alpha_j(t_1)}{(\text{dim}_q t_1)^{2n-2}} M_{t_1} & &  & \\
  & \dfrac{\prod_{j=m+1}^{n-1} \alpha_j(t_2)}{(\text{dim}_q t_2)^{2n-2}} M_{t_2} &  & \\
  &  & \ddots & \\
 &  &  & \dfrac{\prod_{j=m+1}^{n-1} \alpha_j(t_{\textsf{d}})}{(\text{dim}_q t_{\textsf{d}})^{2n-2}} M_{t_{\textsf{d}}}
\end{array}
\right) ~.
\label{rho-red-Psi3}
\end{align}
Here $M_{t_1}, M_{t_2}, \ldots, M_{t_{\textsf{d}}}$ are in general matrices and the factor $\alpha_j(t_\ell)$ is defined as,
\begin{equation}
\alpha_j(t_\ell) = \sum_{x_{2j-1},\, x_{2j}} \left| F_j(t_\ell, x_{2j-1}, x_{2j}, p_j) \right|^2 ~.
\end{equation}
The matrix $M_{t_\ell}$ for the representation $t_\ell$ can be written as Kronecker product:
\begin{equation}
M_{t_\ell} = X(t_\ell) \otimes Y_1(t_\ell) \otimes Y_2(t_\ell) \otimes \ldots \otimes Y_{m-1}(t_\ell) \otimes Z(t_\ell) ~.
\label{Mt-matrix}
\end{equation}
The elements of $X(t_\ell)$ and $Z(t_\ell)$ are given as,
\begin{align}
X_{x_{2n},\, x_{2n}'} &= \sum_{x_{2n-1}} F_n(t_\ell, x_{2n-1}, x_{2n}, p_n) \, F_n^{*}(t_\ell, x_{2n-1}, x_{2n}', p_n) \nonumber \\
Z_{x_{2m-1},\, x_{2m-1}'} &= \sum_{x_{2m}} F_m(t_\ell, x_{2m-1}, x_{2m}, p_m) \, F_m^{*}(t_\ell, x_{2m-1}', x_{2m}, p_m) ~.
\end{align} 
The structure of $Y_i(t_\ell)$ is more complicated which can be written as a block matrix:
\begin{equation}
(Y_i)_{x_{2i-1},\, x_{2i-1}'} = (B_i)^{x_{2i-1},\, x_{2i-1}'} ~,
\end{equation}
where $x_{2i-1}$ and $x_{2i-1}'$ label the row and column of $Y_i(t_\ell)$ respectively. Each element of $Y_i(t_\ell)$ is a matrix $(B_i)^{x_{2i-1},\, x_{2i-1}'}$ and thus $Y_{i}(t_\ell)$ will take the following form:
\begin{equation}
Y_{i}(t_\ell) = \left(
\begin{array}{cccc}
 (B_i)^{1,1} & (B_i)^{1,2} & \ldots & (B_i)^{1, a_{2i-1}} \\
 (B_i)^{2,1} & (B_i)^{2,2} & \ldots & (B_i)^{2, a_{2i-1}} \\
\vdots & \vdots & \ddots & \vdots \\
 (B_i)^{a_{2i-1},1} & (B_i)^{a_{2i-1},2} & \ldots & (B_i)^{a_{2i-1}, a_{2i-1}}
\end{array}
\right) ~.
\end{equation}
The rows and columns of $(B_i)^{\alpha,\, \beta}$ are specified by the values of $x_{2i}$ and $x_{2i}'$ respectively and its elements are,
\begin{equation}
(B_i)^{\alpha,\, \beta}_{x_{2i},\, x_{2i}'} = F_i(t_\ell, \alpha, x_{2i}, p_i) \, F_i^{*}(t_\ell, \beta, x_{2i}', p_i) ~.
\end{equation}
Thus $M_{t_\ell}$ will be a square matrix of order:
\begin{equation}
\text{order}(M_{t_\ell}) = \prod_{j=0}^{2m-1} a_j(t_\ell) ~,
\end{equation}
where $a_0 = a_{2n}$. For example, if the representation $t=t_0$ appears in each of $(R_j \otimes \bar{R}_j)$ with multiplicity two, then each multiplicity label will take two values 1 and 2. Thus for $m=2$ case, the matrix $M_{t_0}$ will be a square matrix of order 16 which will be given as,
\begin{equation}
M_{t_0} = \left(
\begin{array}{cc}
 X_{11} & X_{12} \\
 X_{21} & X_{22} \\
\end{array}
\right) \otimes \left(
\begin{array}{cc}
 \left(
\begin{array}{cc}
 B_{11}^{11} & B_{12}^{11} \\
 B_{21}^{11} & B_{22}^{11} \\
\end{array}
\right) & \left(
\begin{array}{cc}
 B_{11}^{12} & B_{12}^{12} \\
 B_{21}^{12} & B_{22}^{12} \\
\end{array}
\right) \\[15pt]
 \left(
\begin{array}{cc}
 B_{11}^{21} & B_{12}^{21} \\
 B_{21}^{21} & B_{22}^{21} \\
\end{array}
\right) & \left(
\begin{array}{cc}
 B_{11}^{22} & B_{12}^{22} \\
 B_{21}^{22} & B_{22}^{22} \\
\end{array}
\right) \\
\end{array}
\right) \otimes \left(
\begin{array}{cc}
 Z_{11} & Z_{12} \\
 Z_{21} & Z_{22} \\
\end{array}
\right) ~.
\end{equation}
Having written $M_{t_\ell}$ as Kronecker product, its eigenvalues can be computed as:
\begin{equation}
\lambda_i(M_{t_\ell}) = \lambda^{X}_{i_0}\, \lambda^{Y_1}_{i_1}\, \lambda^{Y_2}_{i_2}\, \ldots\, \lambda^{Y_{m-1}}_{i_{m-1}}\, \lambda^{Z}_{i_m} ~,
\end{equation} 
which means that the eigenvalues of $M_{t_\ell}$ are all the possible products of various eigenvalues of matrices $X, Y_1, \ldots, Y_{m-1}, Z$. Thus the eigenvalues of the reduced density matrix corresponding to the block of representation $t_{\ell}$ can be computed as,
\begin{equation}
\lambda_{i}(t_\ell) = \frac{1}{\braket{\Psi_4}}\left(\dfrac{\prod_{j=m+1}^{n-1} \alpha_j(t_\ell)}{(\text{dim}_q t_\ell)^{2n-2}} \, \lambda_i(M_{t_\ell}) \right) ~.
\end{equation} 
The entanglement entropy can now be computed:
\begin{equation}
\boxed{\text{EE} = -\sum_{\ell=1}^{\textsf{d}}\left(\sum_i \lambda_{i}(t_\ell)\, \ln \lambda_{i}(t_\ell) \right)} ~.
\end{equation} 
\subsubsection{Multiplicity free case}
When there is no multiplicity in the representation $t_\ell$ in any of the fusion (for example, if we are working with SU(2) gauge group), we do not need additional multiplicity labels. In such a case, each of the matrix $M_{t_\ell}$ in eq.(\ref{rho-red-Psi3}) is simply a number given as,
\begin{equation}
M_{t_\ell} = \prod_{j=0}^m F_j(t_\ell, p_j) \, F_j^{*}(t_\ell, p_j) ~.
\end{equation}
Thus the reduced density matrix becomes,
\begin{equation}
\rho_A = \frac{1}{\braket{\Psi_4}} \left(
\begin{array}{cccc}
 \dfrac{\prod_{j=1}^{n} \left| F_j(t_1, p_j) \right|^2}{(\text{dim}_q t_1)^{2n-2}} & & & \\
 & \dfrac{\prod_{j=1}^{n} \left| F_j(t_2, p_j) \right|^2}{(\text{dim}_q t_2)^{2n-2}} & & \\
 & & \ddots & \\
 & & & \dfrac{\prod_{j=1}^{n} \left| F_j(t_{\textsf{d}}, p_j) \right|^2}{(\text{dim}_q t_{\textsf{d}})^{2n-2}}
\end{array}
\right) ~.
\end{equation}
This is a diagonal matrix and the eigenvalues, labeled by representation $t_\ell$, can be given as:
\begin{equation} \lambda(t_\ell) = \frac{(\text{dim}_q t_\ell)^{2-2n} \prod_{j=1}^{n} \left| F_j(t_\ell, p_j) \right|^2}{\sum_{\ell'=1}^{\textsf{d}} \left( (\text{dim}_q t_\ell')^{2-2n} \prod_{j=1}^{n} \left| F_j(t_\ell', p_j) \right|^2 \right)}
\end{equation}
and the entanglement entropy will be,
\begin{equation} 
\text{EE} = -\sum_{\ell=1}^{\textsf{d}} \lambda(t_\ell) \ln \lambda(t_\ell) ~.
\label{no-mul}
\end{equation}
\subsubsection{Separability criteria of the reduced density matrix}
Let us study the reduced density matrix in eq.(\ref{rho-red-Psi3}) which acts on the Hilbert space $\mathcal{H}_A$ to get more insight into the entanglement structure of the state $|{\Psi_4}\rangle$. We want to see if this reduced density matrix is separable or not for which we would like to calculate entanglement negativity. Consider the further bi-partitioning of the Hilbert space $\mathcal{H}_A$ into two parts:
\begin{equation}
\mathcal{H}_A = \mathcal{H}_{A_1} \otimes \mathcal{H}_{A_2} = \left(\bigotimes_{i=1}^{r} \mathcal{H}_i \right)  \otimes \left(\bigotimes_{i=r+1}^{m} \mathcal{H}_i \right) ~.
\end{equation}
The partial transpose of the reduced density matrix with respect to $\mathcal{H}_{A_2}$ can be given as,
\begin{align} 
\rho^{\Gamma} &= \frac{1}{\braket{\Psi_4}} \left(
\begin{array}{cccc}
 \dfrac{\prod_{j=m+1}^{n-1} \alpha_j(t_1)}{(\text{dim}_q t_1)^{2n-2}} M_{t_1}^{\Gamma} & & & \\
  & \dfrac{\prod_{j=m+1}^{n-1} \alpha_j(t_2)}{(\text{dim}_q t_2)^{2n-2}} M_{t_2}^{\Gamma} & & \\
 &  & \ddots & \\
&  & & \dfrac{\prod_{j=m+1}^{n-1} \alpha_j(t_{\textsf{d}})}{(\text{dim}_q t_{\textsf{d}})^{2n-2}} M_{t_{\textsf{d}}}^{\Gamma}
\end{array}
\right) ~,
\label{par-trans}
\end{align}
where $M_{t_\ell}^{\Gamma}$ denotes the partial transpose of $M_{t_\ell}$. To see how $M_{t_\ell}^{\Gamma}$ differs from $M_{t_\ell}$, we need to find which multiplicity labels (corresponding to $t_\ell$) are exchanged while partial transposing. In the following, we have explicitly shown the breaking of the multiplicity labels of eq.(\ref{rho-red-Psi3-op}) on the Hilbert spaces $\mathcal{H}_{A_1}$ and $\mathcal{H}_{A_2}$: 
\begin{align}
&|\,\,{\underbrace{x_0\, x_1,\, x_2\, x_3,\, \ldots,\, x_{2r-2}\, {x_{2r-1}}}_{A_1},\,\, \underbrace{{x_{2r}}\, x_{2r+1},\, \ldots,\, x_{2m-2}\, x_{2m-1}}_{A_2}}\,\,\rangle \nonumber \\
& \langle \,\,{\underbrace{x_0'\, x_1',\, x_2'\, x_3',\, \ldots, x_{2r-2}'\, {x_{2r-1}'}}_{A_1},\,\, \underbrace{{x_{2r}'}\, x_{2r+1}',\, \ldots,\, x_{2m-2}'\, x_{2m-1}'}_{A_2}}\,\, | ~.
\label{}
\end{align} 
For the partial transpose with respect to $\mathcal{H}_{A_2}$, we need to swap the basis labels corresponding to $\mathcal{H}_{A_2}$ in the `ket' and `bra', i.e. $x_{2r} \leftrightarrow x_{2r}',\, x_{2r+1} \leftrightarrow x_{2r+1}',\, \ldots,\, x_{2m-1} \leftrightarrow x_{2m-1}'$. In the process of partial transpose, the matrices which are constructed from the basis labels of only $\mathcal{H}_{A_1}$ will be unaffected, while those matrices which are constructed from the basis labels of only $\mathcal{H}_{A_2}$ will be transposed. Only those matrices will transform non-trivially which are made up from the labels of both $\mathcal{H}_{A_1}$ and $\mathcal{H}_{A_2}$. In eq.(\ref{Mt-matrix}), we have given $M_{t_\ell}$ as Kronecker product of various matrices. Thus it is easy to see that its partial transpose will be given as,
\begin{equation}
M_{t_\ell}^{\Gamma} = X \otimes Y_1 \otimes \ldots \otimes Y_{r-1} \otimes Y_{r}^{\Gamma} \otimes Y_{r+1}^T \otimes \ldots \otimes Y_{m-1}^T \otimes Z^T ~,
\end{equation}
where the superscript $T$ denotes the usual transpose of a matrix which does not change its eigenvalues. The matrix $Y_{r}^{\Gamma}$ can be obtained from $Y_{r}$ as,
\begin{align} 
 Y_{r}^{\Gamma}(t_\ell) = \left(
\begin{array}{cccc}
 [(B_r)^{1,1}]^T & [(B_r)^{1,2}]^T & \ldots & [(B_r)^{1, a_{2r-1}}]^T \\
 {[(B_r)^{2,1}]^T} & [(B_r)^{2,2}]^T & \ldots & [(B_r)^{2, a_{2r-1}}]^T \\
\vdots & \vdots & \ddots & \vdots \\
 {[(B_r)^{a_{2r-1},1}]^T} & [(B_r)^{a_{2r-1},2}]^T & \ldots & [(B_r)^{a_{2r-1}, a_{2r-1}}]^T
\end{array}
\right) ~,
\end{align}
where each block of $Y_{r}$ has been transposed. If the matrix $Y_{r}^{\Gamma}(t_\ell)$ has negative eigenvalue for any $t_\ell$, we get a non-zero negativity and the reduced density matrix is non-separable. In fact, we will give an example in later section which shows that the negativity is non-zero but currently we do not have a proof whether $Y_{r}^{\Gamma}(t_\ell)$ will always have negative eigenvalues. On the other hand, when $a_{2r}(t_\ell) = 1$, each of the $(B_r)^{i,j}$ is just a number and thus $Y_{r}^{\Gamma}(t_\ell) = Y_{r}(t_\ell)$. In such a case, there are no negative eigenvalues of $Y_{r}^{\Gamma}(t_\ell)$ and the entanglement negativity vanishes. Moreover, when the multiplicities in the decomposition of the tensor product of representations for the conformal blocks associated with the remaining $S^2$ boundaries $1,2,\ldots,m$ are unit, i.e. when $a_0(t_\ell) = a_1(t_\ell) = \ldots= a_{2m-1}(t_\ell) = 1$ for all $t_\ell$, the reduced density matrix can be written as,
\begin{align} 
\rho = \sum_{\ell=1}^{\textsf{d}} \dfrac{g(t_\ell)\, \prod_{j=m+1}^{n-1} \alpha_j(t_\ell)}{\braket{\Psi_4} (\text{dim}_q t_{\ell})^{2n-2}} \,\, \rho_{t_\ell}^{A_1}  \otimes \rho_{t_\ell}^{A_2} ~,
\label{sep-Psi3}
\end{align}
where
\begin{equation}
g(t_\ell) = \left(\sum_{x_{2m}} \left| F_m(t_{\ell},x_{2m}, p_m) \right|^2 \right)   \left(\sum_{x_{2n-1}} \left| F_n(t_{\ell},x_{2n-1}, p_m) \right|^2 \right)
\end{equation}
and $\rho_{t_{\ell}}^{A_1}$ and $\rho_{t_{\ell}}^{A_2}$ are the density matrices acting on Hilbert spaces $\mathcal{H}_{A_1}$ and $\mathcal{H}_{A_2}$ respectively which correspond to the pure states:
\begin{align}
\rho_{t_{\ell}}^{A_1} &= |{{t_{\ell}}^{(1)}, {t_{\ell}}^{(2)}, \ldots, {t_{\ell}}^{(r)}}\rangle\langle{{t_{\ell}}^{(1)}, {t_{\ell}}^{(2)}, \ldots, {t_{\ell}}^{(r)}} | \nonumber \\
\rho_{t_{\ell}}^{A_2} &= |{{t_{\ell}}^{(r+1)}, {t_{\ell}}^{(r+2)}, \ldots, {t_{\ell}}^{(m)}}\rangle\langle{{t_{\ell}}^{(r+1)}, {t_{\ell}}^{(r+2)}, \ldots, {t_{\ell}}^{(m)}} | ~.
\end{align}
One can check that the coefficients are positive and add up to 1:
\begin{equation}
\braket{\Psi_4} = \sum_{\ell'=1}^{\textsf{d}} \left( \frac{g(t_{\ell'})\, \prod_{j=m+1}^{n-1} \alpha_j(t_{\ell'})}{(\text{dim}_q t_{\ell'})^{2n-2}} \right) \quad \Longrightarrow \quad \sum_{\ell=1}^{\textsf{d}} \left( \dfrac{g(t_\ell)\, \prod_{j=m+1}^{n-1} \alpha_j(t_\ell)}{\braket{\Psi_4} (\text{dim}_q t_{\ell})^{2n-2}} \right) = 1 ~.
\end{equation}
Thus the reduced density matrix is separable. In summary, we can say the following which is similar to our proposition in eq.(\ref{prop-1}):
\begin{empheq}[box=\fbox]{align}
  a_0(t_\ell) = a_1(t_\ell) = \ldots = a_{2m-1}(t_\ell) = 1, \,\, \forall t_\ell \quad &\Longrightarrow \quad \rho = \text{separable} \nonumber \\
	a_{2r}(t_\ell) > 1, \,\, \text{for some } t_\ell \quad \overset{?}{\Longrightarrow} \quad \mathcal{N} \neq 0 \quad &\Longrightarrow \quad \rho = \text{non-separable}
	\label{}
\end{empheq}
Note that we have put a `$?$' on one of the arrows because we do not have a formal proof for a generic case. In the later section, we will study an example which gives non-zero negativity and supports this statement. Thus, once again we see that indeed the multiplicity plays a crucial role in determining whether the state $|{\Psi_4}\rangle$ has a GHZ-like or W-like entanglement structure.

Just like the state $\ket{\Psi_3}$, the entanglement structure of the state $|{\Psi_4}\rangle$ also shows a periodic behavior as we increase the number of twists in the braiding, which we will discuss in the following.
\subsubsection{Periodic entanglement structure}
Here we will show the periodic behavior of entanglement structure for the state $|{\Psi_4}\rangle$. Consider the function $F_j$ given in eq.(\ref{F-function}). Following the previous arguments, when $p_j$ is 0 for some value of $j$, the exponential factor becomes unit and the function $F_j$ becomes:
\begin{equation} 
F_j = \frac{\sqrt{\text{dim}_q R_{2j-1}} \sqrt{\text{dim}_q R_{2j}}}{\left\{ R_{2j} \right\}} \, \delta_{t_\ell,\,0} \, \delta_{x_{2j-1},\,1} \, \delta_{x_{2j},\,1} ~.
\end{equation}
This sets $t_\ell=0$ and hence all the multiplicities become one. In such a case, the quantum state of eq.(\ref{Psi3-nbdy}), after normalization, will be given as
\begin{equation} 
|{\Psi_4}\rangle = |{\phi_{0}^{(1)},\, \phi_{0}^{(2)},\,\ldots,\, \phi_{0}^{(n)}}\rangle
\end{equation}
which is a non-entangled state and the entanglement entropy (as well as negativity) vanishes. Specifically for $n=2$, this analysis tells that the following states are non-entangled states:
\begin{equation}
\begin{array}{c}
\includegraphics[width=0.21\linewidth]{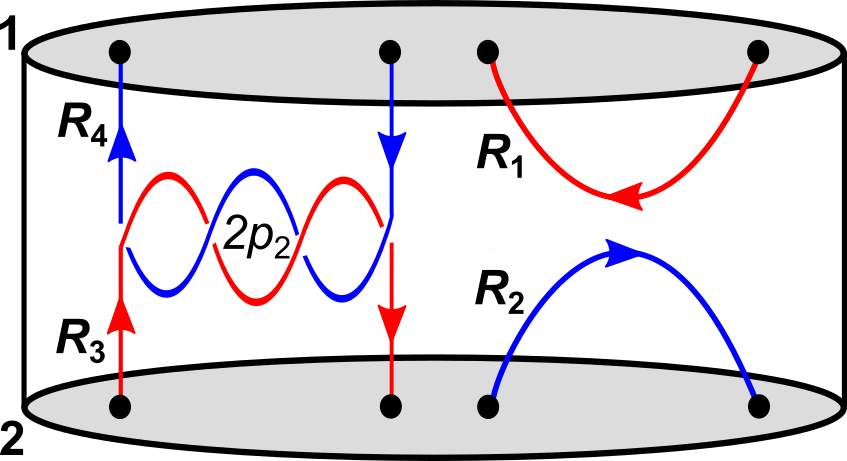}
\end{array} = \begin{array}{c}
\includegraphics[width=0.21\linewidth]{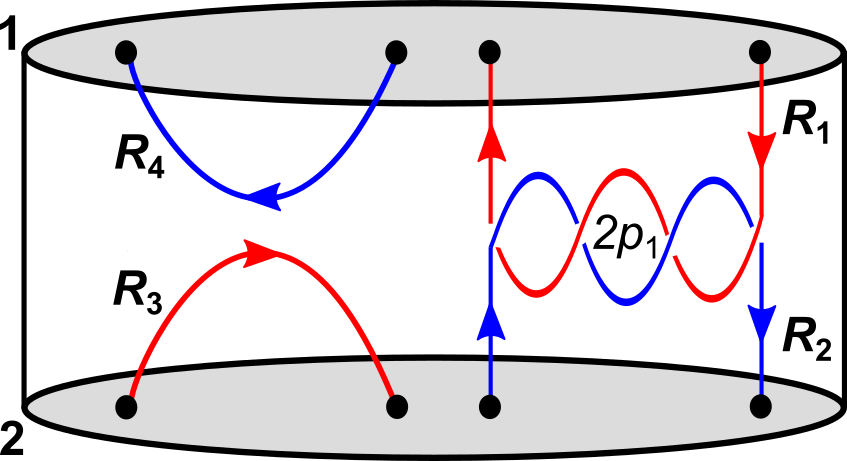} 
\end{array} = \begin{array}{c}
\includegraphics[width=0.21\linewidth]{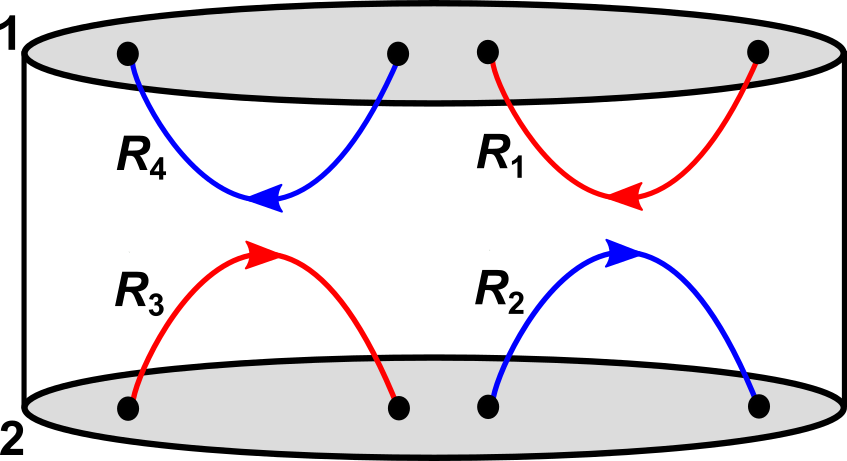} 
\end{array} = |\phi_{0}^{(1)} \rangle \otimes |\phi_{0}^{(2)} \rangle ~.
\end{equation}
Further, the state becomes product state for non-zero values of $p_j$. For example, the exponential factor in eq.(\ref{F-function}) becomes unit when $p_j$ is a multiple of $2N(k+N)$ and $\ket{\Psi_4}$ will be non-entangled. In general, this happens whenever $p_j$ is a multiple of $\textsf{p}$ such that $\textsf{p}$ is a divisor of $2N(k+N)$. The entanglement structure repeats itself in a periodic way when we increase any of the twist number as following: 
\begin{alignat}{2}
\text{EE}(p_j) &= \text{EE}\left(p_j+\textsf{p}\mathbb{Z}\right) &\quad;\quad \text{EE}\left(p_j=0\right) &= 0 \nonumber \\ 
\mathcal{N}(p_j) &= \mathcal{N}\left(p_j+\textsf{p}\mathbb{Z}\right) &\quad;\quad \mathcal{N}\left(p_j=0\right) &= 0 ~.
\label{}
\end{alignat}
Further, there is a symmetry around $p_j=\textsf{p}$:
\begin{equation} 
\text{EE}(p_j) = \text{EE}\left(\textsf{p} - p_j \right) \quad;\quad \mathcal{N}(p_j) = \mathcal{N}\left(\textsf{p} - p_j \right) ~.
\end{equation}
We denote \textsf{p} as `periodicity' which is the fundamental period of this periodic behavior of the entanglement structure. In the following sections, we will compute the entanglement entropy and negativity for various examples showing this behavior and give the explicit values of \textsf{p}.
\subsubsection{Wilson lines carrying symmetric representations of SU($N$)}
In this section we will study the entanglement structure when all the representations are symmetric representations of SU($N$). Since we will be dealing with symmetric representations and their conjugates, we will use the following notation to write them:  
\begin{align}
&\sym{a} = a \Lambda_1 \equiv [a;0] \quad;\quad \symC{b} = b \Lambda_{N-1} \equiv [0;b] \nonumber \\[2pt]
&\mixed{b}{a} = a \Lambda_1 + b \Lambda_{N-1} \equiv [a;b] ~, \\[0.5pt] \nonumber
\end{align} 
where $\Lambda_1, \Lambda_2, \ldots, \Lambda_{N-1}$ are the fundamental weights of SU$(N)$. The notation $[a;0]$ will denote a symmetric representation whose Young diagram has one row with $a$ number of boxes. Its conjugate is denoted as $[0;a]$ whose Young tableau is given in the above equation where $\ytableausetup{centertableaux, boxsize=1.1em}\begin{ytableau} \bullet \end{ytableau}$ represents a vertical row with $(N-1)$ boxes. 
From the group theory we know the following decomposition, which will be helpful in the computation:
\begin{align}
a \leq b :& \quad [a;0] \otimes [0;b] = \bigoplus_{\ell=0}^a \, [\ell;b-a+\ell] \nonumber \\
a \geq b :& \quad [a;0] \otimes [0;b] = \bigoplus_{\ell=0}^b \, [a-b+\ell;\ell] ~.
\end{align}
The quantum dimension and the quadratic Casimir of the representation $[x;y]$ is given as,
\begin{align}
\text{dim}_q [x;y] &= \frac{[N+x-2]!\,[N+y-2]!\,[N+x+y-1]}{[x]!\,[y]!\,[N-1]!\,[N-2]!} \nonumber \\[5pt]
C_{[x;y]} &= \frac{(N-1) \left(Nx+Ny+x^2+y^2\right)+2 x y}{2 N} ~.
\label{quadCas}
\end{align}
Since the decomposition of the tensor product of a symmetric representation and its conjugate does not give representations with multiple occurrences, we are dealing with a multiplicity free case and we can omit all the multiplicity labels. The 3j-phase for a representation $t_\ell \in [a;0] \otimes [0;b]$ can be taken as \cite{Zodinmawia:2011osq}:
\begin{equation}
\left\{[a;0] \,,\, [0;b] \,,\, t_\ell \right\} = (-1)^{\abs{a-b}+\,\ell} ~,
\end{equation}
where $t_\ell$ is either $[a-b+\ell;\ell]$ or $[\ell;b-a+\ell]$ depending on whether $a \geq b$ or $a \leq b$. 

Let us first consider the simplest case where all the representations $R_j$ are fundamental representations of SU($N$), i.e.,
\begin{equation}
R_1 = R_2 = \ldots = R_{2n}= \text{$\tiny\yng(1)$} \equiv [1;0] ~.
\end{equation}
Both representations $t_\ell$ and $s_\ell$ take two values:
\begin{equation}
t_\ell, s_\ell \in (R \otimes \bar{R}) = [0;0] \oplus [1;1] ~.
\end{equation}
We assume $k \geq 2$ so that both the representations are integrable. The Racah matrix required for this computation is given in eq.(\ref{Rac-fund}). Using various information, we can compute the function $F_j$ as:
\begin{align}
F_j\left(t_0,p_j\right) &= -\exp(\frac{2 i \pi  p_j}{k N+N^2}) \left( \frac{\exp(-\frac{2 i \pi  N p_j}{k+N}) + [N+1][N-1]}{[N]} \right) \nonumber \\
F_j\left(t_1,p_j\right) &= -\exp(\frac{2 i \pi  p_j}{k N+N^2}) \left(1-\exp(-\frac{2 i \pi  N p_j}{k+N}) \right) \frac{\sqrt{[N+1]\,[N-1]}}{[N]} ~.
\end{align} 
The total Hilbert space is $2^n$ dimensional and we break it into two parts as $\mathcal{H} = \mathcal{H}_A \otimes \mathcal{H}_B$ corresponding to the bi-partition $(m|n-m)$. Tracing out $\mathcal{H}_B$ assigns an entanglement structure to the state $|{\Psi_4}\rangle$ and we get the following entanglement entropy:
\begin{align} 
\text{EE} &= \frac{\left(\prod_{j=1}^nf_{p_j}\right)}{\left(\prod_{j=1}^n f_{p_j}\right) + [N-1]^2[N+1]^2\left(\prod_{j=1}^n g_{p_j}\right)} \ln(\frac{\left(\prod_{j=1}^n f_{p_j}\right) + [N-1]^2[N+1]^2\left(\prod_{j=1}^n g_{p_j}\right)}{\left(\prod_{j=1}^nf_{p_j}\right)}) \nonumber \\
&+ \frac{[N-1]^2[N+1]^2\left(\prod_{j=1}^n g_{p_j}\right)}{\left(\prod_{j=1}^n f_{p_j}\right) + [N-1]^2[N+1]^2\left(\prod_{j=1}^n g_{p_j}\right)} \ln(\frac{\left(\prod_{j=1}^n f_{p_j}\right) + [N-1]^2[N+1]^2\left(\prod_{j=1}^n g_{p_j}\right)}{[N-1]^2[N+1]^2\left(\prod_{j=1}^n g_{p_j}\right)}),
\label{EE-a=1}
\end{align}
where we have defined the following:
\begin{align} 
f_p &= [N+1]^3[N-1]^3 + 2\cos\left(\frac{2 \pi  N p}{k+N}\right)[N+1]^2[N-1]^2 + [N+1][N-1] \nonumber \\
g_p &= 4\sin ^2\left(\frac{\pi  N p}{k+N}\right) .
\label{}
\end{align}
The entanglement structure is periodic in any of the twist number $p_j$ with a fundamental period $\textsf{p}$ i.e. 
\begin{equation} 
\text{EE}(p_j) = \text{EE}(p_j+\textsf{p}\mathbb{Z}) \quad;\quad \text{EE}(0)=0  ~,
\label{}
\end{equation}
where the period is given as,\footnote{For SU(2), the entanglement entropy vanishes at $k=2$.}
\begin{equation}
\textsf{p} = \frac{k+N}{\text{gcd}(k,N)} ~.
\label{}
\end{equation}
\begin{figure}[t]
\centerline{\includegraphics[width=5.8in]{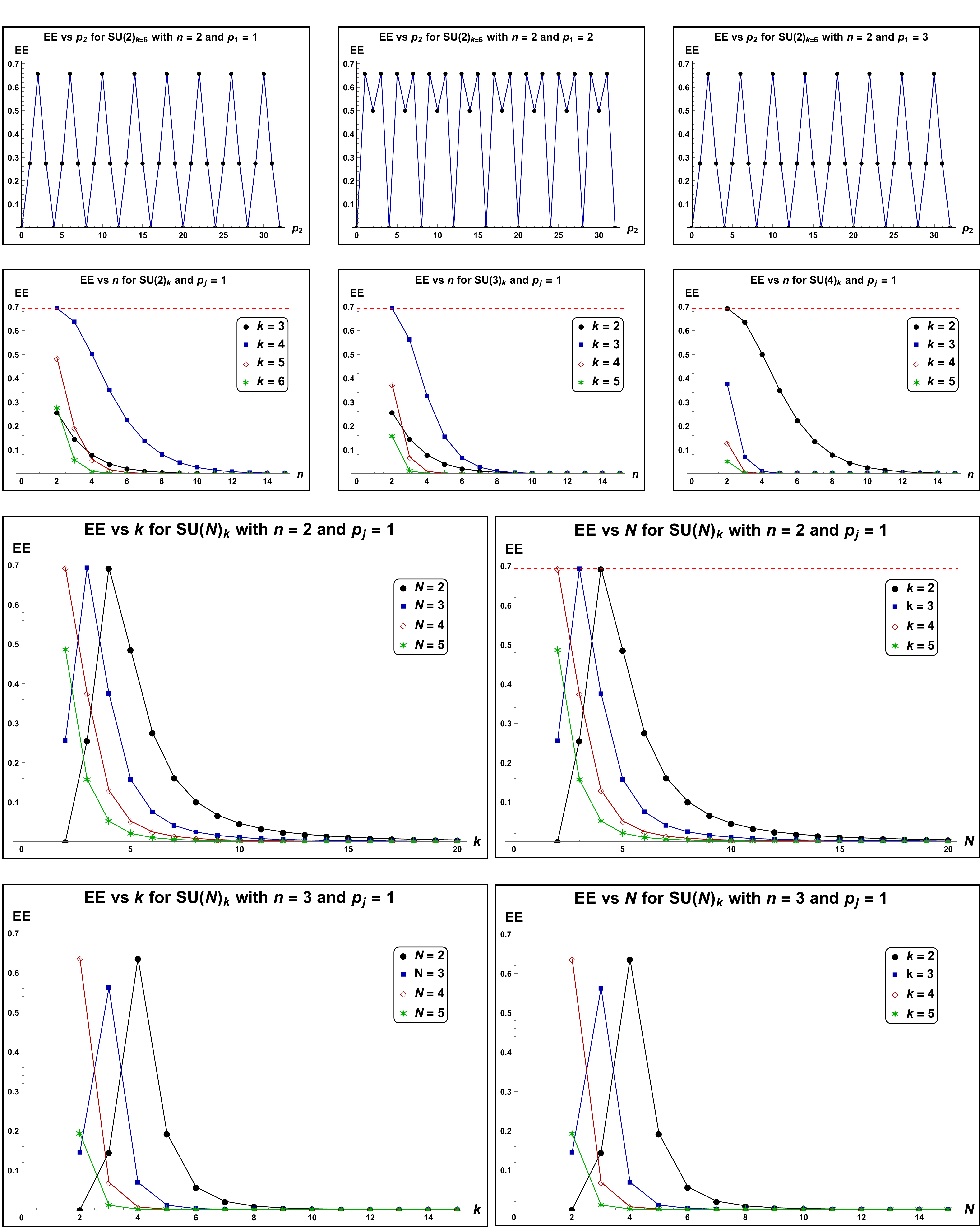}}
\caption[]{Variation of SU$(N)_k$ entropy for the state $\ket{\Psi_4}$ when all the Wilson lines carry fundamental representation. The horizontal dotted line represents the maximum value ($\ln 2$) of entropy. }
\label{EEvsk}
\end{figure}
This behavior can be seen in the first three plots of figure \ref{EEvsk}. The last four plots show that the entropy remains same under the exchange $N\longleftrightarrow k$ just like the earlier case of the state $\ket{\Psi_3}$. This can be explicitly seen from the eq.(\ref{EE-a=1}) noting that $f_p$ and $g_p$ are invariant under $N\longleftrightarrow k$. Since the reduced density matrix has only two non-zero eigenvalues, $\text{EE} \leq \ln 2$. We also did the numerical analysis to find the values of $N$, $k$ and $p_j$ at which the entropy is maximum ($\ln 2$). We find that it is possible only for $n = 2$ in which case the state will be maximally entangled Bell state. The values at which this can happen are given below:
\begin{equation}
\begin{rcases}
    \text{SU(2)}_{k=4K} & \text{with twists $p_j=K, K+1$ (mod $2K+1$)} \\
		\text{SU(3)}_{k=3} & \text{with twists $p_j=1$ (mod 2)} \\
		\text{SU($4M$)}_{k=2} & \text{with twists $p_j=M, M+1$ (mod $2M+1$)} 
\end{rcases} \,\, \Longrightarrow |{\Psi_4^{(n=2)}}\rangle = |{\Psi_{\text{Bell}}}\rangle ~,
\label{Bell-state-a=1}
\end{equation}
where $K\geq 1$ and $M \geq 1$ are integers and the twist values are given modulo the periodicity. Note that these are the same values at which the state $\ket{\Psi_3}$ was found to be a Bell state (see eq.(\ref{Bell-state-Psi3})).

For large values of $k$ or $N$ (with other parameters fixed), the asymptotic behavior of the entanglement entropy is given as,
\begin{align} 
\text{EE}(k \gg 1) &\sim \frac{4^n \pi^{2n} \left(\prod_{j=1}^n p_j^2\right)}{k^{2n} N^{2n}(N^2-1)^{n-2}} \left(1 - \ln\left[ \frac{4^n \pi^{2n} \left(\prod_{j=1}^n p_j^2\right)}{k^{2n} N^{2n}(N^2-1)^{n-2}} \right]  \right) \nonumber \\  
\text{EE}(N \gg 1) &\sim \frac{4^n \pi^{2n} \left(\prod_{j=1}^n p_j^2\right)}{k^{2n} N^{2n}(k^2-1)^{n-2}} \left(1 - \ln\left[ \frac{4^n \pi^{2n} \left(\prod_{j=1}^n p_j^2\right)}{k^{2n} N^{2n}(k^2-1)^{n-2}} \right]  \right)
\label{}
\end{align}
and thus the state is no longer entangled ($\text{EE} \to 0$) as $k \to \infty$ or $N \to \infty$. 

Next consider the case where the Wilson lines carrying fundamental representation braid with the Wilson lines carrying symmetric representation, i.e. our choice is:
\begin{equation}
R_1 = R_3 = \ldots = R_{2n-1}= [1;0] \quad;\quad R_2 = R_4 = \ldots = R_{2n}= [a;0] ~,
\end{equation}
where $a\geq 1$ with $a=1$ corresponding to the fundamental representation which has been already discussed above. The Hilbert space in this case is still two dimensional with the representations $t_\ell$ and $s_\ell$ taking the following values:
\begin{align}
t_\ell &\in (R_{2j-1} \otimes \bar{R}_{2j-1}) \cap (R_{2j} \otimes \bar{R}_{2j}) = [0;0] \oplus [1;1] \nonumber \\
s_\ell &\in (\bar{R}_{2j-1} \otimes R_{2j}) = [a-1;0] \oplus [a;1] ~.
\end{align}
Moreover since we want all the representations appearing in the tensor product to be integrable, we must have $k \geq (a+1)$. The required fusion matrix for our computation can be taken from \cite{Zodinmawia:2011osq}:
\begin{equation}
a_{ts}\left[
\begin{array}{cc}
 R_{2j-1} & \bar{R}_{2j-1} \\
 R_{2j} & \bar{R}_{2j}
\end{array}
\right] = \left(
\begin{array}{c|cc}
  & s_0=[a-1;0] & s_1=[a;1] \\ \hline
t_0=[0;0] & -\sqrt{\frac{[a]}{[N]\,[N+a-1]}} & \sqrt{\frac{[N+a]\,[N-1]}{[N]\,[N+a-1]}} \\[8pt]
t_1=[1;1] & \sqrt{\frac{[N+a]\,[N-1]}{[N]\,[N+a-1]}} & \sqrt{\frac{[a]}{[N]\,[N+a-1]}}
\end{array}
\right) ~.
\end{equation}
The entropy for fundamental representation given in eq.(\ref{EE-a=1}) can be generalized for this case and the result is given as,
\begin{align} 
\text{EE} &= \frac{\left(\prod_{j=1}^nf_{p_j}^{(a)}\right)}{\left(\prod_{j=1}^n f_{p_j}^{(a)}\right) + [N-1]^2[N+1]^2\left(\prod_{j=1}^n g_{p_j}^{(a)}\right)} \ln(\frac{\left(\prod_{j=1}^n f_{p_j}^{(a)}\right) + [N-1]^2[N+1]^2\left(\prod_{j=1}^n g_{p_j}^{(a)}\right)}{\left(\prod_{j=1}^nf_{p_j}^{(a)}\right)}) \nonumber \\
&+ \frac{[N-1]^2[N+1]^2\left(\prod_{j=1}^n g_{p_j}^{(a)}\right)}{\left(\prod_{j=1}^n f_{p_j}^{(a)}\right) + [N-1]^2[N+1]^2\left(\prod_{j=1}^n g_{p_j}^{(a)}\right)} \ln(\frac{\left(\prod_{j=1}^n f_{p_j}^{(a)}\right) + [N-1]^2[N+1]^2\left(\prod_{j=1}^n g_{p_j}^{(a)}\right)}{[N-1]^2[N+1]^2\left(\prod_{j=1}^n g_{p_j}^{(a)}\right)}),
\label{EE-a}
\end{align}
where we have defined the following:
\begin{align} 
f_p^{(a)} &= [N+a]^2[N+1][N-1]^3 + 2\cos\left(\frac{2 \pi p(N+a-1)}{k+N}\right)[N+a][N+1][N-1]^2[a] + [N+1][N-1][a]^2 \nonumber \\
g_p^{(a)} &= \frac{4[N+a][a]}{[N+1]}\sin ^2\left(\frac{\pi\, p\,(N+a-1)}{k+N}\right) .
\label{}
\end{align}
Again the entanglement structure is periodic in any of the twist number $p_j$ with a fundamental period $\textsf{p}$ with the following properties  
\begin{equation} 
\text{EE}(p_j) = \text{EE}(p_j+\textsf{p}\mathbb{Z}) \quad;\quad \text{EE}(0)=0 \quad;\quad \text{EE}(p_j) = \text{EE}(\textsf{p}-p_j)
\label{}
\end{equation}
where the period is given as,
\begin{equation}
\textsf{p} = \frac{k+N}{\text{gcd}(k-a+1,\,N+a-1)} ~.
\label{}
\end{equation}
For fixed values of $p_j$, the asymptotic behavior of the entropy for large $k$ (finite $N$) and large $N$ (finite $k$) are given below:
\begin{align} 
\text{EE}(k \gg 1) &\sim \frac{4^n \pi^{2n} a^n (N+a)^n  \left(\prod_{j=1}^n p_j^2\right)}{k^{2n} N^{2n}(N-1)^{n-2}(N+1)^{2n-2}} \left(1 - \ln\left[ \frac{4^n \pi^{2n} a^n (N+a)^n  \left(\prod_{j=1}^n p_j^2\right)}{k^{2n} N^{2n}(N-1)^{n-2}(N+1)^{2n-2}} \right]  \right) \nonumber \\  
\text{EE}(N \gg 1) &\sim \frac{4^n \pi^{2n} a^n (k-a)^n  \left(\prod_{j=1}^n p_j^2\right)}{k^{2n} N^{2n}(k+1)^{n-2}(k-1)^{2n-2}} \left(1 - \ln\left[ \frac{4^n \pi^{2n} a^n (k-a)^n  \left(\prod_{j=1}^n p_j^2\right)}{k^{2n} N^{2n}(k+1)^{n-2}(k-1)^{2n-2}} \right]  \right) ~.
\label{}
\end{align}
Just like the earlier case, we find that the entropy may obtain its maximum value $\ln 2$ only when $n=2$ and this can happen for the following values of $a$, $N$, $k$ and $p_j$:
\begin{alignat}{3}
&\text{SU(2)}_{k=4K}&&: a=1 \,\, \& \,\, a=(4K{-}1) \,&&;\quad p_j=K \,\, \& \,\, K{+}1 \, (\text{mod } \textsf{p}) \nonumber \\
&\text{SU(3)}_{k=2K+1}&&: a=K&&;\quad p_j=1 \, (\text{mod } \textsf{p})   \nonumber \\
&\text{SU($4M$)}_{k=4K-2}&&: a=(4K{-}3)&&;\quad 	p_j=(K{+}M{-}1) \,\, \& \,\, (K{+}M) \, (\text{mod } \textsf{p}) ~,
\label{Bell-state-a}
\end{alignat}
where $K\geq 1$ and $M \geq 1$ are integers and the twist numbers are given modulo appropriate periodicity values \textsf{p}. Thus these values of the parameters will describe the state $\ket{\Psi_4}$ as a Bell state.

For the examples in this section, we have restricted to the symmetric representations of SU($N$), the tensor decompositions of which are always multiplicity free. Unfortunately the Racah matrices are not known in general and we are limited in our choice of representations. In the next example, we will consider the \text{$\tiny\yng(2,1)$} representation of SU$(N)$, the Racah matrices of which were obtained in \cite{Gu:2014nva}. In this case, we can get $t_\ell$ with multiplicity which enables us to show the explicit results of entanglement negativity.
\subsubsection{Wilson lines carrying mixed representation of SU($N$)}
Let us consider the case when all the Wilson lines carry $(2,1) = \text{$\tiny\yng(2,1)$}$ representation of SU($N$), i.e. we will set
\begin{equation}
R_1 = R_2 = \ldots = R_{2n} = R = \text{$\tiny\yng(2,1)$} ~.
\end{equation}
In this case, the representations $t_\ell, s_\ell \in (R \otimes \bar{R})$ which is given as:
\begin{align}
R \otimes \bar{R} = t_1 \oplus t_2 \oplus t_3 \oplus t_4 \oplus t_5 \oplus t_6 \oplus 2\,t_7 ~,
\end{align}
where various representations are given as following:
\begin{alignat}{3}
t_1 = \bullet, \quad t_2 &= (2^2, 1^{N-4}), &\quad t_3 &= (3, 1^{N-3}), &\quad t_4 &= (3^2, 2^{N-3}) \nonumber \\
t_5 &= (4, 2^{N-2}), & t_6 &= (4, 3, 2^{N-4}, 1), & t_7 &= (2, 1^{N-2}) ~.
\end{alignat}
The representation $t_7$ occurs with multiplicity 2 while others appear once. The Racah matrix elements are given in the appendix B. The multiplicity labels will be omitted for $t_1, t_2, t_3, t_4, t_5, t_6$ since they occur only once. The 3j-phases required for the computation can be taken as following:
\begin{alignat}{4}
\{\bar{R}, R, t_1\} &= 1, &\quad \{\bar{R}, R, t_2\} &= 1, &\quad \{\bar{R}, R, t_3\} &= -1, &\quad \{\bar{R}, R, t_4\} &= -1 \nonumber \\
\{\bar{R}, R, t_5\} &= 1, & \{\bar{R}, R, t_6\} &= 1, &\quad \{\bar{R}, R, t_7, 1\} &= 1, &\quad \{\bar{R}, R, t_7, 2\} &= -1 ~.
\end{alignat}
The quantum dimensions of various representations are:
\begin{alignat}{2}
\text{dim}_q t_1 = 1, \quad \text{dim}_q t_2 &= \frac{[N-3][N]^2[N+1]}{[2]^2}, &\quad \text{dim}_q t_3 &= \text{dim}_q t_4 = \frac{[N-2][N-1][N+1][N+2]}{[2]^2}, \nonumber \\
\text{dim}_q t_5 &= \frac{[N-1][N]^2[N+3]}{[2]^2}, &\quad \text{dim}_q t_6 &= \frac{[N-3][N-1]^2[N+1]^2[N+3]}{[3]^2}, \nonumber \\
\text{dim}_q t_7 &= [N-1][N+1] ~.
\end{alignat}
We also require the quadratic Casimirs:
\begin{equation}
C_{t_1} = 0,\, C_{t_2} = 2(N-1),\, C_{t_3} = 2N,\, C_{t_4} = 2N,\, C_{t_5} = 2(N+1),\, C_{t_6} = 3N,\, C_{t_7} = N ~.
\end{equation}
The function $F$ in the eq.(\ref{F-function}) for various representations will be given as,
\begin{align} 
F_j(t_y, p_j) &= \sum_{\ell=1}^6 \{ \bar{R}, R, t_{\ell} \} \sqrt{\text{dim}_q t_{\ell}}\, \mathcal{R}_{\ell y}\,\, \Omega_j(t_{\ell}) + \sqrt{\text{dim}_q t_{7}}(\mathcal{R}_{7y}-\mathcal{R}_{8y})\, \Omega_j(t_{7}) \nonumber \\
F_j(t_7,1,1,p_j) &= \sum_{\ell=1}^6 \{ \bar{R}, R, t_{\ell} \} \sqrt{\text{dim}_q t_{\ell}}\, \mathcal{R}_{\ell 7}\,\, \Omega_j(t_{\ell}) + \sqrt{\text{dim}_q t_{7}}\,(\mathcal{R}_{77}-\mathcal{R}_{87})\, \Omega_j(t_{7}) \nonumber \\
F_j(t_7,2,2,p_j) &= \sum_{\ell=1}^6 \{ \bar{R}, R, t_{\ell} \} \sqrt{\text{dim}_q t_{\ell}}\, \mathcal{R}_{\ell 8}\,\, \Omega_j(t_{\ell}) + \sqrt{\text{dim}_q t_{7}}\,(\mathcal{R}_{78}-\mathcal{R}_{88})\, \Omega_j(t_{7}) \nonumber \\
F_j(t_7,1,2,p_j) &= \sum_{\ell=1}^6 \{ \bar{R}, R, t_{\ell} \} \sqrt{\text{dim}_q t_{\ell}}\, \mathcal{R}_{\ell 9}\,\, \Omega_j(t_{\ell}) + \sqrt{\text{dim}_q t_{7}}\,(\mathcal{R}_{79}-\mathcal{R}_{89})\, \Omega_j(t_{7}) \nonumber \\
F_j(t_7,2,1,p_j) &= \sum_{\ell=1}^6 \{ \bar{R}, R, t_{\ell} \} \sqrt{\text{dim}_q t_{\ell}}\, \mathcal{R}_{\ell,10}\, \Omega_j(t_{\ell}) + \sqrt{\text{dim}_q t_{7}}\,(\mathcal{R}_{7,10}-\mathcal{R}_{8,10})\, \Omega_j(t_{7}) ,
\label{F-fun-(2,1)}
\end{align}
where $t_y$ in the first equation can take values $t_y = t_1, t_2, t_3, t_4, t_5, t_6$. The $\mathcal{R}_{ij}$ denotes the element of the Racah matrix $\mathcal{R}$ given in the appendix B in eq.(\ref{Rac-(2,1)}) and we have defined\footnote{We have discarded the terms $\Omega_j(R)^{-1}$ and $\Omega_j(\bar{R})^{-1}$ in $F_j$ because these are overall factors and will be canceled out while computing $F_jF_j^{*}$.}:
\begin{equation}
\Omega_j(t) = \exp\left[\frac{2 i\pi \, p_j\, C_{t}}{k+N} \right] ~.
\end{equation}
Each of the Hilbert space in this case is ten dimensional and hence the total Hilbert space has dimension: $\text{dim}(\mathcal{H}) = 10^n$. We divide this Hilbert space into $\mathcal{H}_A \otimes \mathcal{H}_B$ corresponding to the bi-partition $(m|n-m)$. Tracing out $\mathcal{H}_B$ gives the reduced density matrix acting on $\mathcal{H}_A$ ($\text{dim}(\mathcal{H}_A) = 10^m$) whose block diagonal form is given as:
\begin{equation}
\rho_A = \frac{1}{\braket{\Psi_4}} \left(
\begin{array}{ccccc}
 \dfrac{\prod_{j=1}^{n} \left| F_j(t_1, p_j) \right|^2}{(\text{dim}_q\,t_1)^{2n-2}} & & & &  \\
 & \ddots & & &  \\
 & & & \dfrac{\prod_{j=1}^{n} \left| F_j(t_6, p_j) \right|^2}{(\text{dim}_q \, t_6)^{2n-2}} &  \\
 & & & & \dfrac{\prod_{j=m+1}^{n-1} \alpha_j(t_7)}{(\text{dim}_q\, t_7)^{2n-2}} M
\end{array}
\right) ~,
\label{rho-Psi3-(2,1)}
\end{equation}
where 
\begin{equation}
\alpha_j(t_7) = \left| F_j(t_7, 1, 1, p_j) \right|^2 + \left| F_j(t_7, 1, 2, p_j) \right|^2 + \left| F_j(t_7, 2, 1, p_j) \right|^2 + \left| F_j(t_7, 2, 2, p_j) \right|^2
\end{equation}
and $M$ is a square matrix of order $2^{2m}$ given as Kronecker product:\footnote{Note that since $\text{dim}(\mathcal{H}_A) = 10^m$, the density matrix $\rho_A$ is also of the order $10^m$. However many of its blocks are 0 and we have only written the non-zero blocks in eq.(\ref{rho-Psi3-(2,1)}).}
\begin{equation}
M = X \otimes Y_1 \otimes Y_2 \otimes \ldots \otimes Y_{m-1} \otimes Z ~.
\end{equation}
Here $X$ and $Z$ are $2 \times 2$ matrices given as,
\begin{align}
X &= \left(
\begin{array}{cc}
 \sum_{i=1}^2 \left| F_n(t_7, i, 1, p_n) \right|^2 & \sum_{i=1}^2 F_n(t_7, i, 1, p_n)\, F_n^{*}(t_7, i, 2, p_n) \\
 \sum_{i=1}^2 F_n(t_7, i, 2, p_n)\, F_n^{*}(t_7, i, 1, p_n) & \sum_{i=1}^2 \left| F_n(t_7, i, 2, p_n) \right|^2
\end{array} \right) \nonumber \\ 
Z &= \left(
\begin{array}{cc}
 \sum_{i=1}^2 \left| F_m(t_7, 1, i, p_m) \right|^2 & \sum_{i=1}^2 F_m(t_7, 1, i, p_m)\, F_m^{*}(t_7, 2, i, p_m) \\
 \sum_{i=1}^2 F_m(t_7, 2, i, p_n)\, F_m^{*}(t_7, 1, i, p_m) & \sum_{i=1}^2 \left| F_m(t_7, 2, i, p_m) \right|^2
\end{array} \right) ~.
\label{X&Z-matrix}
\end{align} 
Also the matrix $Y_i$ is a $4\times 4$ matrix:
\begin{align}
Y_i = \left(
\begin{array}{cccc}
 F_i^{*}(t_7, 1, 1, p_i) & F_i^{*}(t_7, 1, 2, p_i) & F_i^{*}(t_7, 2, 1, p_i) & F_i^{*}(t_7, 2, 2, p_i) \\
\end{array}
\right) \otimes \left(
\begin{array}{c}
 F_i(t_7, 1, 1, p_i) \\
 F_i(t_7, 1, 2, p_i) \\
 F_i(t_7, 2, 1, p_i) \\
 F_i(t_7, 2, 2, p_i) \\
\end{array}
\right)  
\label{Y-matrix} ~.
\end{align}

Let us first do the computation for $n=2$ with the bi-partition $(1|1)$, i.e. $m=1$. The reduced density matrix in this case will be,
\begin{equation}
\rho_A = \frac{1}{\braket{\Psi_4}} \left(
\begin{array}{ccccc}
 \dfrac{\left| F_1(t_1, p_1) \right|^2 \left| F_2(t_1, p_2) \right|^2}{(\text{dim}_q\,t_1)^{2}} & & & &  \\
 & \ddots & & &  \\
 & & & \dfrac{\left| F_1(t_6, p_1) \right|^2 \left| F_2(t_6, p_2) \right|^2}{(\text{dim}_q\,t_6)^{2}} &  \\
 & & & & \dfrac{X \otimes Z}{(\text{dim}_q\, t_7)^{2}} 
\end{array}
\right) ~,
\end{equation}
where $X$ and $Z$ can be obtained from eq.(\ref{X&Z-matrix}) by setting $m=1$ and $n=2$. The entanglement structure will be invariant under the exchange $p_1 \leftrightarrow p_2$ and is periodic in $p_1$ or $p_2$ with properties:
\begin{equation} 
\text{EE}_A(p_j) = \text{EE}_A(p_j+\textsf{p}\mathbb{Z}) \quad;\quad \text{EE}_A(0)=0 \quad;\quad \text{EE}_A(p_j) = \text{EE}_A(\textsf{p}-p_j) ~,
\label{}
\end{equation}
where the period is given as,
\begin{equation}
\textsf{p} = \begin{cases}
(k+N),\,\, \text{ for $N=$ odd} \\
(k+N),\,\, \text{ for $N=$ even \& gcd$(k,N)=1$} \\
\dfrac{(k+N)}{2},\,\, \text{ for $N=$ even \& gcd$(k,N)>1$}
\end{cases} ~.
\label{period-(2,1)-n=2}
\end{equation}
This periodic behavior can be seen from the first two plots of figure \ref{EEPlotsPsi3-(2,1)-n=2}. We have also given various other plots in figure \ref{EEPlotsPsi3-(2,1)-n=2} showing the variation of SU$(N)_k$ entanglement entropy as a function of $k$ for fixed $N$ and vice versa. We can see that $\text{EE} \rightarrow 0$ as $k \to \infty$ or $N \to \infty$ and the state $|{\Psi_4^{(n=2)}}\rangle$ becomes a non-entangled state in this limit.
\begin{figure}[t]
\centerline{\includegraphics[width=6.0in]{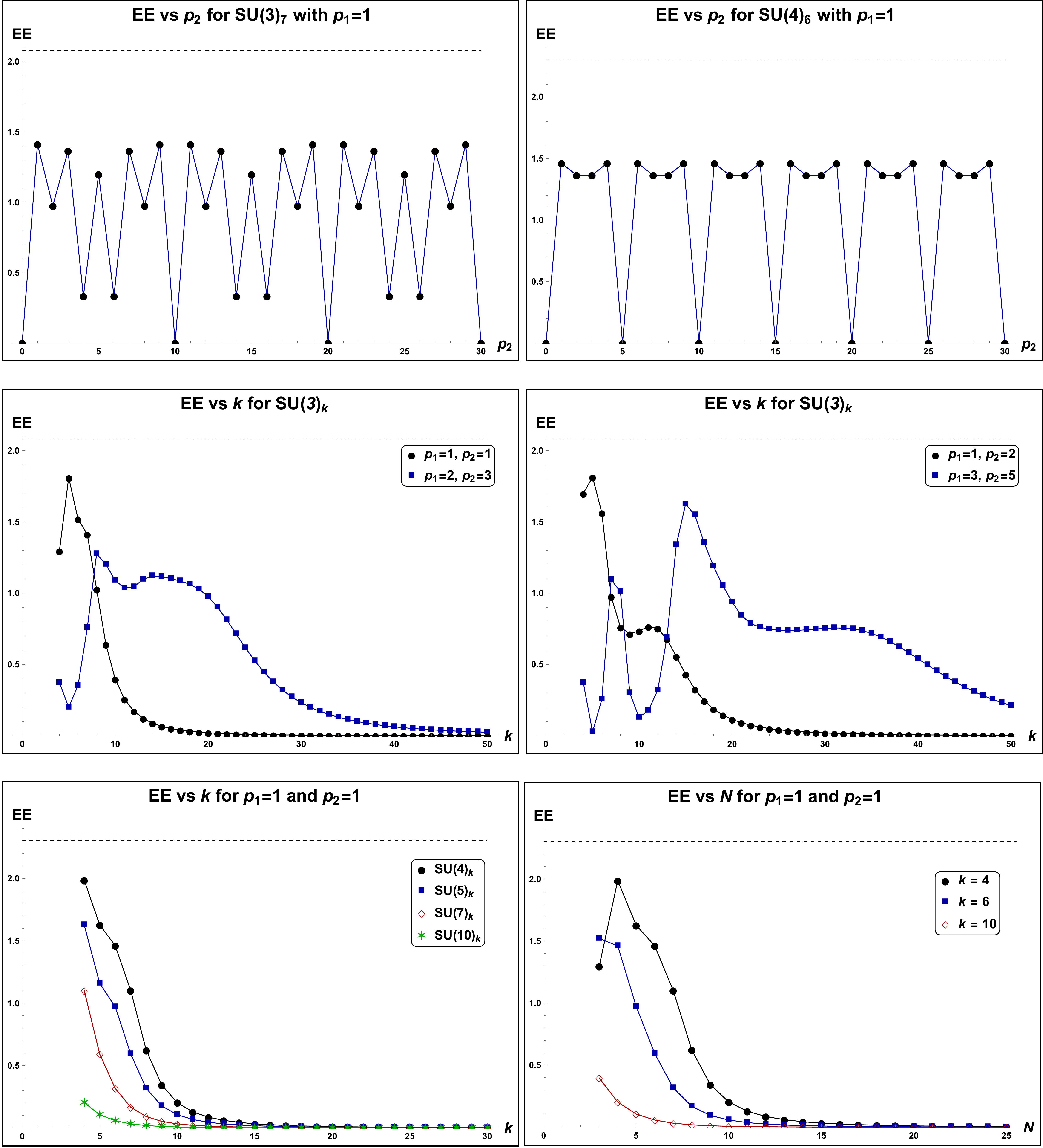}}
\caption[]{The variation of SU$(N)_k$ entanglement entropy for $(1|1)$ bi-partition of the state $|{\Psi_4^{(n=2)}}\rangle$ when all the Wilson lines carry $\text{$\tiny\yng(2,1)$}$ representation. The horizontal dotted line corresponds to maximum possible value of EE which is $\ln8$ for SU(3) and $\ln10$ for SU($N \geq 4$).}
\label{EEPlotsPsi3-(2,1)-n=2}
\end{figure}

Next, let us study the entanglement structure of the state $|{\Psi_4^{(n=3)}}\rangle$. The total Hilbert space is $\mathcal{H} = \mathcal{H}_1 \otimes \mathcal{H}_2 \otimes \mathcal{H}_3$ with $\text{dim}(\mathcal{H}) = 1000$ where each of $\mathcal{H}_i$ is the space associated with $S^2$ with four point conformal block carrying representations $\{ \bar{R}, R, \bar{R}, R \}$ with $R = \text{$\tiny\yng(2,1)$}$. There are two inequivalent ways of tracing out Hilbert spaces. In the first case, we trace out $\mathcal{H}_2 \otimes \mathcal{H}_3$ which correspond to $m=1$. The reduced density matrices can be obtained from the general form given in eq.(\ref{rho-Psi3-(2,1)}) by substituting $n=3$ and $m=1$. The entanglement entropy has the same periodicity and follows the same pattern as in the case of $(1|1)$ bi-partition for the $n=2$ case. The only difference is that the entanglement entropies in this case are smaller than the corresponding eigenvalues of the $(1|1)$ bi-partition. This is in fact a general trend, where the entanglement entropy for $(1|n-1)$ bi-partition decreases as $n$ increases (keeping other parameters fixed).

The second way of tracing out the Hilbert space is to trace out only $\mathcal{H}_3$ which correspond to $m=2$ and gives a reduced density matrix defined on $\mathcal{H}_1 \otimes \mathcal{H}_2$ which is:\footnote{For SU(3), the representations $t_2$ and $t_6$ are no longer present in the tensor decomposition of $R \otimes \bar{R}$ and we must set the corresponding blocks to 0 in eq.(\ref{rhoAB-n=3}).}
\begin{equation}
\rho = \frac{1}{\braket{\Psi_4}} \left(
\begin{array}{ccccc}
 \dfrac{\prod_{j=1}^{3} \left| F_j(t_1, p_j) \right|^2}{(\text{dim}_q\,t_1)^{4}} & & & &  \\
 & \ddots & & &  \\
 & & & \dfrac{\prod_{j=1}^{3} \left| F_j(t_6, p_j) \right|^2}{(\text{dim}_q\,t_6)^{4}} &  \\
 & & & & \dfrac{X \otimes Y_1 \otimes Z}{(\text{dim}_q\, t_7)^{4}} 
\end{array}
\right) ~.
\label{rhoAB-n=3}
\end{equation}
To keep the analysis simple, let us set $p_1 = p_2 = p_3 = p$. The explicit form of matrices $X$, $Y_1$ and $Z$ can be obtained from eq.(\ref{X&Z-matrix}) and eq.(\ref{Y-matrix}) by setting $m=2$ and $n=3$ where we note that $F_1 = F_2 = F_3 = F$ since we have set $p_1 = p_2 = p_3$. We are also interested in computing the entanglement negativity. The partial transpose of $\rho$ with respect to $\mathcal{H}_2$ can be obtained from the general form given in eq.(\ref{par-trans}). The matrices $X$ and $Z$ for this case are given as,
\begin{equation}
X = \left(
\begin{array}{cc}
 \sum_{i=1}^2 F_{i1}\,F_{i1}^{*} & \sum_{i=1}^2 F_{i1}\,F_{i2}^{*} \\
 \sum_{i=1}^2 F_{i2}\,F_{i1}^{*} & \sum_{i=1}^2 F_{i2}\,F_{i2}^{*}
\end{array} \right) \quad;\quad Z = \left(
\begin{array}{cc}
 \sum_{i=1}^2 F_{1i}\,F_{1i}^{*} & \sum_{i=1}^2 F_{1i}\,F_{2i}^{*} \\
 \sum_{i=1}^2 F_{2i}\,F_{1i}^{*} & \sum_{i=1}^2 F_{2i}\,F_{2i}^{*}
\end{array} \right) ~.
\end{equation}
Also the matrix $Y_1$ and its partial transpose are:
\begin{equation}
Y_1 = \left(
\begin{array}{cccc}
 F_{11}F_{11}^{*} & F_{11}F_{12}^{*} & F_{11}F_{21}^{*} & F_{11}F_{22}^{*} \\
 F_{12}F_{11}^{*} & F_{12}F_{12}^{*} & F_{12}F_{21}^{*} & F_{12}F_{22}^{*} \\
 F_{21}F_{11}^{*} & F_{21}F_{12}^{*} & F_{21}F_{21}^{*} & F_{21}F_{22}^{*} \\
 F_{22}F_{11}^{*} & F_{22}F_{12}^{*} & F_{22}F_{21}^{*} & F_{22}F_{22}^{*} \\
\end{array}
\right), Y_1^{\Gamma} = \left(
\begin{array}{cccc}
 F_{11}F_{11}^{*} & F_{12}F_{11}^{*} & F_{11}F_{21}^{*} & F_{12}F_{21}^{*} \\
 F_{11}F_{12}^{*} & F_{12}F_{12}^{*} & F_{11}F_{22}^{*} & F_{12}F_{22}^{*} \\
 F_{21}F_{11}^{*} & F_{22}F_{11}^{*} & F_{21}F_{21}^{*} & F_{22}F_{21}^{*} \\
 F_{21}F_{12}^{*} & F_{22}F_{12}^{*} & F_{21}F_{22}^{*} & F_{22}F_{22}^{*} \\
\end{array}
\right) ~,
\end{equation} 
where we have defined $F_{\alpha \beta} \equiv F(t_7, \alpha, \beta, p)$ which are given in eq.(\ref{F-fun-(2,1)}). Both the entanglement entropy and negativity are periodic in twist number $p$:
\begin{alignat}{3}
\text{EE}_A(p) &= \text{EE}_A(p_j+\textsf{p}\mathbb{Z}) &\quad;\quad \text{EE}_A(0) &= 0 &\quad;\quad \text{EE}_A(p) &= \text{EE}_A(\textsf{p}-p) \nonumber \\
\mathcal{N}(p) &= \mathcal{N}(p+\textsf{p}\mathbb{Z}) &\quad;\quad \mathcal{N}(0) &= 0 &\quad;\quad \mathcal{N}(p) &= \mathcal{N}(\textsf{p}-p) ~,
\label{}
\end{alignat}
where the period \textsf{p} is the same as given in eq.(\ref{period-(2,1)-n=2}). 
\begin{figure}[t]
\centerline{\includegraphics[width=7.0in]{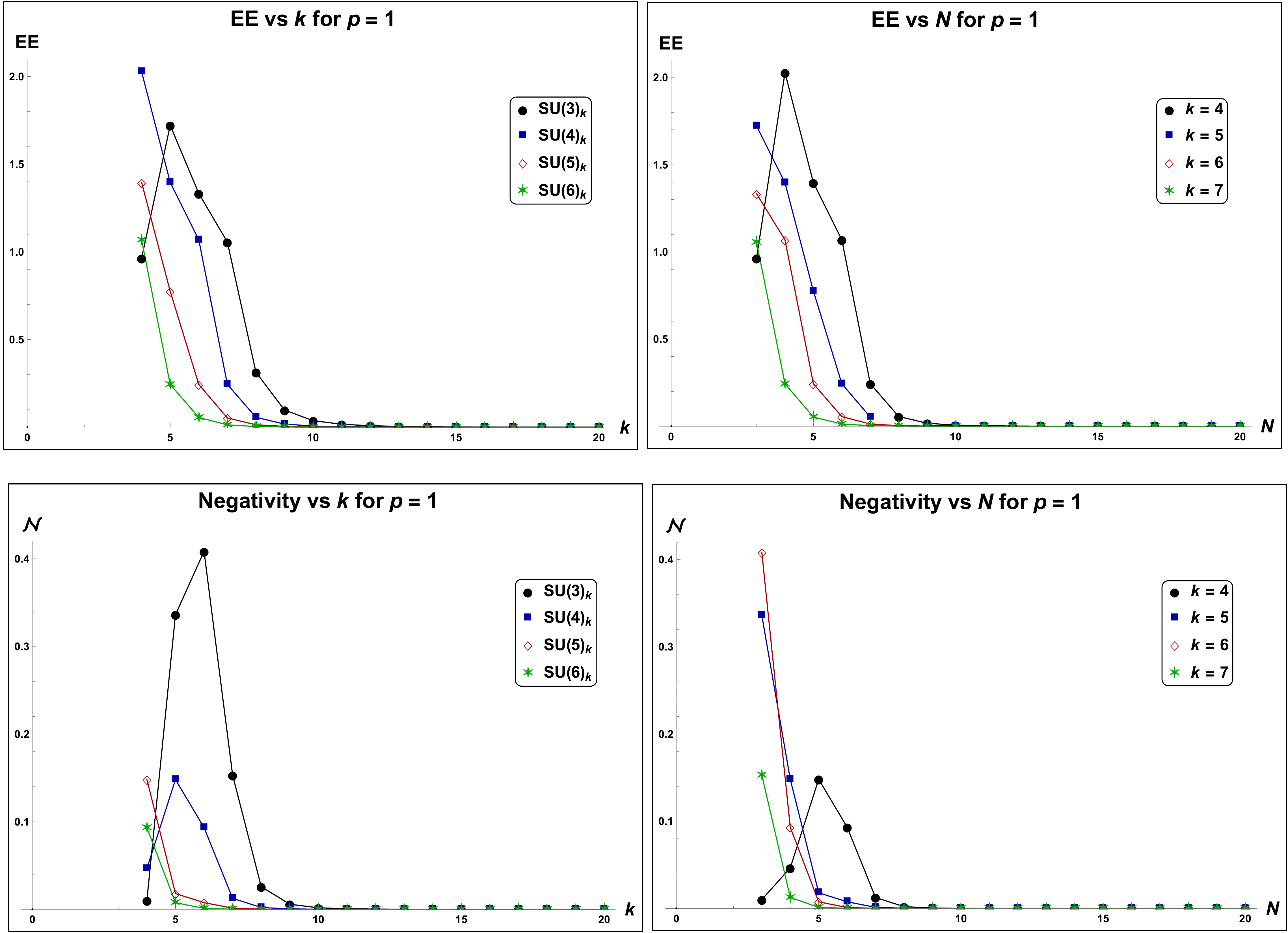}}
\caption[]{Plots showing variation of SU$(N)_k$ entanglement entropy and entanglement negativity obtained for the reduced density matrix of eq.(\ref{rhoAB-n=3}) acting on the Hilbert space $\mathcal{H}_1 \otimes \mathcal{H}_2$. The non-zero values of $\mathcal{N}$ implies that it is non-separable and the state $|{\Psi_4^{(n=3)}}\rangle$ has a $W$-like entanglement structure.}
\label{PlotsPsi3-(2,1)-n=3}
\end{figure}
We have given some plots in figure \ref{PlotsPsi3-(2,1)-n=3} showing the variation of SU$(N)_k$ entanglement entropy and entanglement negativity as a function of $k$ for fixed $N$ and vice versa. We can see that the entanglement negativity is non-zero. Thus the reduced density matrix is non-separable on $\mathcal{H}_1 \otimes \mathcal{H}_2$ and the state $|{\Psi_4^{(n=3)}}\rangle$ has a $W$-like entanglement structure.
\section{Conclusion and discussion}
\label{sec5}
The theme of this paper was to understand the entanglement structure between $n$ disjoint $S^2$ boundaries of a three-manifold described by $SU(N)$ Chern-Simons theory. We have focused on $S^2$ with four punctures corresponding to the four point conformal blocks carrying the integrable representations of SU($N$) gauge group with Chern-Simons level $k$. The punctures are connected through the Wilson lines embedded in the bulk of the three-manifold. The quantum states considered in this work live in the total Hilbert space which is a tensor product of the Hilbert spaces associated with each $S^2$. The states are labeled by those representations `$t$' of SU$(N)$ which appear in the fusion of the representations associated with various conformal blocks. Further, since $t$ may appear multiple times in the fusion, the states also carry extra labels to keep track of the multiple occurrences of $t$. In fact, the multiplicity of $t$ plays a very crucial role in determining the entanglement structure of these quantum states. When there is no multiplicity, the states have a GHZ-like entanglement structure, i.e., the reduced density matrices (obtained by tracing one or more $\mathcal{H}_{S^2}$) are fully separable. We proved this by decomposing the reduced density matrix into a convex combination of pure product states (see eqn(\ref{sep-Psi12}) and eqn(\ref{sep-Psi3})). On the other hand, when there is multiplicity, we show that the entanglement negativity is non-zero for the states $|{\Psi_1}\rangle$ and $|{\Psi_2}\rangle$ (eq.(\ref{neg-Psi1}) and eq.(\ref{neg-Psi2}) respectively). The entanglement negativity for the state $\ket{\Psi_4}$ for $n=3$ is worked out for the $\text{$\tiny\yng(2,1)$}$ representations placed on the Wilson lines and is shown to be non-zero in the plots of figure \ref{PlotsPsi3-(2,1)-n=3}. This exercise shows that the multiplicity is responsible for the  reduced density matrix to be non-separable, i.e. entangled and the states will have a W-like entanglement structure. 

The entanglement structure of the states $|{\Psi_1}\rangle$ and $|{\Psi_2}\rangle$ remains invariant under the braiding between the Wilson lines. We explicitly show that any braiding can be seen as a unitary transformation on the basis states and does not change the entropy. Our computation for general $n$ is consistent with the diagrammatic replica method described in \cite{Melnikov:2018zfn} for $n=2$. We also consider the two boundary states $\ket{\Psi_3}$ and $\ket{\Psi_4}$ which have necklace braiding in which the Wilson lines connecting the punctures of $j^{\text{th}}$ boundary braid $2p_j$ times (i.e. $p_j$ number of full twists) with the Wilson lines connecting the punctures of $(j+1)^{\text{th}}$ boundary. This type of braiding affects the entanglement structure and we give various examples showing its effect on the entropy. In fact we have generalized $\ket{\Psi_4}$ to $n$ boundaries and gave various plots showing the variation of entropy with $n$ in figure \ref{EEvsk}. We find that these states have a periodic structure in any of the twist number $p_j$, i.e. the spectrum of the reduced density matrix remains same if we increase $p_j$ to $(p_j + \textsf{p} \mathbb{Z})$.  Moreover when any of the $p_j$ is 0 or a multiple of $\textsf{p}$, the state becomes a non-entangled state and the entropy vanishes. We call the integer $\textsf{p}$ as `periodicity' which is the fundamental period of this periodic behavior and its value depends on $N$ and $k$. We give various examples showing this periodic behavior in the entropy and explicitly obtained the values of $\textsf{p}$.

For two boundaries ($n=2$), we find that the states $|{\Psi_1}\rangle$ and $|{\Psi_2}\rangle$ are Bell states (which is in agreement with \cite{Melnikov:2018zfn}). Further we show that it may be possible for the states $\ket{\Psi_3}$ and $\ket{\Psi_4}$ to be the Bell states for particular values of $N$ and $k$ as listed in eq.(\ref{Bell-state-Psi3}) and eq.(\ref{Bell-state-a}) for symmetric representations. For $n \geq 3$ boundaries, we find that it is possible for the states $|{\Psi_1}\rangle$ and $|{\Psi_2}\rangle$ to be maximally entangled GHZ states for a specific choice of rank and level of the gauge group as given in eq.(\ref{GHZ-fund}). For example, when all the Wilson lines carry fundamental representation, $|{\Psi_1}\rangle$ will be a GHZ state for SU$(N)_{k=N}$ and $|{\Psi_2}\rangle$ will be a GHZ state for SU$(2)_{k=2}$.

For each of the examples considered in this paper, we also analyzed the large $k$ and large $N$ behavior of the entanglement entropies. For the states $|{\Psi_1}\rangle$ and $|{\Psi_2}\rangle$, the large $k$ limit of the entropy can be obtained by simply replacing $\text{dim}_q(t) \to \text{dim}(t)$. We also gave a prescription on how to compute the $N \to \infty$ (for finite value of $k$) limit of the entropy using the identities of $q$-numbers. In these limits, the entropy converges to a finite value. We also obtained the entropy values when both $N \to \infty$ and $k \to \infty$ (in no particular order). We find that for symmetric representations and $\text{$\tiny\yng(2,1)$}$ representations, the entropy for the state $|{\Psi_1}\rangle$ converges to a non-zero value but the entropy of $|{\Psi_2}\rangle$ (for $n>2$)  goes to 0 (eq.(\ref{EElargekN-con}) and eq.(\ref{EE-mixed-con})). In the case of adjoint representation (where $|{\Psi_1}\rangle = |{\Psi_2}\rangle$), the entropy vanishes in this limit (eq.(\ref{EEAdj-con})). Moreover the large $k$ and large $N$ values of entropy for the states $|{\Psi_3}\rangle$ and $|{\Psi_4}\rangle$ tends to 0 as $k \to \infty$ or $N \to \infty$ which is clear from the plots shown in figures \ref{EEPlots1Necklace}, \ref{EEvsk}, \ref{EEPlotsPsi3-(2,1)-n=2} and \ref{PlotsPsi3-(2,1)-n=3}.

Entanglement entropy for three-manifolds with two $S^2$ boundaries having more than four punctures, related by cobordism,  needs to be investigated. There could be local braiding and necklaces of Wilson lines inside such a three-manifold. The computation of quantum states will involve six or higher point conformal blocks. Further, we need to define the quantum states for three-manifolds with $n$ copies of $S^2$ boundaries with six or more punctures to determine the  necklace states. It appears that in such cases, the local braiding will give entanglement entropy involving the partition function on $S^2 \times S^1$ in the presence of Wilson lines (similar to the eq.(\ref{EE-Z-fn})) \cite{Melnikov:2018zfn}. However, such a relation between the entanglement entropy and the partition function on $S^2 \times S^1$ in the presence of necklaces needs to be explored. We will pursue these aspects in future.

In \cite{Dong:2008ft}, the entanglement entropy by partial tracing within an $S^2$ boundary divided into two spatial regions was discussed. It would be interesting to study the entanglement structure obtained by the additional division of each $S^2$ into sub-regions for the three-manifolds with $n$ copies of $S^2$. We hope to report progress in future work.

Recently, the entanglement structure in Chern-Simons theory is discussed within topological string context \cite{Hubeny:2019bje} exploiting Gopakumar-Vafa duality. The implication of our results on the dual closed topological strings is another plausible direction to pursue.
 \vspace{0.5cm}

\textbf{Acknowledgements}
We would like to thank Abhijit Gadde, Shiraz Minwalla, R.K. Kaul, Piotr Sulkowski, T.R. Govindarajan and Lata Kh Joshi for useful discussion and correspondence. The work of VKS is supported by the ERC Starting Grant no. 335739 ``Quantum fields and knot homologies'' funded by the European Research Council under the European Union's Seventh Framework Programme. PR and Saswati would like to acknowledge DST-RFBR grant (INT/RUS/RFBR/P-309) for support. Saswati would like to thank CSIR for the research fellowship. PR would like to thank ICTP where part of this work was done during her visit as senior associate.

\appendix
\section{Computing the states $|{\Psi_3}\rangle$ and $|{\Psi_4}\rangle$}
\label{appendix-A}
\begin{figure}[h]
\centerline{\includegraphics[width=6.0in]{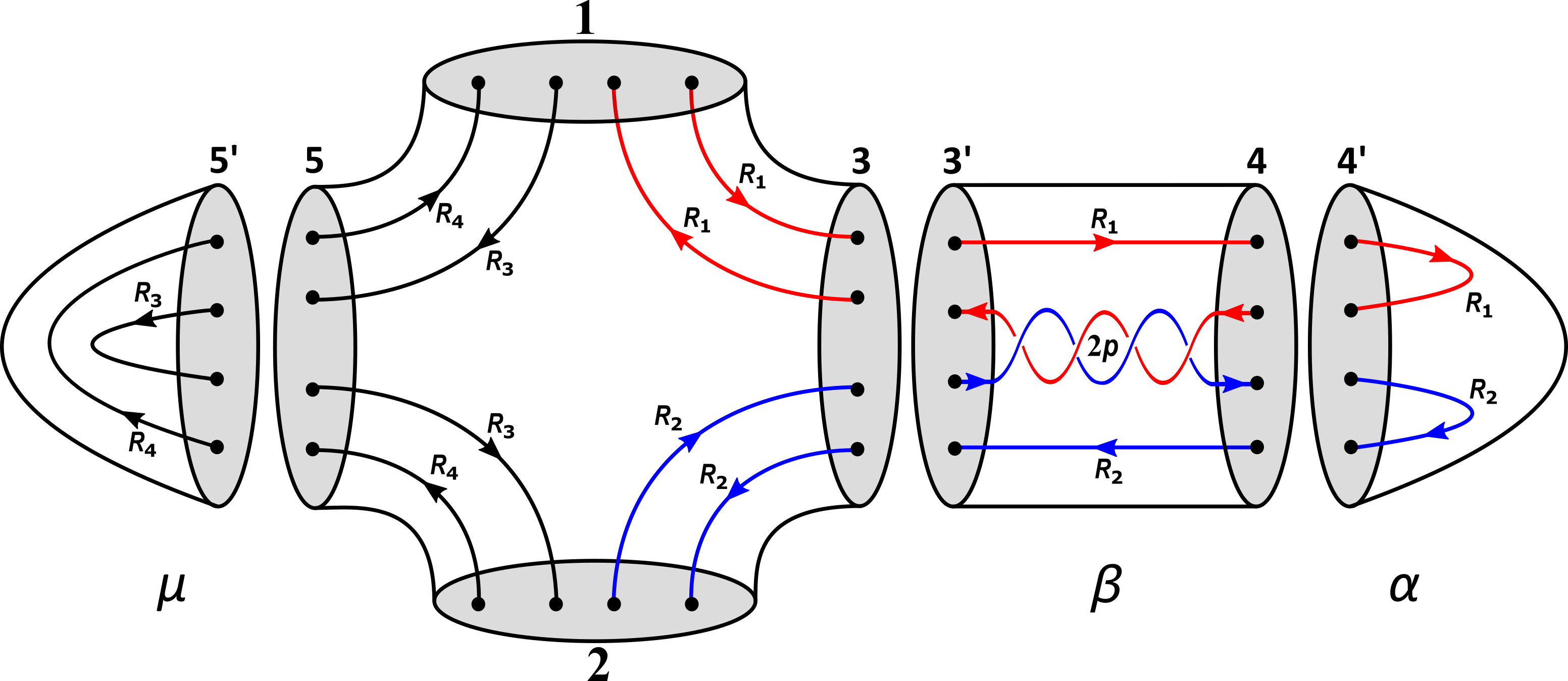}}
\caption[]{Figure showing the construction of the state $|{\Psi_3}\rangle$. The $S^2$ boundaries labeled $j'$ are oppositely oriented compared to the boundaries labeled $j$ (where $j=3, 4, 5$) and can be glued together resulting in a two boundary state.}
\label{Psi3-app}
\end{figure}

The state $|{\Psi_3}\rangle$ given in eq.(\ref{Psi3state}) can be obtained by considering the state $|{\Psi_2}\rangle$ of eq.(\ref{psi1}) on four $S^2$ boundaries and gluing the extra two copies of $S^2$ with appropriate quantum states as shown in figure \ref{Psi3-app}. The boundaries labeled $3'$, $4'$ and $5'$ have opposite orientations to that of boundaries labeled $3$, $4$ and $5$ respectively. Gluing these oppositely oriented boundaries as shown in the figure will give the final state which is a two boundary state.
The quantum state $\beta$ in figure \ref{Psi3-app} can be obtained by acting the $b_2$ braiding operator $2p$ times on one of the basis states:
\begin{align}
\beta &=  \sum_{t,\,r_1,\,r_2} \langle{\phi_{t,\,r_1 r_2}^{(3')}} | \otimes \left(b_2^{(-)}\right)^{2p} |{\phi_{t,\,r_1 r_2}^{(4)}}\rangle  \nonumber \\
&= \sum_{t,\,r_1,\,r_2} \sum_{s, u, v} \, a_{s,\,u v}^{t,\,r_1 r_2}\left[
\begin{array}{cc}
 R_{1} & \bar{R}_{1} \\
 R_{2} & \bar{R}_{2}
\end{array}
\right] \langle{\phi_{t,\,r_1 r_2}^{(3')}} | \otimes \left(b_2^{(-)}\right)^{2p_1} |{\hat{\phi}_{s,\,u v}^{(4)}}\rangle ~,
\end{align}
where we have used the transformation rules of eq.(\ref{basis-transform}) in the second equation to change the basis, so that the $b_2$ operator can be applied. Here $t \in (R_1 \otimes \bar{R}_1) \cap (R_2 \otimes \bar{R}_2)$ and $r_1, r_2$ are the corresponding multiplicity labels. Similarly $s \in (\bar{R}_1 \otimes {R}_2)$ with $u, v$ denoting the multiplicity labels. Using the braiding eigenvalues from eq.(\ref{B-eigenvalue}), we obtain:\footnote{Since the braiding operator is acting even number of times, the phase factor $\{\bar{R}_1, R_2, s, u\}$ in the eigenvalue of $b_2$ given in eq.(\ref{B-eigenvalue}) is raised to even power and hence becomes 1.}
\begin{align}
\beta = \sum_{\substack{t,\,r_1,\,r_2\\s,\,u,\,v}} a_{s,\,uv}^{t,\,r_1 r_2}\left[
\begin{array}{cc}
 R_{1} & \bar{R}_{1} \\
 R_{2} & \bar{R}_{2}
\end{array}
\right] \exp\left(\frac{2\pi i\, p\, (C_{s} - C_{\bar{R}_1} - C_{R_2})}{k+N}\right) \langle{\phi_{t,\,r_1 r_2}^{(3')}} | \otimes |{\hat{\phi}_{s,\,u v}^{(4)}}\rangle ~.
\end{align}
The quantum state $\alpha$ in figure \ref{Psi3-app} can be simply given as:
\begin{equation}
\langle{\alpha} | = \langle{\phi_{0,\,11}^{(4')}} | = \sum_{s', u', v'} a_{s',\,u'v'}^{0,\,11}\left[
\begin{array}{cc}
 R_{1} & \bar{R}_{1} \\
 R_{2} & \bar{R}_{2}
\end{array}
\right]^{*} \langle{\hat{\phi}_{s',\,u' v'}^{(4')}} | ~.
\end{equation}
Gluing of the boundaries $4$ and $4'$ can be achieved by taking the inner products of quantum states $\alpha$ and $\beta$ which gives, 
\begin{align}
\langle{\gamma} | \equiv \langle \alpha | \beta \rangle = \sum_{t,\,r_1,\,r_2} F(t, r_1, r_2, p) \, \langle{\phi_{t,\,r_1 r_2}^{(3')}} | ~,
\end{align}
where we have defined:
\begin{align}
F(t, r_1, r_2, p) = \sum_{s,u,v} \, a_{s,\,uv}^{0,\,11}\left[
\begin{array}{cc}
 R_{1} & \bar{R}_{1} \\
 R_{2} & \bar{R}_{2}
\end{array}
\right]^{*} a_{s,\,uv}^{t,\,r_1 r_2}\left[
\begin{array}{cc}
 R_{1} & \bar{R}_{1} \\
 R_{2} & \bar{R}_{2}
\end{array}
\right] \exp\left(\frac{2\pi i\, p (C_{s} - C_{\bar{R}_1} - C_{R_2})}{k+N}\right) \nonumber ~.
\end{align}
The state $\langle \mu |$ in figure \ref{Psi3-app} is written as,
\begin{equation}
\langle{\mu} | = \langle{\hat{\phi}_{0,\,1 1}^{(5')}} | = \sum_{t', r_3', r_4'} a_{0,\,11}^{t',\,r_3' r_4'}\left[
\begin{array}{cc}
 R_{4} & \bar{R}_{3} \\
 R_{3} & \bar{R}_{4}
\end{array}
\right] \langle{{\phi}_{t',\,r_3' r_4'}^{(5')}} | ~.
\end{equation}
The four boundary state in figure \ref{Psi3-app} is the $|{\Psi_2}\rangle$ of eq.(\ref{psi1}) for $n=4$ with appropriate representations and can be written as:
\begin{align}
|{\Psi_2}\rangle &= \sum_{w,\,y_1,y_2,y_3,y_4} \frac{ \{R_1, \bar{R}_1, w, y_1\} \{R_2, \bar{R}_2, w, y_2\} \{R_4, \bar{R}_3, w, y_3\} \{R_4, \bar{R}_3, w, y_4\}}{\text{dim}_q w} \nonumber \\
&\times \,\,  |{\phi_{w,\,y_4 y_1}^{(1)},\, \phi_{w,\,y_1 y_2}^{(3)},\, \phi_{w,\,y_2 y_3}^{(2)},\, \phi_{w,\,y_3 y_4}^{(5)}}\rangle ~,
\end{align}
where $w \in (R_1 \otimes \bar{R}_1) \cap (R_2 \otimes \bar{R}_2) \cap (R_4 \otimes \bar{R}_3) \cap (R_4 \otimes \bar{R}_3)$ and $y_1, y_2, y_3, y_4$ are corresponding multiplicity labels. The final state can be obtained by gluing the boundary $3'$ with 3 and $5'$ with 5 which can be achieved by taking the inner product $\langle \gamma, \mu | \Psi_2 \rangle$. Using the following property of the Racah matrix
\begin{align}
a_{s,\,uv}^{0,\,11}\left[
\begin{array}{cc}
 P & \bar{P} \\
 Q & \bar{Q}
\end{array}
\right] &= \frac{\{Q\} \, \{\bar{P},Q, s, u \}}{\sqrt{(\text{dim}_q \, P) (\text{dim}_q \, Q)}} \,\, \sqrt{\text{dim}_q \, s} \,\,\delta_{uv} \nonumber \\
a_{0,\,11}^{t,\,ij}\left[
\begin{array}{cc}
 P & Q \\
 \bar{Q} & \bar{P}
\end{array}
\right] &= \frac{\{Q\} \, \{P,Q,t,i \}}{\sqrt{(\text{dim}_q \, P) (\text{dim}_q \, Q)}} \,\, \sqrt{\text{dim}_q \, t} \,\,\delta_{ij} ~,
\label{Racah-prop}
\end{align}
the inner product gives the required state: 
\begin{equation}
\boxed{|{\Psi_3}\rangle = \sum_{t,\, r_1,r_2,r_3} \frac{ \{R_1, \bar{R}_1, t, r_1\} \{R_2, \bar{R}_2, t, r_2\} \{R_4, \bar{R}_3, t, r_3\} \{R_2\}\{R_3\}}{\sqrt{\text{dim}_q t} \,\, \sqrt{\text{dim}_q R_1 \, \text{dim}_q R_2\,\text{dim}_q R_3 \, \text{dim}_q R_4} } F(t, r_1, r_2, p) |{\phi_{t,\,r_3 r_1}^{(1)},\, \phi_{t,\,r_2 r_3}^{(2)}}\rangle} \nonumber ~
\end{equation}
and the function $F$ can be further simplified using eq.(\ref{Racah-prop}) as,
\begin{align}
F(t, r_1, r_2, p) = \sum_{s,u}\, \{\bar{R}_1, R_2, s, u\}\sqrt{\text{dim}_q s}\, a_{s,\,uu}^{t,\,r_1 r_2}\left[
\begin{array}{cc}
 R_{1} & \bar{R}_{1} \\
 R_{2} & \bar{R}_{2}
\end{array}
\right] \exp\left(\frac{2\pi i\, p\, (C_{s} - C_{\bar{R}_1} - C_{R_2})}{k+N}\right) ~. \nonumber
\end{align}
This is the state given in eq.(\ref{Psi3state}) where the factors like $\{R_{2}\}$, $\{R_{3}\}$ and $(\text{dim}_q \, R_{j})$ are constant terms and will be canceled out while normalizing $|{\Psi_3}\rangle$.

\begin{figure}[h]
\centerline{\includegraphics[width=6.2in]{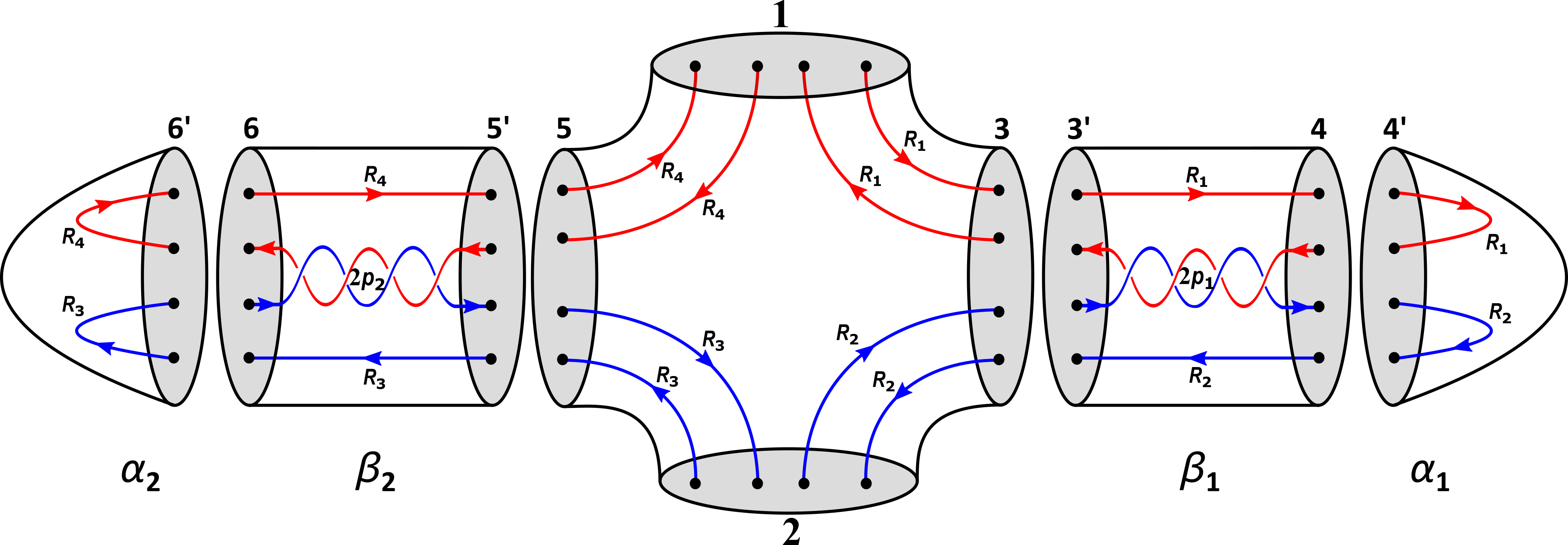}}
\caption[]{Figure showing how to construct the state $|{\Psi_4}\rangle$ on $S^2 \cap S^2$. The $S^2$ boundaries labeled $j'$ are oppositely oriented compared to the boundaries labeled $j$ (where $j=3, 4, 5, 6$) and can be glued together resulting in a two boundary state.}
\label{Psi3-2bdy}
\end{figure}

\begin{figure}[h]
\centerline{\includegraphics[width=5.85in]{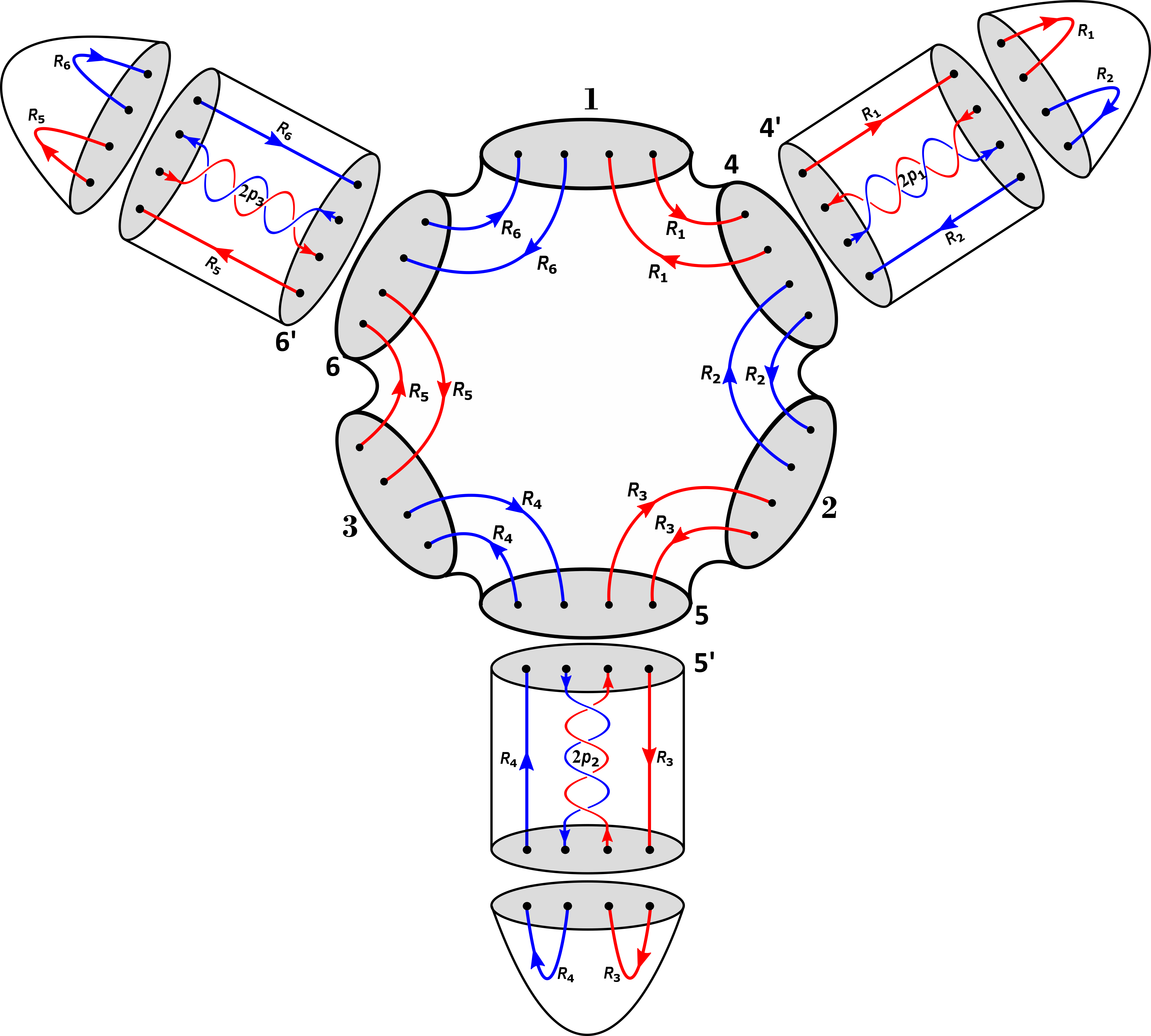}}
\caption[]{Figure showing the construction of the state $|{\Psi_4}\rangle$ for $n=3$. The $S^2$ boundaries labeled $4', 5'$ and $6'$ are oppositely oriented compared to $4, 5$ and $6$ respectively and can be glued together resulting in a three boundary state.}
\label{Psi3-3bdy}
\end{figure}

Next we want to compute the state $|{\Psi_4}\rangle$ given in eq.(\ref{Psi3-nbdy}) on $n$ number of boundaries. For this, we need to consider the state $|{\Psi_2}\rangle$ of eq.(\ref{psi1}) on $2n$ copies of $S^2$ and glue the extra $n$ copies with $S^2$ boundaries of appropriate quantum states. We will show the computation of $|{\Psi_4}\rangle$ for $n=2$ and $n=3$ from where it can be generalized to any $n$. The figure \ref{Psi3-2bdy} gives the construction of $|{\Psi_3}\rangle$ for $n=2$. The boundaries labeled $3'$, $4'$, $5'$, and $6'$ have opposite orientations to that of boundaries labeled $3$, $4$, $5$, and $6$ respectively. Gluing these oppositely oriented boundaries as shown in the figure will give the final state which is a two boundary state.
The quantum state $\beta_1$ is similar to the state $\beta$ shown in the figure \ref{Psi3-app} and can be directly written:
\begin{equation}
\beta_1 = \sum_{\substack{t,\,r_1,\,r_2\\s_1,\,u_1,\,v_1}} a_{s_1,\,u_1v_1}^{t,\,r_1 r_2}\left[
\begin{array}{cc}
 R_{1} & \bar{R}_{1} \\
 R_{2} & \bar{R}_{2}
\end{array}
\right] \exp\left(\frac{2\pi i\, p_1 (C_{s_1} - C_{\bar{R}_1} - C_{R_2})}{k+N}\right) \langle{\phi_{t,\,r_1 r_2}^{(3')}} | \otimes |{\hat{\phi}_{s_1,\,u_1 v_1}^{(4)}}\rangle ~,
\end{equation}
where $t \in (R_1 \otimes \bar{R}_1) \cap (R_2 \otimes \bar{R}_2)$ and $r_1, r_2$ are the corresponding multiplicity labels. Similarly $s_1 \in (\bar{R}_1 \otimes {R}_2)$ with $u_1, v_1$ denoting the multiplicity labels. The quantum state $\alpha_1$ is same as the state $\alpha$ in the figure \ref{Psi3-app}:
\begin{equation}
\langle{\alpha_1} | = \langle{\phi_{0,\,11}^{(4')}} | = \sum_{s_1, u_1, v_1} a_{s_1,\,u_1v_1}^{0,\,11}\left[
\begin{array}{cc}
 R_{1} & \bar{R}_{1} \\
 R_{2} & \bar{R}_{2}
\end{array}
\right]^{*} \langle{\hat{\phi}_{s,\,u v}^{(4')}} | ~.
\end{equation}
Gluing of the boundaries $4$ and $4'$ by taking the inner products of states $\alpha_1$ and $\beta_1$ gives, 
\begin{align}
\langle{\gamma_1} | \equiv \langle \alpha_1 | \beta_1 \rangle = \sum_{t,\,r_1,\,r_2} F_1(t, r_1, r_2, p_1) \langle{\phi_{t,\,r_1 r_2}^{(3')}} | ~,
\end{align}
where we have defined:
\begin{align}
F_1(t, r_1, r_2, p_1) = \sum_{s_1,u_1,v_1} \, a_{s_1,\,u_1v_1}^{0,\,11}\left[
\begin{array}{cc}
 R_{1} & \bar{R}_{1} \\
 R_{2} & \bar{R}_{2}
\end{array}
\right]^{*} a_{s_1,\,u_1v_1}^{t,\,r_1 r_2}\left[
\begin{array}{cc}
 R_{1} & \bar{R}_{1} \\
 R_{2} & \bar{R}_{2}
\end{array}
\right] \exp\left(\frac{2\pi i\, p_1 (C_{s_1} - C_{\bar{R}_1} - C_{R_2})}{k+N}\right) \nonumber ~.
\end{align}
Similarly, the state $\bra{\gamma_2}$ can be obtained by gluing boundaries $6$ and $6'$:
\begin{align}
\langle{\gamma_2} | \equiv \langle \alpha_2 | \beta_2 \rangle = \sum_{t',\,r_3,\,r_4} F_2(t', r_3, r_4, p_2) \langle{\phi_{t',\,r_3 r_4}^{(5')}} | ~,
\end{align}
where $t' \in (R_3 \otimes \bar{R}_3) \cap (R_4 \otimes \bar{R}_4)$ with $r_3, r_4$ as the corresponding multiplicity labels and the function $F_2$ is defined as:
\begin{align}
F_2(t', r_3, r_4, p_2) = \sum_{s_2,u_2, v_2} \, a_{s_2,\,u_2v_2}^{0,\,11}\left[
\begin{array}{cc}
 R_{3} & \bar{R}_{3} \\
 R_{4} & \bar{R}_{4}
\end{array}
\right]^{*} a_{s_2,\,u_2v_2}^{t',\,r_3 r_4}\left[
\begin{array}{cc}
 R_{3} & \bar{R}_{3} \\
 R_{4} & \bar{R}_{4}
\end{array}
\right] \exp\left(\frac{2\pi i\, p_2 (C_{s_2} - C_{\bar{R}_3} - C_{R_4})}{k+N}\right) \nonumber ~.
\end{align}
Here representation $s_2 \in (\bar{R}_3 \otimes R_4)$ with $u_2$ and $v_2$ as the multiplicity labels. The four boundary state in figure \ref{Psi3-2bdy} can be written using eq.(\ref{psi1}) as:
\begin{equation}
|{\Psi_2}\rangle = \sum_{w,\,y_1,y_2,y_3,y_4} \frac{\prod_{j=1}^4 \{R_j, \bar{R}_j, w, y_j\}}{\text{dim}_q w} |{\phi_{w,\,y_4 y_1}^{(1)},\, \phi_{w,\,y_1 y_2}^{(3)},\, \phi_{w,\,y_2 y_3}^{(2)},\, \phi_{w,\,y_3 y_4}^{(5)}}\rangle ~,
\end{equation}
where $w \in (R_1 \otimes \bar{R}_1) \cap (R_2 \otimes \bar{R}_2) \cap (R_3 \otimes \bar{R}_3) \cap (R_4 \otimes \bar{R}_4)$ and $y_1, y_2, y_3, y_4$ are corresponding multiplicity labels. The final state can be obtained by gluing the boundary $3'$ with 3 and $5'$ with 5 which can be achieved by taking the inner product $\langle \gamma_1, \gamma_2 | \Psi_2 \rangle$. Thus we will get the required state:  
\begin{align}
|{\Psi_4^{\text{2bdy}}}\rangle = \sum_{t,\, r_1,r_2,r_3,r_4} \frac{\prod_{j=1}^4 \{R_j, \bar{R}_j, t, r_j\}}{\text{dim}_q t} F_1(t, r_1, r_2, p_1)\, F_2(t, r_3, r_4, p_2) |{\phi_{t,\,r_4 r_1}^{(1)},\, \phi_{t,\,r_2 r_3}^{(2)}}\rangle \nonumber ~.
\end{align}
In a similar way, one can also obtain the state $|{\Psi_4}\rangle$ for $n=3$ as shown in the figure \ref{Psi3-3bdy} by gluing the boundaries of various states. 
The six boundary state is the $|{\Psi_2}\rangle$ of eq.(\ref{psi1}) for $n=6$ and is given as:
\begin{equation}
|{\Psi_2}\rangle = \sum_{t,\,r_1,\ldots,r_6} \frac{\prod_{j=1}^6 \{R_j, \bar{R}_j, t, r_j\}}{(\text{dim}_q t)^2} |{\phi_{t,\,r_6 r_1}^{(1)},\, \phi_{t,\,r_1 r_2}^{(4)},\, \phi_{t,\,r_2 r_3}^{(2)},\, \phi_{t,\,r_3 r_4}^{(5)},\, \phi_{t,\,r_4 r_5}^{(3)},\, \phi_{t,\,r_5 r_6}^{(6)}}\rangle ~.
\end{equation}
To get the desired state, we need to compute the inner product $\langle \gamma_1, \gamma_2, \gamma_3 | \Psi_2 \rangle$ where,
\begin{align}
\langle{\gamma_1, \gamma_2, \gamma_3} | = \sum_{\substack{t_1,\,t_2,\,t_3 \\ x_1,\ldots,\,x_6}} \,\, \prod_{j=1}^3 F_j(t_j, x_{2j-1}, x_{2j}, p_j)\,\, \langle{\phi_{t_1,\,x_1 x_2}^{(4')},\, \phi_{t_2,\,x_3 x_4}^{(5')},\, \phi_{t_3,\,x_5 x_6}^{(6')}} | ~.
\end{align}
Computing the inner product gives the required state:
\begin{align}
|{\Psi_4^{\text{3bdy}}}\rangle = \sum_{t,\, r_1,\ldots,r_6} \frac{\prod_{j=1}^6 \{R_j, \bar{R}_j, t, r_j\}}{(\text{dim}_q t)^2} \prod_{j=1}^3 F_j(t, r_{2j-1}, r_{2j}, p_j) \,\, \bigotimes_{i=1}^3 |{\phi_{t,\,r_{2i-2}\, r_{2i-1}}^{(i)}}\rangle ~.
\end{align}
This procedure can be generalized to $n$ boundaries and we get:
\begin{equation}
\boxed{|{\Psi_4}\rangle = \sum_{t,\, r_1,\ldots,r_{2n}} \frac{\prod_{j=1}^{2n} \{R_j, \bar{R}_j, t, r_j\}}{(\text{dim}_q t)^{n-1}} \prod_{j=1}^n F_j(t, r_{2j-1}, r_{2j}, p_j) \,\, \bigotimes_{i=1}^n |{\phi_{t,\,r_{2i-2}\, r_{2i-1}}^{(i)}}\rangle} ~,
\end{equation}
where the function $F$ is given as (after using eq.(\ref{Racah-prop})),
\begin{align}
F_j(t, r_{2j-1}, r_{2j}, p_j) &= \sum_{s_j,u_j} \, \{\bar{R}_{2j-1}, R_{2j}, s_j, u_j \} \sqrt{\text{dim}_q \, s_j}\,\, a_{s_j,\,u_j v_j}^{t,\,r_{2j-1} r_{2j}}\left[
\begin{array}{cc}
 R_{2j-1} & \bar{R}_{2j-1} \\
 R_{2j} & \bar{R}_{2j}
\end{array}
\right] \nonumber \\
& \times \exp\left(\frac{2\pi i\, p_j (C_{s_j} - C_{\bar{R}_{2j-1}} - C_{R_2j})}{k+N}\right) ~.
\end{align}
Here $s_j \in (\bar{R}_{2j-1} \otimes R_{2j})$ with corresponding multiplicity tracking label as $u_j$.
\section{Racah matrix for $(2,1)$ representation of SU($N$)}
\label{appendix-B}
In this section, we tabulate the type I Racah matrix elements for $R = \text{$\tiny\yng(2,1)$}$ which were obtained in \cite{Gu:2014nva} and are denoted as,
\begin{equation}
a_{t_i,\, r_i, r_i'}^{t_j,\, r_j, r_j'}\left[
\begin{array}{cc}
 R & \bar{R} \\
 R & \bar{R}
\end{array}
\right] ~,
\end{equation}
where $t_i,\, r_i,\, r_i'$ labels the row and $t_j,\,r_j,\, r_j'$ labels the column of the Racah matrix (see \cite{Gu:2014nva} for more details). The representations $t_\ell \in (R \otimes \bar{R}) = t_1 \oplus t_2 \oplus t_3 \oplus t_4 \oplus t_5 \oplus t_6 \oplus 2\,t_7$ and are given as,
\begin{alignat}{3}
t_1 = \bullet, \quad t_2 &= (2^2, 1^{N-4}), &\quad t_3 &= (3, 1^{N-3}), &\quad t_4 &= (3^2, 2^{N-3}) \nonumber \\
t_5 &= (4, 2^{N-2}), & t_6 &= (4, 3, 2^{N-4}, 1), & t_7 &= (2, 1^{N-2}) ~.
\end{alignat}
Since only the representation $t_7$ appears twice, the Racah matrix (which we shall denote by $\mathcal{R}$) will be of order 10, where the ten rows are labeled as: 
\begin{equation}
t_1,\, t_2,\, t_3,\, t_4,\, t_5,\, t_6,\, (t_7)_{11},\, (t_7)_{22},\, (t_7)_{12},\, (t_7)_{21} ~.
\end{equation}
We will also use the same ordering for the columns. The elements of the Racah matrix $\mathcal{R}$ are given as,
\begin{equation}
\mathcal{R}_{\alpha \beta} = \sqrt{\text{dim}_q t_{\alpha}} \, \sqrt{\text{dim}_q t_{\beta}}\,\, \mathcal{Q}_{\alpha \beta} ~,
\label{Rac-(2,1)}
\end{equation}
where $\alpha$ and $\beta$ take values from 1 to 10 with the identification $t_8 = t_9 = t_{10} = t_7$. The elements $\mathcal{Q}_{\alpha \beta}$ can be organized into a $10 \times 10$ matrix which is given as:
\begin{equation}
\mathcal{Q} = \left(
\begin{array}{c|cccccccccc}
 &  t_1 & t_2 & t_3 & t_4 & t_5 & t_6 & (t_7)_{11} & (t_7)_{22} & (t_7)_{12} & (t_7)_{21} \\ \hline
t_1 &  \Lambda_1 & \Lambda_1 & -\Lambda_1 & -\Lambda_1 & \Lambda_1 & \Lambda_1 & \Lambda_1 & -\Lambda_1 & 0 & 0 \\
t_2 & \Lambda_1 & \Lambda_{14} & \Lambda_{16} & \Lambda_{16} & \Lambda_{11} & \Lambda_{18} & \Lambda_{23} & \Lambda_{10} & -\Lambda_6 & -\Lambda_6 \\
t_3 & -\Lambda_1 & \Lambda_{16} & \Lambda_{15} & \Lambda_{15} & \Lambda_{12} & 0 & \Lambda_{22} & \Lambda_3 & -\Lambda_7 & -\Lambda_8 \\
t_4 & -\Lambda_1 & \Lambda_{16} & \Lambda_{15} & \Lambda_{15} & \Lambda_{12} & 0 & \Lambda_{22} & \Lambda_3 & -\Lambda_8 & -\Lambda_7 \\
t_5 & \Lambda_1 & \Lambda_{11} & \Lambda_{12} & \Lambda_{12} & \Lambda_{13} & \Lambda_{17} & \Lambda_{21} & \Lambda_9 & -\Lambda_5 & -\Lambda_5 \\
t_6 & \Lambda_1 & \Lambda_{18} & 0 & 0 & \Lambda_{17} & \Lambda_{19} & \Lambda_{11} & 0 & 0 & 0 \\
(t_7)_{11} & \Lambda_1 & \Lambda_{23} & \Lambda_{22} & \Lambda_{22} & \Lambda_{21} & \Lambda_{11} & \Lambda_{20} & \Lambda_4 & \Lambda_2 & \Lambda_2 \\
(t_7)_{22} & -\Lambda_1 & \Lambda_{10} & \Lambda_3 & \Lambda_3 & \Lambda_9 & 0 & \Lambda_4 & \Lambda_3 & 0 & 0 \\
(t_7)_{12} & 0 & \Lambda_6 & \Lambda_8 & \Lambda_7 & \Lambda_5 & 0 & -\Lambda_2 & 0 & \Lambda_3 & \Lambda_3 \\
(t_7)_{21} & 0 & \Lambda_6 & \Lambda_7 & \Lambda_8 & \Lambda_5 & 0 & -\Lambda_2 & 0 & \Lambda_3 & \Lambda_3 \\
\end{array}
\right) ~.
\end{equation}
The various terms used here are given in the following:
\begin{align}
\Lambda_1 &= \tfrac{[3]}{\omega},\, \Lambda_2 = \tfrac{i[2 N]}{\omega\sqrt{\mu}[N]},\, \Lambda_3 = \tfrac{-1}{\omega},\, \Lambda_4 =\tfrac{-[4]}{[2]\omega},\, \Lambda_5 = \tfrac{i\sqrt{[N-2]}[N+1]}{\omega [N]\sqrt{[N+2]}},\, \Lambda_6 = \tfrac{-i[N-1]\sqrt{[N+2]}}{\omega \sqrt{[N-2]}[N]} \nonumber \\
\Lambda_7 &= \tfrac{i[N+1]}{\omega \sqrt{\mu}},\, \Lambda_8 = \tfrac{-i[N-1]}{\omega \sqrt{\mu}},\, \Lambda_9 = \tfrac{-[N-2]}{\omega [N]} ,\, \Lambda_{10} = \tfrac{-[N+2]}{\omega [N]},\, \Lambda_{11} = \tfrac{-[3]^2}{\omega \mu},\, \Lambda_{12} = \tfrac{[3][N-2]}{\omega \mu} \nonumber \\
\Lambda_{13} &= \tfrac{[2]^2[N-1]+[2]^2[N+1]-[N-4][N][N+3]}{[N-2][N-1][N]^3[N+1][N+2][N+3]},\, \Lambda_{14} = \tfrac{[2]^2[N-1]+[2]^2[N+1]-[N+4][N][N-3]}{[N-3][N-2][N-1][N]^3[N+1][N+2]},\, \Lambda_{15} = \tfrac{[3]}{\omega \mu} \nonumber \\
\Lambda_{16} &= \tfrac{[3][N+2]}{\omega \mu [N]},\, \Lambda_{17} = \tfrac{[2][3]^2}{\omega \mu [N][N+3]},\, \Lambda_{18} = \tfrac{[2][3]^2}{\omega \mu [N][N-3]},\, \Lambda_{19} = \tfrac{-[3]^3[N]}{\omega^2 \mu [N+3][N-3]} \nonumber \\
\Lambda_{20} &= \tfrac{[N][N+1][N+2] \, X_{20}}{\omega^3 \mu^2 [2][N][N+1]^2+\omega^4 \mu [2][N+2]^2},\, \Lambda_{21} = \tfrac{[N][N+1] \, X_{21}}{\omega^3 \mu [2]},\, \Lambda_{22} = \tfrac{X_{22}}{\omega^2 \mu [2]},\, \Lambda_{23} = \tfrac{X_{23}}{\omega^2 \mu [2][N+1]}
\end{align}
where we have defined $\omega = [N-1][N][N+1]$ and $\mu = [N-2][N+2]$, with $[x]$ denoting the $q$-number defined in eq.(\ref{q-number}). The other factors involved in above equation are given below: 
\begin{align}
X_{20} &= [2]^2 [3]^2 [2N]^2 - \left( \tfrac{2 \omega^4 \mu [3]}{[N]^3 [N+1]^3} + \tfrac{2 \omega^4 \mu [3]}{[N]^3 [N-1]^3} - \tfrac{\omega^3 \mu^2 [2][N+2]}{[N]^2 [N+1]^2} - \tfrac{\omega^3 \mu^2 [2][N-2]}{[N]^2 [N-1]^2} \right) \left( \tfrac{[N+2]}{[N+1]} + \tfrac{[N-2]}{[N-1]} \right) \nonumber \\
&+ \omega^2 \mu [2] [3]^2 \left( \tfrac{[N-3]}{[N-2]} + \tfrac{[N+3]}{[N+2]} \right) \nonumber \\
X_{21} &= -[2]^2 [3]^2 - \tfrac{2 \omega^2 \mu [3]^2}{[N]^2 [N+1]^2} - \tfrac{2 \omega^2 \mu [3]}{[N]^2 [N-1]^2} + \tfrac{\mu^3 [2] [N+1]}{[N+2]} + \tfrac{\omega \mu [3]^2 [N-3]}{[N+1] [N-2]} + \tfrac{\omega \mu [3]^2 [N+3]}{[N+1] [N+2]} \nonumber \\
X_{22} &= \mu [2][N]([3]^2 - \mu) + \mu [3][N-3]\left( 1 - \tfrac{[3][N]}{[N-2]} \right) + \mu [3][N+3]\left( 1 - \tfrac{[3][N]}{[N+2]} \right) \nonumber \\
X_{23} &= -[2]^2 [3]^2 - \tfrac{2 \omega^2 \mu [3]^2}{[N]^2 [N-1]^2} - \tfrac{2 \omega^2 \mu [3]}{[N]^2 [N+1]^2} + \tfrac{\mu^3 [2] [N-1]}{[N-2]} + \tfrac{\omega \mu [3]^2 [N-3]}{[N-1] [N-2]} + \tfrac{\omega \mu [3]^2 [N+3]}{[N-1] [N+2]} ~.
\end{align}
\bibliographystyle{JHEP}
\bibliography{S2+S2}

\end{document}